\documentclass[journal=ancham,manuscript=article]{achemso}

\usepackage{chemformula} 
\usepackage[T1]{fontenc} 
\usepackage{amsmath}
\usepackage{upgreek}
\usepackage[symbol]{footmisc}
\usepackage{color}
\usepackage{wasysym}
\usepackage[version=3]{mhchem}
\usepackage{xr-hyper}
\usepackage{siunitx}

\title{\large The Acoustophotoelectric Effect: Efficient Phonon--Photon--Electron Coupling in Zero-Voltage-Biased 2D SnS\textsubscript{2} for Broadband Photodetection}
\author{Hossein Alijani}
\affiliation{Micro/Nanophysics Research Laboratory, RMIT University, Melbourne, VIC 3001, Australia}
\author{Philipp Reineck}
\affiliation{ARC Centre of Excellence for Nanoscale BioPhotonics, School of Science, RMIT University, Melbourne, VIC 3001, Australia}
\author{Robert Komljenovic}
\affiliation{Micro/Nanophysics Research Laboratory, RMIT University, Melbourne, VIC 3001, Australia}
\author{Salvy Russo}
\affiliation{ARC Centre of Excellence in Exciton Science, School of Science, RMIT University, Melbourne, VIC 3001, Australia}
\author{Mei Xian Low}
\affiliation{School of Engineering, RMIT University, Melbourne, VIC 3001, Australia}
\author{Sivacarendran Balendhran}
\affiliation{School of Physics, The University of Melbourne, Parkville, VIC 3010, Australia}
\author{Kenneth Crozier}
\affiliation{School of Physics, The University of Melbourne, Parkville, VIC 3010, Australia}
\alsoaffiliation{Department of Electrical and Electronic Engineering, The University of Melbourne, Parkville, VIC 3010, Australia}
\alsoaffiliation{ARC Centre of Excellence for Transformative Meta-Optical Systems, The University of Melbourne, VIC 3010, Australia}
\author{Sumeet Walia}
\affiliation{School of Engineering, RMIT University, Melbourne, VIC 3001, Australia}
\author{Geoff. R. Nash}
\affiliation{Natural Sciences, Faculty of Environment, Science and Economy, University of Exeter, Exeter EX4 4QF, United Kingdom}
\author{Leslie Y. Yeo}
\affiliation{Micro/Nanophysics Research Laboratory, RMIT University, Melbourne, VIC 3001, Australia}
\author{Amgad R. Rezk}
\affiliation{Micro/Nanophysics Research Laboratory, RMIT University, Melbourne, VIC 3001, Australia}
\email{amgad.rezk@rmit.edu.au}

\keywords{Acoustophotoelectric effect, acoustophotocurrent, surface acoustic waves, tin disulfide, photodetection}
        
\begin{document}

\newpage
\begin{abstract}
Two-dimensional (2D) layered metal dichalcogenides constitute a promising class of materials for photodetector applications due to their excellent optoelectronic properties. The most common photodetectors, which work on the principle of photoconductive or photovoltaic effects, however, require either the application of external voltage biases or built-in electric fields, which makes it challenging to simultaneously achieve high responsivities across broadband wavelength excitation---especially beyond the material's nominal band gap---while producing low dark currents. In this work, we report the discovery of an intricate phonon--photon--electron coupling---which we term the {\em acoustophotoelectric} effect---in SnS$_2$ that facilitates efficient photodetection through the application of 100-MHz-order propagating surface acoustic waves (SAWs). This effect not only reduces the band gap of SnS$_2$, but also provides the requisite momentum for indirect band gap transition of the photoexcited charge carriers, to enable broadband photodetection beyond the visible light range, whilst maintaining pA-order dark currents---remarkably without the need for any external voltage bias. More specifically, we show in the infrared excitation range that it is possible to achieve up to eight orders of magnitude improvement in the material's photoresponsivity compared to that previously reported for SnS\textsubscript{2}-based photodetectors, in addition to exhibiting superior performance compared to most other 2D materials reported to date for photodetection. 

\end{abstract}
\newpage

Two-dimensional (2D) layered metal chalcogenides have drawn significant attention for optoelectronic applications due to their excellent electronic and optical properties.\citep{wang2012electronics,mak2016photonics} Specifically, 2D group IV metal chalcogenides have emerged as strong candidates for photodetectors because of their high absorption coefficient and relatively narrow band gap.\citep{zhou2016booming,wang2019broadband,wang20192d} A number of mechanisms exist by which the absorbed light is converted to electrical signals in these 2D materials, the most common of these being based on photoconductive and photovoltaic effects.\citep{qiu2021photodetectors} While photodetectors based on the photoconductive mechanism can be constructed from a single monolithic material, they require an external DC bias for transfer of the photoexcited charge carriers, resulting in high dark currents, which lower their detectivity. On the other hand, photovoltaic photodetectors are based on heterostructures of p--n junctions with built-in electric fields such that an external DC bias is not required. As such, they produce lower dark currents, but nevertheless involve more complex architectures and suffer from lower responsivities due to their limited active areas.\citep{sett2021engineering} In addition, the photodetection range is limited by the band gap of the materials used. Therefore, a single-material photodetector that is capable of generating low dark currents while maintaining large responsivity, along with the ability to detect broadband illumination beyond the material's nominal band gap, would be of significant interest.\citep{li2021recent}

Recently, SnS\textsubscript{2},\citep{huang2014tin} which is an earth-abundant and environmentally-friendly intrinsic n-type semiconductor possessing a relatively wide band gap (2.1--2.8~eV)\citep{su2015chemical,ahn2015deterministic} and high absorption coefficient ($\approx ~$10$^{4}$~cm$^{-1}$),\citep{giri2019balancing} has shown considerable promise for nanoelectronics,\citep{de2012high,xia2015large,yang2017van} and, in particular, photodetectors,\citep{huang2015highly,su2015chemical,tao2015flexible,xia2015large,wu2016ultrathin,yang2016controllable,zhou2016large,fan2016wavelength,gao2016broadband,li2017two,jia2018thickness,liu2019tunable,tian2020visible,yu2020giant,lei2020thermal,fan2021enhanced,fu2021controllable,shooshtari2021ultrafast,luo2022phase} 
 even in comparison to other 2D materials such as MoS\textsubscript{2} and WS\textsubscript{2}.\cite{jia2018thickness,yu2020giant} In order to improve its photodetection performance, various strategies have been explored, including the application of gate voltages,\citep{zhou2016large,yang2016controllable,ying2019high}  chemical doping,\citep{li2017two,liu2019tunable,fan2021enhanced} sensitization with PbS quantum dots,\citep{gao2016broadband} oxygen plasma treatment,\citep{yu2020giant} surface coating,\citep{jia2018thickness} alloy engineering,\citep{khimani2019alloy,shu2020growth} vacuum operation,\citep{tao2015flexible,zhou2016large,wu2016ultrathin} and heterojunction formation.\citep{tian2020visible,shooshtari2021ultrafast} Among these, chemical doping has been shown to facilitate considerable performance improvements in SnS\textsubscript{2}-based photodetectors. Such schemes however commonly involve the use of aggressive chemical or electrochemical processes that result in poor stability,\citep{liu2019tunable} in addition to irreversibly deteriorating the crystal structure by introducing defects into the material,\citep{yuan2019enhanced} although some doping techniques may not be suitable for wafer-scale 2D semiconductors.\citep{yu2020giant} Moreover, it is important to note that SnS\textsubscript{2}, as with many other semiconductors, is not expected to produce photocurrents when excited by wavelengths with lower energies than its nominal band gap.\citep{lei2020thermal} Furthermore, given that SnS\textsubscript{2} is an indirect semiconductor, an additional momentum source is required, usually in the form of phonon energy, to facilitate the indirect transition across the valence and conduction bands.\citep{wang2012electronics} As such, SnS\textsubscript{2}-based photodetectors, as well as most other photodetectors based on other 2D materials, have been limited to date to operation in the visible region; operation at wavelengths beyond 600~nm, where they exhibit very low photoresponsivities, have therefore rarely been reported.\citep{lei2020thermal}

\begin{figure*}[!ht]
\centering
  \includegraphics[width=\textwidth]{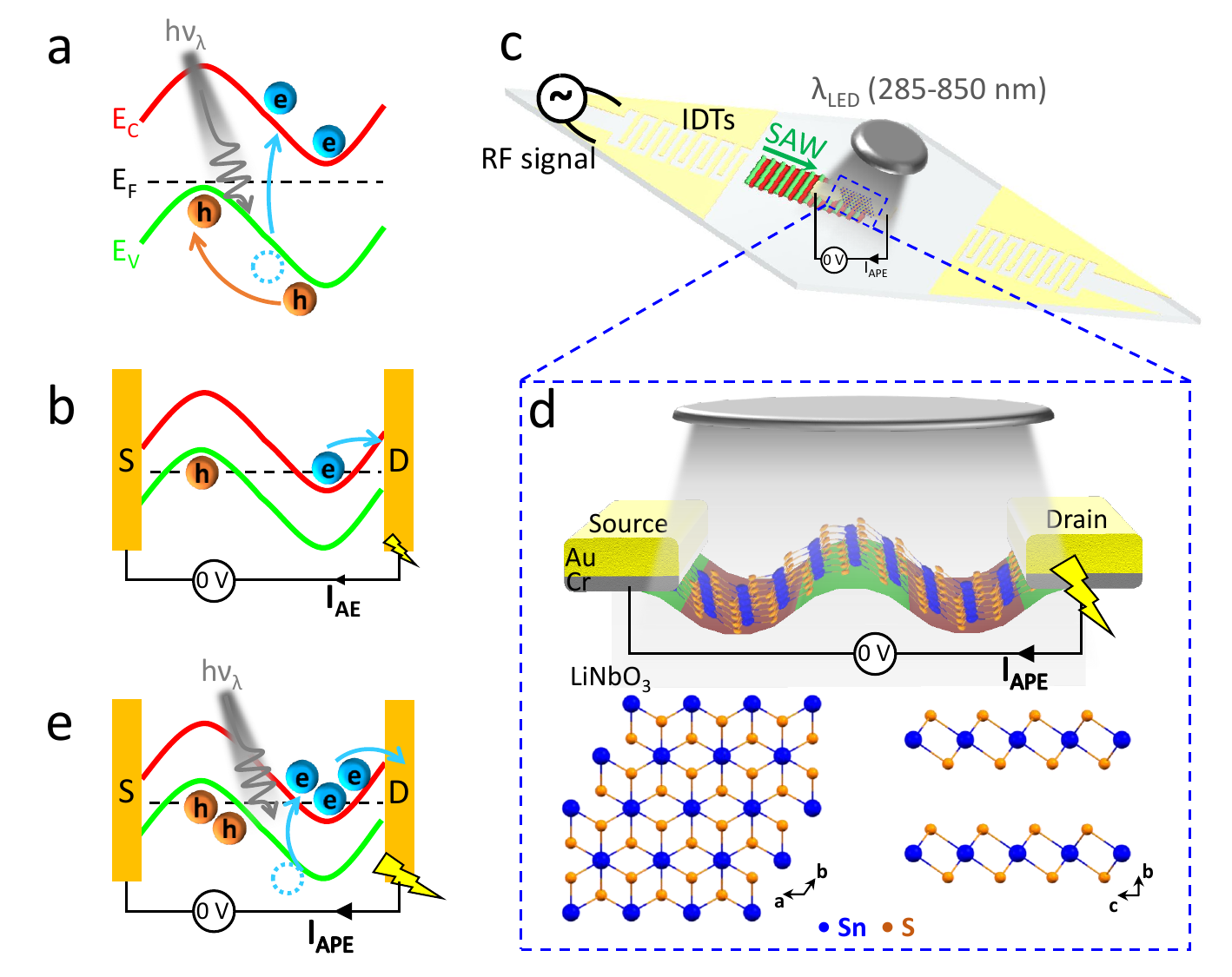}
  \caption{Schematics illustrating (a) the acoustic modulation of photoexcited charge carriers in a semiconducting material, in which the SAW alters the energy bands and hinders the recombination of photoexcited electron--hole pairs ($E_{\rm C}$, $E_{\rm F}$, and $E_{\rm V}$ denote the conduction band, Fermi level, and valence band, respectively), and, (b) the acoustoelectric (AE) effect, in which the SAWs generate an AE current $I_{\rm AE}$ that supports momentum transfer to electrons in the conduction band. (c) Schematic depiction of the experimental setup facilitating the coupling of the SAW, which is generated on and which propagates along a single crystal piezoelectric substrate (LiNbO$_3$), into a 50-nm-thick (approx.) SnS\textsubscript{2} nanoflake. The magnification in (d) shows the deformation of the nanoflake under the SAW excitation in relation to the position of the source and drain contact pads (not to scale); the inset shows top and side views of the SnS\textsubscript{2} atomic structure. (e) Schematic demonstration of the acoustophotoelectric effect (APE), wherein the SAW lowers the band gap, and the momentum transfer from the SAW facilitates transport of the photoexcited charge carriers to the drain to produce an acoustophotoelectric current $I_{\rm APE}$.}
\label{fig:schem}
\end{figure*}

An alternative approach that we postulate could improve the photoresponsivity of SnS\textsubscript{2}, while maintaining its pristine condition, is through the use of MHz-order electromechanical coupling in the form of surface acoustic waves (SAWs)---nanometer-amplitude electromechanical waves that are generated on a single-crystal piezoelectric substrate, which have found application in telecommunication devices, superconducting circuits and acoustofluidics, among others.\citep{delsing2019roadmap,zhang2020acoustic,rezk2021high} In addition to their demonstration as an efficient means for the synthesis and manipulation of 2D transition metal dichalcogenides and carbides/nitrides,\citep{marqus2018increasing,ahmed2018ultrafast,ahmed2019rapid,alijani2021acoustomicrofluidic,ghazaly2021ultrafast,ahmed2022recovery} SAWs have also been shown to facilitate the control and manipulation of charge carriers in nanostructures.\citep{rotter1998giant,kinzel2011directional,zheng2016acoustic} Broadly, SAWs have been used to either (i) manipulate photoexcited charge carriers in 2D semiconductors, to prevent their recombination and therefore allow their transfer over long distances (Fig.~\ref{fig:schem}a),\citep{rezk2015acoustic,rezk2016acoustically,peng2022long} or, (ii) to enhance the electrical output in {\em non-photoexcited} 2D systems (Fig.~\ref{fig:schem}b).\citep{rotter1998giant,rotter1999nonlinear,govorov2000nonlinear,esslinger1992acoustoelectric,rocke1994voltage,ebbecke2004acoustoelectric,bandhu2013macroscopic,lane2018flip,zhao2022acoustically,zheng2018anomalous,preciado2015scalable} The latter effect arises as a consequence of the native SAW electromechanical coupling that facilitates interaction of the wave with the mobile carriers. In a two-dimensional electron system (2DES), for example, this interaction has been shown to alter the SAW velocity and attenuate the wave intensity.\citep{rotter1998giant} The momentum from the wave is then transferred as a force onto the 2DES---a phenomenon known as the acoustoelectric (AE) effect, which has been investigated for a variety of other materials, including GaAs,\citep{rotter1998giant,rotter1999nonlinear} InGaAs,\citep{govorov2000nonlinear} AlGaAs/GaAs heterojunctions,\citep{esslinger1992acoustoelectric,rocke1994voltage} single-walled carbon nanotubes,\citep{ebbecke2004acoustoelectric} graphene,\citep{bandhu2013macroscopic,lane2018flip,zhao2022acoustically} black phosphorus (BP),\citep{zheng2018anomalous} and MoS\textsubscript{2}.\citep{preciado2015scalable} However, enhancement in carrier mobility due to the intricate tripartite coupling of phonons, photons and electrons in a {\em photoexcited} 2D material under the influence of the SAW has yet to be explored. 

In this work, we report the discovery of unique  phonon--photon--electron interactions, which we refer to as the {\em acoustophotoelectric} effect, that arises when the SAW is coupled into a 2D photoexcited SnS\textsubscript{2} nanoflake on a piezoelectric substrate (Fig.~\ref{fig:schem}(c,d)). Such an effect is observed to dramatically enhance the photocurrent output while operating at 0~V external DC bias, yielding a very low pA-order dark current and high photoresponsivity over a broad range of illumination wavelengths from the ultraviolet through to near-infrared regions. Moreover, the acoustophotoelectric effect not only lowers the nominal band gap of SnS\textsubscript{2} from 2.3 to 1.4~eV, as confirmed by density functional theory (DFT) calculations, but also provides the requisite momentum to effect indirect transition of charge carriers into the conduction band, thereby extending photodetection capability considerably beyond the nominal band gap of SnS\textsubscript{2} (approx.~550~nm). Specifically, photoresponsivities up to eight orders of magnitude higher than those of SnS\textsubscript{2}-based photodetectors reported in the literature were achieved in the infrared excitation range. We show too that thinner SnS\textsubscript{2} flakes (approx.~${\cal O}$(10~nm)) exhibit larger acoustophotoelectric effect and hence acoustically-induced photoresponsivity $R$ values compared to thicker ones (approx.~${\cal O}$(100~nm)), across all wavelengths excitations across the UV, visible, and infrared ranges. This can be attributed to the enhanced acoustophotoelectric effect  due to larger phonon--photon--electron coupling as a result of the material's higher innate piezoelectricity (as inferred from its  electromechanical coupling coefficient, $d_{33}$) in thinner films that lead to more efficient interactions between the SAW and the charge carriers in the material. With thicker films, however, the weaker coupling of the SAW into the material results in a diminished acoustophotoelectric effect. In addition, we briefly demonstrate that a similar acoustophotoelectric effect can be induced in other 2D materials, including  MoS\textsubscript{2} and SnS, therefore highlighting the promise of this technique as a practical, reversible, and non-destructive approach to significantly improve the performance of photodetectors based on 2D nanomaterials.

\section{Results and Discussion}
\label{sec:res}

Although the focus of the present work encompasses the interplay between the electromechanical coupling of the SAW with a photoexcited material, \textit{i.e.}, the {\em acoustophotoelectric effect}, we first begin by baselining our observations of the electric current produced under the acoustic forcing but in the {\em absence} of light excitation, \textit{i.e.}, the {\em acoustoelectric effect}; these results will therefore be used as the control case to benchmark the results obtained for the photoexcited cases. Parenthetically, we note though that while the acoustoelectric effect has previously been reported in other 2D materials, such as graphene \citep{poole2015acoustoelectric,poole2018acoustoelectric,bandhu2013macroscopic,lane2018flip,zhao2022acoustically} GaAs,\citep{rotter1998giant} black phosphorus,\citep{zheng2018anomalous} and MoS\textsubscript{2},\citep{preciado2015scalable} it has not been demonstrated to date in any group IV A metal dichalcogenides.

\begin{figure}[!t]
\centering
  \includegraphics[width=0.4\textwidth]{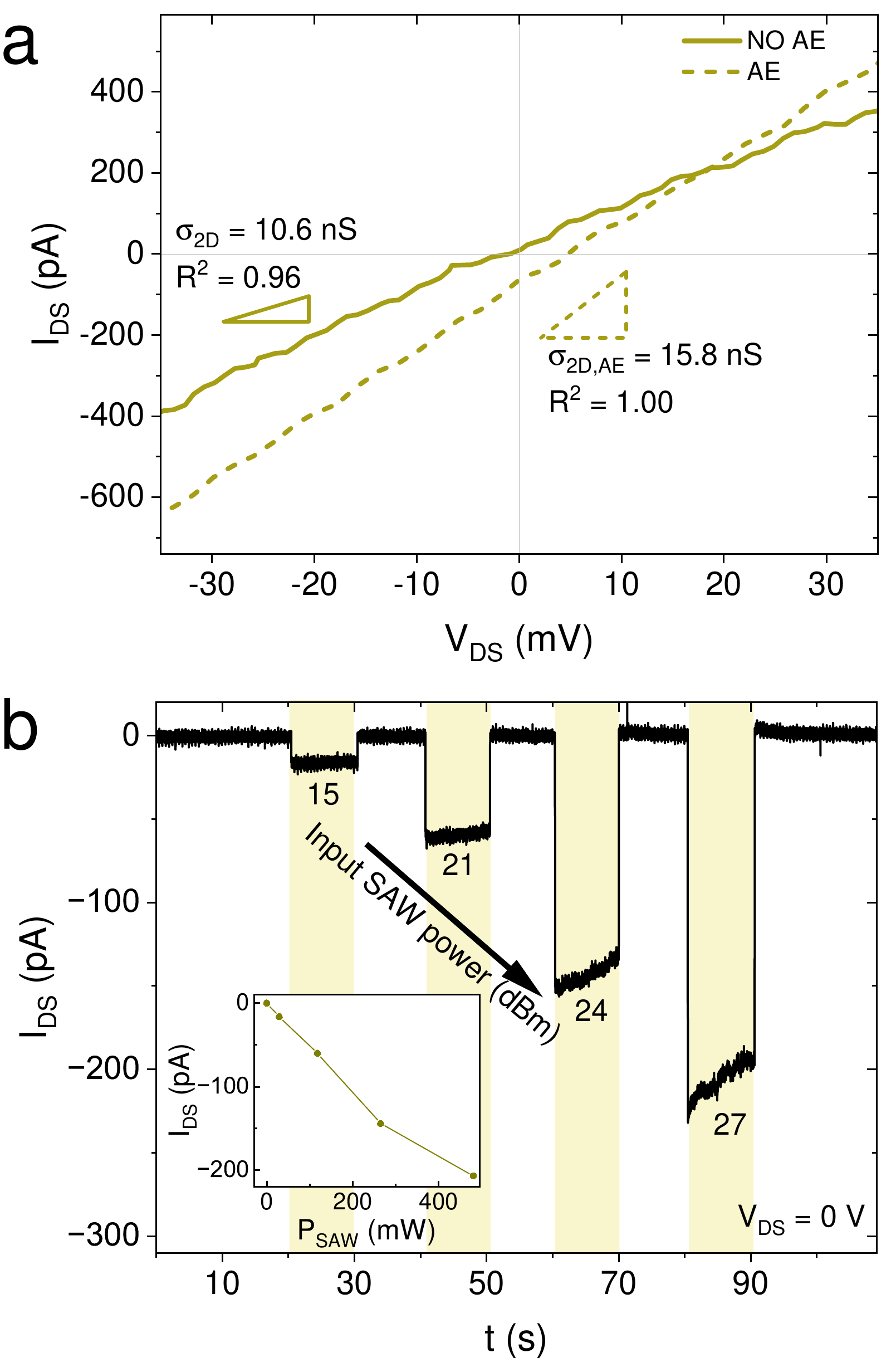}
  \caption{(a) Output current $I_{\rm DS}$ as a function of the applied voltage $V_{\rm DS}$ for a SnS\textsubscript{2} nanoflake, both in the absence and in the presence (21~dBm SAW power) of the acoustoelectric effect (AE). (b) Zero volt DC bias ($V_{\rm DS} = 0$ V) transient response ($I_{\rm DS}$--$t$)  of the SnS\textsubscript{2} nanoflake to the SAW excitation at different input powers; the areas shaded in yellow show the excitation periods and the inset is a plot of the acoustoelectric current $I_{\rm AE}$ as a function of the SAW power $P_{\rm SAW}$ calculated from the $I_{\rm DS}$--$t$ response.}
  \label{fig:ae}
\end{figure}

Figure~\ref{fig:ae}a shows the current--voltage characteristics of a pristine (as verified from the x-ray photoelectron spectra (XPS)  in Fig.~S1) SnS\textsubscript{2} nanoflake with thickness 48.7~$\pm$~1.3~nm (Fig.~S2; considering a monolayer to be 0.6~nm thick, we estimate the flake to be composed of around 80 layers),\cite{song2013high,huang2014tin} from which we calculate the flake's conductivity to be 10.6~nS. The linearity in the profile over the -30 to 30~mV applied voltage range $V_{\rm DS}$ verifies the good Ohmic contact between the flake and the Cr/Au electrical contact pads.\citep{xia2015large,zhou2016large,fan2016wavelength,ying2019high,liu2019tunable,fan2021enhanced} Upon excitation of the SAW, it can be seen that the conductivity increases to 15.8~nS due to the momentum transfer from the acoustic waves to the charge carriers. \citep{bandhu2013macroscopic} More importantly, the application of the SAW can be seen to generate an offset acoustoelectric current of 79.5~pA at $V_{\rm DS}=0$ V. 

The transient acoustoelectric response of the SnS\textsubscript{2} nanoflake to the SAW, obtained by cycling the SAW over four input powers from 15 to 27~dBm in 10~s on/off intervals with  $V_{\rm DS}=0$ V, is plotted in Fig.~\ref{fig:ae}b. It can be seen that the acoustoelectric current increases with increasing SAW power (see the inset of Fig.~\ref{fig:ae}b), consistent with that predicted by the Weinreich equation\citep{weinreich1956acoustodynamic,rotter1998giant,rotter1999charge} (Equation~S.1 in the Supplementary Information) together with the mobility of SnS\textsubscript{2}, calculated from the transfer curves in Fig.~S3. For example, for a given input SAW power of 21~dBm, the theoretically estimated current is 210~pA, which is the same order of the experimental value of 60~pA (obtained from Fig.~\ref{fig:ae}b). 

\begin{figure*}[!ht]
\centering
  \includegraphics[width=\textwidth]{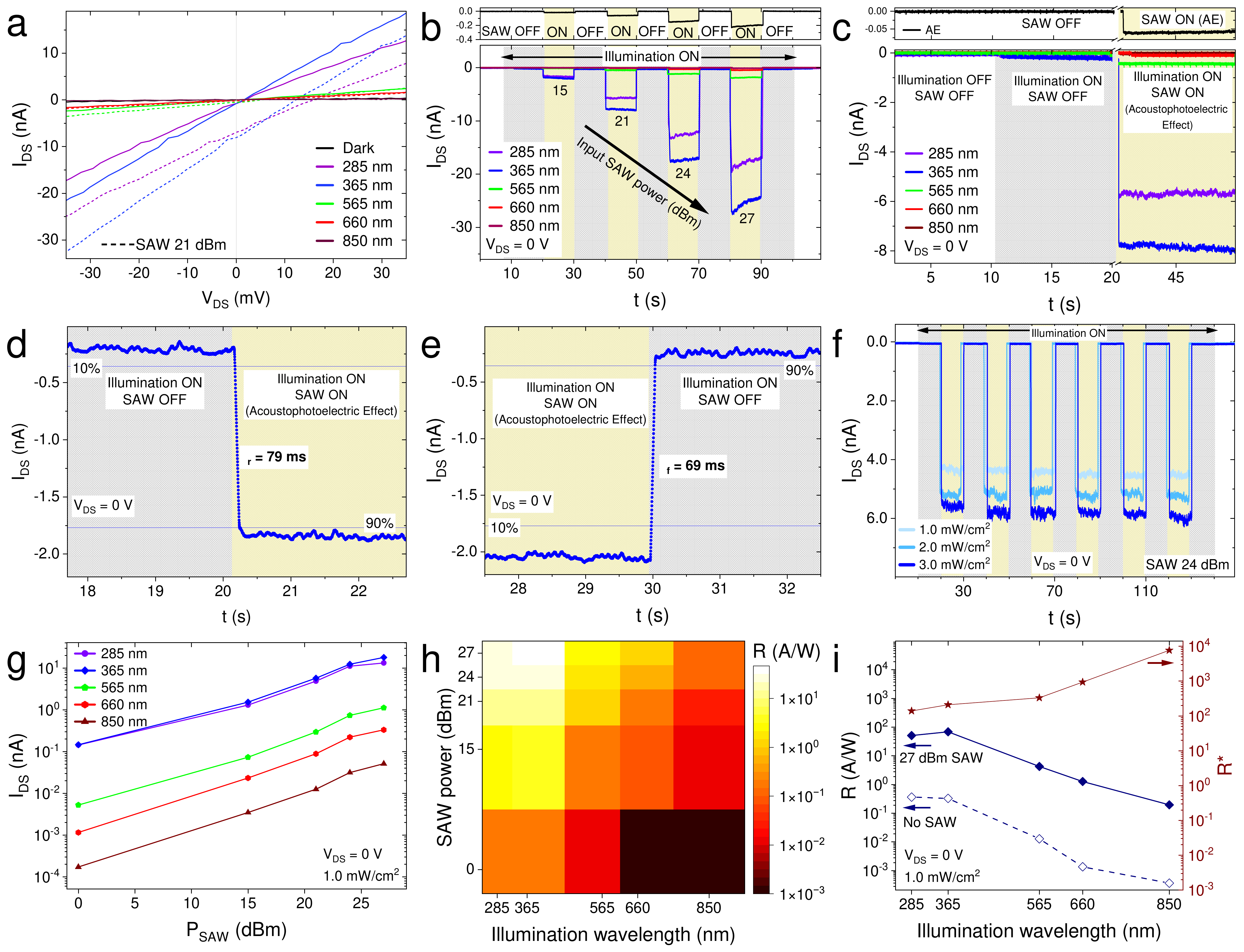}
  \caption{(a) Output current $I_{\rm DS}$ as a function of the applied voltage $V_{\rm DS}$ for a SnS\textsubscript{2} nanoflake under light illumination (3.0~mW/cm$^{2}$ intensity) with different wavelengths, both in the absence (solid lines) and in the presence (21~dBm) of the SAW coupling (dashed lines). (b) Zero volt DC bias ($V_{\rm DS} = 0$ V) transient response ($I_{\rm DS}$--$t$) of the SnS\textsubscript{2} nanoflake under different illumination wavelengths (3.0~mW/cm$^{2}$ intensity) to the SAW excitation at varying powers; the shaded areas show the excitation periods whereas the top panel shows the acoustoelectric (AE) response (no illumination) for comparison. (c) Magnified view of the transient response in (b) corresponding to the start of the illumination period and subsequently the initiation of the SAW excitation (24~dBm). (d) Rise ($\tau_{\rm r}$) and (e) fall ($\tau_{\rm f}$) response times of the  SnS\textsubscript{2} nanoflake to 15~dBm SAW excitation under 365~nm illumination with 3.0~mW/cm$^{2}$ intensity. (f) Cycling characteristics of the SnS\textsubscript{2} nanoflake under 365~nm illumination at different illumination intensities and 24~dBm SAW. (g) Absolute current output I\textsubscript{DS} (with a representative standard error of $\pm$~0.1~nA), and, (h) photoresponsivity $R$ as a function of the input SAW power for different illumination wavelengths with 1.0~mW/cm$^{2}$ intensity. (i) Variation of the photoresponsivity 
in the absence and in the presence of 27~dBm SAW excitation ($R$) and the relative enhancement ($R^{*}$) with illumination wavelength.}
  \label{fig:pae}
\end{figure*}

In contrast, it can be seen from Fig.~\ref{fig:pae}a that a significantly larger current is produced when the SnS\textsubscript{2} nanoflake is illuminated with light (365~nm), even in the absence of the SAW excitation (-20~nA at -30~mV bias compared with -600~pA in the absence of light), attesting to the strong photosensitivity of SnS\textsubscript{2},\citep{jia2018thickness,xia2015large} which we note is the strongest under 285 and 365~nm illumination. The photosensitivity of SnS\textsubscript{2} can be attributed to the commonly known photoconductive effect in which photon absorption in a semiconducting material results in the generation of excess free carriers, thereby leading to an increase in conductivity (\textit{i.e.}, the slope of the $I_{\rm DS}$--$V_{\rm DS}$ curve). A further indication of the transport of photoinduced charge carriers over the electrical contact barriers is also evident from the linearity of the curves. \citep{su2015chemical} 

When excited with the SAW, however, we observe a significant offset in the $I_{\rm DS}$--$V_{\rm DS}$  profile for the same illumination wavelength (365~nm) and intensity (3.0~mW/cm$^{2}$), resulting in a current output of -8~nA at \emph{zero-bias} voltage (Fig.~\ref{fig:pae}a; cf.~-0.1~nA in the absence of the SAW). To further probe the effect of the acoustic coupling, we turn to the current--time ($I_{\rm DS}$--$t$) measurements conducted at zero DC bias in Fig.~\ref{fig:pae}b, wherein the nanoflake is again excited by incrementally increasing SAW powers within 10~s on/off periods, but in the presence of light illumination at different wavelengths. In distinct contrast to the case where the SAW is not present, we see giant enhancements in the photocurrent ($I_{\rm photo} = I_{\lambda} - I_{\rm Dark}$, wherein $\lambda$, $I_{\lambda}$ and $I_{\rm Dark}$ denote the illumination wavelength, and the collected current in the presence and in the absence of light excitation, respectively) output due to the SAW, which can be more evidently seen in Fig.~\ref{fig:pae}c. For comparison, the photocurrent under an illumination wavelength of 365~nm with an intensity of 3.0~mW/cm$^{2}$ is only 135~pA, whereas excitation of the nanoflake with the SAW at 21~dBm input power yields 7.8~nA---an enhancement of almost two orders of magnitude. Importantly, this acoustophotoelectric current is also much larger than the acoustoelectric current, in the absence of light excitation of 60~pA, with the same SAW power (top panels in Fig.~\ref{fig:pae}b,c), alluding to the stronger current generation capacity of the acoustophotoelectric effect over both the typical photoconductive and acoustoelectric effects. 

To verify that the observed acoustophotoelectric current did not arise from the substrate itself, we conducted the same experiment on bare LiNbO\textsubscript{3} in the absence of the SnS\textsubscript{2} flake, in which no change in conductivity was observed (Fig.~S4a) or electrical current produced (Fig.~S4b).

Additionally, it is also possible to rule out the contribution of thermally-generated currents, particularly given that the relatively low applied SAW power over short durations (10~s) did not lead to appreciable temperature rises. Prolonged exposure to the SAW 
at 27~dBm over 30 s yielded a temperature rise in the substrate of only  0.5~$^{\circ}$C, with no observable changes in the acoustophotoelectric current at 365~nm illumination. In any case, we note that the
 thermoelectric performance of SnS\textsubscript{2} is typically hindered by intrinsically low carrier concentrations as a consequence of the material's large band gap and high thermal conductivity.\citep{zhan2020phonon}

We further observe that the SnS\textsubscript{2} nanoflake responds to the SAW excitation with rise and fall times of 79 and 69~ms, respectively, as can be seen from Fig.~\ref{fig:pae}d,e. Moreover, a study of the electrical stability of the nanoflake subject to multiple SAW excitation cycles under illumination (Fig.~\ref{fig:pae}f) indicated no significant deterioration to its performance, at least up to 6 cycles. We note that the acoustophotoelectric effect observed here is not just limited alone to SnS\textsubscript{2}. Similar observations were evident in other 2D materials, for example, MoS\textsubscript{2} and SnS (see Fig.~S5a,b), therefore alluding to the generality of the phenomenon. 

An understanding of how the SAW interacts with the photoexcited carriers is nevertheless crucial to elucidate the underlying mechanisms responsible for the acoustophotoelectric effect. Typically, upon illumination of a semiconducting material, the charge carriers are excited from the valence band (E\textsubscript{V}) to the conduction band (E\textsubscript{C}), following which they tend to recombine to generate a photoluminescent (PL)  signal. It is nevertheless well-known that SAWs are able to prevent the recombination of charge carriers, and quench the photoluminescence in semiconducting materials (Fig.~\ref{fig:schem}a), as has been previously reported, for example, in MoS\textsubscript{2},\citep{rezk2016acoustically,zheng2018acoustically} WSe\textsubscript{2},\citep{peng2022long} GaAs,\citep{volk2010enhanced,kinzel2011directional,hernandez2012acoustically,weiss2016surface} and InGaAs/GaAs.\citep{rocke1997acoustically} Similar PL quenching and spectrum broadening was observed here for the representative SnS\textsubscript{2} nanoflake (Fig.~S6), even beyond its nominal band gap up to 800~nm, thus indicating an alteration of the band structure in the material. In the absence of light excitation, and when a source and drain are present (Fig.~\ref{fig:schem}b), the transfer of momentum from the SAW to the fast conduction electrons generates an acoustoelectric current \citep{parmenter1953acousto} (Eqn.~S.1 (Supplementary Information)), as has been shown for graphene and other 2D materials.\citep{rotter1998giant,lane2018flip,preciado2015scalable,zheng2018anomalous} 
On illumination, however, the photoexcited charge carriers are raised to the conduction band that is altered by the SAW (Fig.~\ref{fig:schem}e), where they accumulate and are transported to the drain. This tripartite phonon--photon--electron coupling is then responsible for the generation of the acoustophotoelectric current, thereby nullifying the need for an external DC bias that inevitably increases the dark currents. This is in contrast to typical photoconductors, where an external DC bias across the source and drain is required to collect the photocurrent. This can, for example, be seen in Fig.~S7 where in the absence of the SAW, a large dark current of 20~nA is introduced with 1~V applied DC bias.  Moreover, long rise and fall times (tens of seconds) can be seen, and little to no change in photocurrent is observed when the SnS\textsubscript{2} nanoflake is excited at longer wavelengths (\textit{e.g.}, 850 nm). In stark contrast, we observe in Fig.~\ref{fig:pae}g low (pA-order) dark currents as well as an increase in the acoustophotoelectric current for all illumination wavelengths, and especially beyond the SnS\textsubscript{2} nominal band gap at 565, 660 and 850~nm, even without an applied DC bias.  

\begin{figure}[!h]
\centering
  \includegraphics[width=0.395\textwidth]{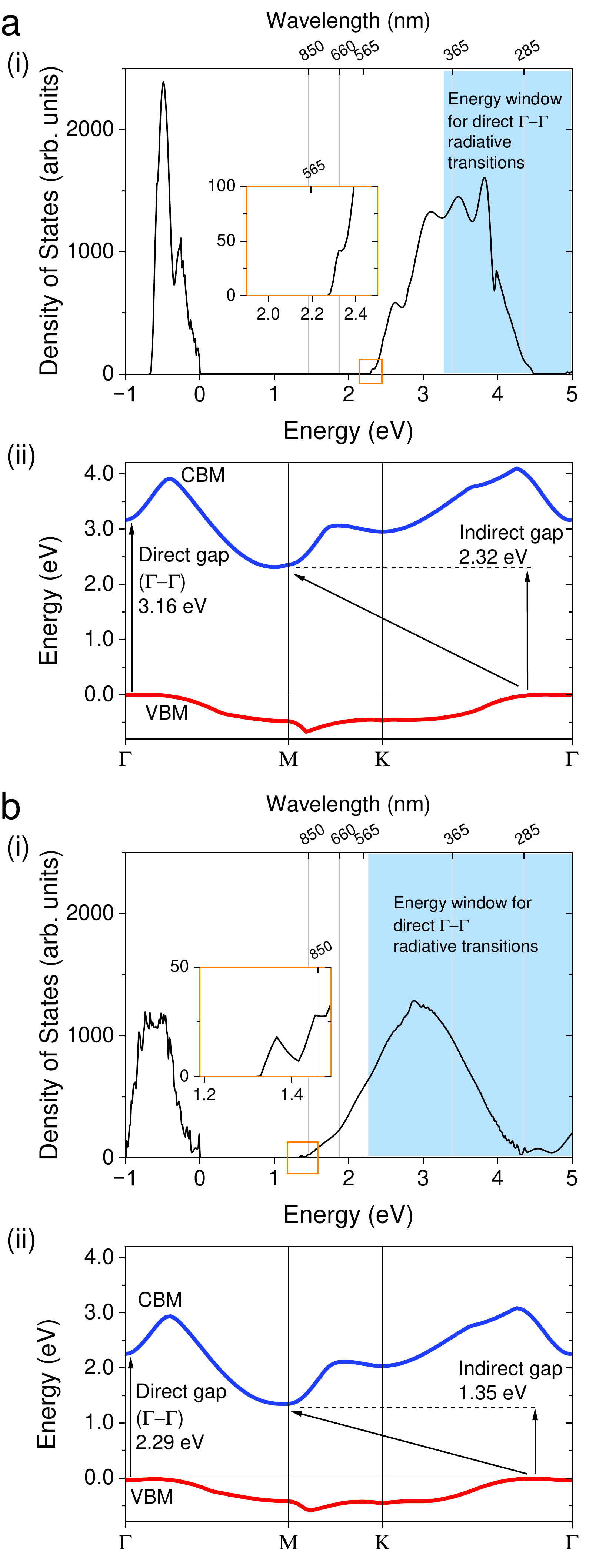}
  \caption{Simulated (i) electronic density of states (eDOS), and, (ii) band structures (BS) of a 20~nm SnS\textsubscript{2} thin film for the case (a) when the electric field and strain are absent ($E = 0$, $\upepsilon = 0$), and, (b) in the presence of both electric field and strain ($E = 3 \times 10^{8}$ V/m, $\upepsilon = 0.1\%$). The insets in the eDOS plots show magnified views of the regions delineated by the orange squares. In the band structures, CBM and VBM denote the conduction band minimum and valence band maximum, respectively.}
  \label{fig:dft}
\end{figure}

Further insight into the acoustophotoelectric enhancement of the current can be gleaned from density functional theory (DFT) simulations\citep{dovesi2014crystal14,dovesi2014crystal142} (see Methods section) of the effect of the electromechanical coupling afforded by the SAW on the electronic properties of a 20 nm SnS\textsubscript{2} (defect-free) thin film. The SAW is approximated in the simulations as combined electric ($E \sim \mathcal{O}$ ($10^{8}$~V/m)) and mechanical strain ($\upepsilon \sim 0.1$\%) fields corresponding to that estimated for the 27 dBm SAW from the experiments.  Figure~\ref{fig:dft} shows the resultant band structure (BS) and total electronic density of states (eDOS) for the case in which the strain and externally applied field are absent (Fig.~\ref{fig:dft}a(i,ii)) and that for an applied external electric field of $3 \times 10^{8}$ V/m and strain of 0.1\% associated with that for 27~dBm SAW power in the experiments (Fig.~\ref{fig:dft}b(i,ii)). A comparison between these plots allows us to conjecture that the electromechanical coupling arising from the SAW enhances the photocurrent at all illumination wavelengths by lowering the band gap (initially 2.32 eV; indirect) in the absence of SAW excitation, as seen from the BS plot in Fig.~\ref{fig:dft}a (which is in good agreement with the experimental optical band gap calculated from the Tauc plot in Fig.~S8), consistent with that observed in Fig.~\ref{fig:pae}g.

To elaborate, it is expected that the degree of photocurrent enhancement is dependent on whether the illumination wavelength corresponds to an energy region for which indirect or direct radiative transition can occur. Both Figs.~\ref{fig:dft} and S8 show that SnS\textsubscript{2} is a semiconductor with an indirect band gap, \textit{i.e.}, it possesses energies lower than the threshold for direct radiative transitions. The enhancement in photocurrent at 285 and 365~nm can be attributed to the efficient transport of the photoexcited carriers since the photon energy is already deeper within the conduction band region, thereby permitting direct radiative transition to occur. For longer wavelength excitation and in contrast to direct semiconductors, additional phonons are required for conservation of energy and crystal momentum in indirect semiconductors; as such, indirect semiconductors usually require larger momentum transfer to effect indirect transitions.\citep{gu2007band,wang2012electronics,bunge2016polymer} Since such transitions involve a two-step process and are strongly dependent on the phonon density of states, the rate at which these transitions occur are typically much slower than those for direct transitions.\citep{cheng2010competitiveness} Nevertheless, we see from  Fig.~\ref{fig:dft}b that even though the photon energy at 565~nm is below that for direct ($\Gamma$--$\Gamma$) radiative transition, detailed analysis of the band structure shows that direct radiative transition is possible along the $\Gamma$--M path of the Brillouin zone due to the effect of the SAW in lowering the band gap of the SnS\textsubscript{2} thin film from 2.32 eV down to 1.35 eV. As such, the enhancement in the photocurrent due to the applied SAW at 565~nm can be attributed to the fact that the photon energy is now sufficient for direct $(\Gamma$--$\Gamma$), as opposed to indirect, transition.

We should note that although band gap tuning of semiconductors has been achieved to date through the application of a static transverse electric field \textit{via} the Franz--Keldysh and Stark effects, the change in the band gap that has been demonstrated through these methods has been confined to only a few hundred meV and at the extra cost of patterning complex electrode structures to include physical top and bottom gates.\citep{chaves2020bandgap} The application of the SAW here therefore not only offers a facile means by which electrode-free electric field perturbation with associated directional strain can be imposed on the material, but also facilitates considerably larger changes in the band gap, in addition to providing the required momentum for indirect transitions. 

In the present case of SnS\textsubscript{2}, the indirect band gap is lowered by 0.97~eV, so much so that the electromechanical coupling of the SAW into the SnS\textsubscript{2} thin film not only provides the momentum required for the generation of the acoustophotoelectric current but also for the photoexcitation of the charge carriers to be extended to illumination wavelengths up to 850~nm.

To contrast the acoustophotoelectric effect observed with the SAW with the typical photoconductive effect that has been discussed to date, we examined the SnS\textsubscript{2} film under different illumination wavelengths from 285 to 850~nm with 1~V applied DC bias in the absence of the SAW excitation. As can be seen in Fig.~S7, high photocurrent values are obtained but at the expense of higher dark currents and much longer response times, the latter due to the slow generation of electron--hole pairs.\citep{yu2020giant} The eDOS and BS plots for SnS\textsubscript{2}  in the absence of the SAW excitation (Fig.~\ref{fig:dft}a) confirm the stronger photocurrent response at 285 and 365~nm since direct radiative transitions occur at these photon energies in comparison with the weaker response at 565~nm where the energy becomes sufficient to permit indirect transitions. At 660~nm, below the band gap in Fig.~\ref{fig:dft}a, we nevertheless observe a very weak photocurrent in the experiments (although this could possibly arise due to defects in the thin film, which could introduce defect states and assist the absorption of photons with energies lower than the intrinsic band gap\citep{yu2020giant}) whereas poor photosensitivity was observed at 850~nm, even in the presence of the 1~V DC bias.

\begin{figure}[!t]
\centering
  \includegraphics[width=0.6\textwidth]{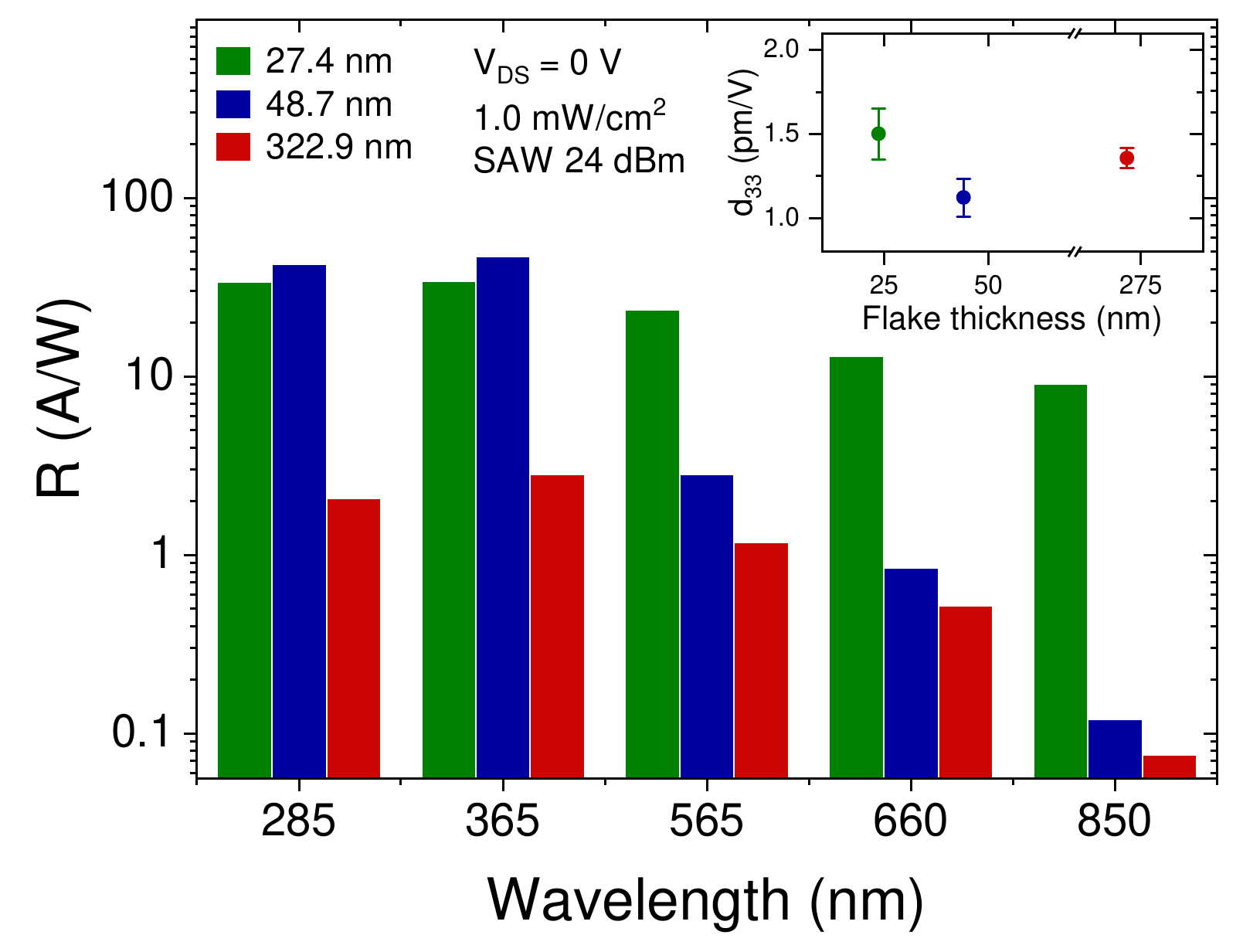}
  \caption{Spectral responsivity $R$ of SnS\textsubscript{2} flakes of varying thicknesses under different illumination wavelengths (24~dBm SAW power). The inset shows the piezoelectric force microscopy measurements of the electromechanical coefficient $d_{33}$ for SnS\textsubscript{2} flakes of similar thicknesses.}
  \label{fig:thickness}
\end{figure}

Finally, to demonstrate the utility of the platform in enhancing the photodetection performance of SnS\textsubscript{2}, we measure the important figure of merit qualities of photodetectors, including the dark current, photoresponsivity ($R_{\lambda} = I_{\rm photo} / (P_{\lambda}  A)$, in which $R_{\lambda}$, $P_{\lambda}$, and $A$ constitute the responsivity under an illumination wavelength $\lambda$, the illumination intensity, and the active area of the photodetector, respectively), external quantum efficiency EQE, noise equivalent power NEP (calculated from the noise characteristics plots in Fig.~S9), specific detectivity $D^*$, and response times (see calculation methods for each parameter in the photodetector characterization section in the Supplementary Information). Values for the dark current extracted from the $I_{\rm DS}$--$t$ plots in Fig.~\ref{fig:pae}b are tabulated in Table~S1, from which we note their pA-order magnitude across the cases examined---an advantage of operating without an external DC bias, which results in lower noise levels and higher $D^*$,\citep{huang2021understanding} in addition to minimizing quiescent power.\citep{satterthwaite2018high} As can be seen from Fig.~\ref{fig:pae}g,h, both the acoustophotocurrent as well as the photoresponsivity are dramatically enhanced with increasing SAW input powers up to $R$$_{365}$$ = 68.9$~A/W at 27~dBm. For the longer illumination wavelengths (565, 660, and 850~nm), we note that the decrease in the band gap due to the electric and strain fields produced by the SAW electromechanical coupling into the material, in addition to momentum transfer from the SAW to the charge carriers to faciliate their indirect transition, lead to higher photoresponsivities, especially $R_{850}$ of 0.2~A/W for a 48.7 nm flake---up to eight orders of magnitude larger than that reported in the literature;\citep{tao2015flexible,wu2016ultrathin,lei2020thermal}; an even larger enhancement of 9.1~A/W for thinner (27.4 nm) flakes was also observed (Fig.~\ref{fig:thickness}). It is worth further noting that although acoustic modulation in photodetection has previously been shown for a heterojunction p--n diode,\citep{zheng2018acoustically} the performance enhancement achieved here is far greater, particularly for a homogeneous material, especially in terms of the other photodetection parameters, namely, $R$, EQE, and $D^*$. 

In a similar manner to that observed with other 2D nanomaterials, the properties of SnS\textsubscript{2}, such as its band gap,\cite{ahn2015deterministic,gonzalez2016layer,seo2017thickness} photodetection performance,\cite{jia2018thickness} or piezoelectric response,\cite{wang2020piezoelectric} are expected to vary with its thickness. Indeed, we observe from Fig.~\ref{fig:thickness} that while the acoustophotoelectric responsivity $R$ of flakes with different thicknesses in general decreases with increasing illumination wavelengths due to the reasons discussed above, the decrease appears to be more pronounced for thicker samples due to the weaker coupling of the SAW into the flake as a consequence of the relative inability of the wave to penetrate into the bulk of the material given the ${\cal O}$(10 nm) vibration displacement amplitude; this was similarly observed in previous work on coupling the SAW into thick\cite{rezk2015acoustic} and thin\cite{rezk2016acoustically} MoS$_2$ flakes. Moreoever, we note that while both the 27.4 nm and 48.7 nm thick flakes exhibit strong responsivities in the UV region due to the high optical absorption of SnS\textsubscript{2} at these wavelengths, it is the thinner (27.4 nm) flake that notably maintains a more uniform photoresponse across the range of illumination wavelengths, decreasing only from 33.9~A/W at 285~nm  to 9.1~A/W at 850~nm. This can be attributed to the increased piezoelectricity of thinner SnS\textsubscript{2} flakes (as observed from the piezoresponse force microscopy (PFM) measurements of the electromechanical  coefficient $d_{33}$ in the inset of Fig.~\ref{fig:thickness}). As has been observed with other piezoelectric 2D materials,\cite{rotter1998giant,rezk2016acoustically} this leads to more effective coupling of the SAW into the material. In this respect, the acoustophotoelectric effect also qualitatively behaves in a similar way to the acoustoelectric effect, wherein the current increases with increasing input SAW power: from Eqn.~S.1, we observe the influence of the piezoelectric coupling coefficient $K_{\rm eff}^2$ (which is related to $d_{33}$), and more specifically, the piezoelectricity of the material on the acoustoelectric (and, in the present case, the acoustophotoelectric) current that is generated. The strong role of the SAW coupling into the material on its acoustophotoelectric response can perhaps be seen in the difference in the responsivities of the 48.7 nm and 322.9 nm thick flakes: on the one hand, the higher $d_{33}$ value that leads to more effective coupling of the SAW for the 322.9 nm thick flake compared to the 48.7 nm thick flake facilitates greater attenuation of the SAW such that more  of its momentum is available to be transferred to the charge carriers to produce the acoustoelectric and acoustophotoelectric current; concurrently, the relatively weaker acoustic penetration into the thicker flake, in contrast, results in weaker coupling of energy into the material, which could explain the poorer responsivity for the 322.9 nm thick flake compared to that for the 48.7 nm flake despite its higher $d_{33}$ value. Conversely, while the piezoelectricity and hence acoustoelectric and acoustophotoelectric effects are expected to improve toward the monolayer limit, thinner flakes below 20 nm possess diminishing optical absorption, and are hence not ideal for photodetection applications.\cite{jia2018thickness}

The present demonstration of the ability to extend the range of SnS\textsubscript{2} photodetectors beyond its nominal band gap without altering its pristine chemical condition for broadband operation has yet to be reported to date. This relative enhancement in the photoresponsivity with the SAW coupling across a wide wavelength range can be seen more clearly through the modified responsivity parameter, $R^* = R_{\lambda + {\rm SAW}}/R_{\lambda}$, in which $R_{\lambda + {\rm SAW}}$ denotes the responsivity in the presence of the SAW coupling. Importantly, two to three orders of magnitude increases across a broad wavelength range, from the ultraviolet (UV-B, UV-A) to visible, and through to near-infrared (NIR) illumination, can be seen in Fig.~\ref{fig:pae}i, noting particularly the two- to eight-decades of improvement in the NIR region at 850~nm compared to existing $R$ values in the literature.\citep{tao2015flexible,wu2016ultrathin,lei2020thermal} Enhancements in the other photodetection performance measures, namely EQE and $D^*$, can also be observed with the SAW (Fig.~S10a,b), with similar two to three order of magnitude improvements seen across all wavelengths (Fig.~S10c,d). Notably, the EQE and $D^*$ values at 850~nm of 42\% (compared to 0.15 $\%$ in the literature)\citep{wu2016ultrathin} and $2.7 \times 10^{8}$~Jones (compared to $9.9\times 10^{6}$~Jones)\citep{wu2016ultrathin}, respectively, were achieved. In addition, we obtain NEP values of $7.4 \times 10^{-13}$~W/Hz$^{1/2}$ at 285~nm to $1.9 \times 10^{-10}$~W/Hz$^{1/2}$ at 850~nm, which are the lowest reported in the literature  for  SnS\textsubscript{2} photodetectors to date. A detailed performance comparison of the current acoustophotoelectric platform against other SnS\textsubscript{2} photodetectors reported in the literature is presented in Supplementary Table~S1. Additionally, we also note that superior photoresponsivity of the present platform in the NIR region compared to the performance of other 2D materials and heterostructures that have been reported to date, with the exception of BP, which is known to be unstable in air, and PdSe\textsubscript{2}, which requires high intensity laser illumination and gate voltage\cite{liang2019high} and is hence not suited for low power light detection (Supplementary Table~S2).

\section{Conclusions}
\label{sec:conc}

We experimentally uncover the existence of, and elucidate through DFT simulations, a novel acoustophotoelectric effect arising from a trifecta of phonon--photon--electron interactions generated through coupled oscillating electric and strain fields associated with the electromechanical coupling of SAWs into a thin SnS\textsubscript{2} film. We show that the effect, and the large acoustophotocurrents that can be generated as a consequence of decreasing the band gap of the material and providing the required momentum for indirect transitions, is particularly advantageous in enhancing the photodetection capability of the material, yielding improvements in its photoresponsivity, external quantum efficiency and specific detectivity by two- to three-orders of magnitude while maintaining low pA-order dark currents---without the need for an external applied DC bias. Critically, the effect is observed over a broad range of illumination wavelengths, way beyond the nominal band gap of SnS\textsubscript{2}, therefore extending the operability of the photodetector into the NIR region where the 9.1~A/W responsivity that can be obtained for 850~nm illumination is up to eight decades in magnitude larger than SnS\textsubscript{2}-based photodetectors reported in the literature to date at similar wavelengths. Given that the effect is not  merely confined to SnS\textsubscript{2}, we expect the potential of the phenomenon, and, more broadly, the SAW platform, in general, can be extended to enhance the  performance of other photodetectors based on 2D materials.

\section{Methods} 

\subsection{Experimental procedures}

\subsubsection{Device fabrication and operation}

{\bf{SAW device}}: The piezoelectric chip, schematically illustrated in Fig.~\ref{fig:schem}c, consists of a single-crystal piezoelectric lithium niobate (LiNbO\textsubscript{3}, 128$^{\circ}$ Y-rotated, X-propagating LiNbO\textsubscript{3}; Roditi Ltd., London, UK) substrate, on which two sets of interdigitated transducer (IDT) electrodes were fabricated by sputter depositing (SPI-Module Sputter Coater, Structure Probe Inc., West Chester, PA, USA) a 10-nm-thick titanium adhesion layer followed by a 200-nm-thick gold layer, and subsequently wet etching IDT patterns comprising 60 straight interleaved finger pairs. To minimize the wave reflections and noise levels during the experiments, the LiNbO\textsubscript{3} chip was cut into a diamond shape and placed on a shock absorption sheet ($\theta$Gel, Taica Corporation, Tokyo, Japan). The SAW is then  generated by applying an oscillating electrical signal at the resonant frequency (80~MHz) using a signal generator (N9310A, Keysight Technologies, Santa Rosa, CA, USA) coupled to an amplifier (LYZ-22+; Mini Circuits, Brooklyn, NY, USA).

\noindent {\bf{Crystal exfoliation}}: SnS\textsubscript{2} bulk crystals (2D Semiconductors, Scottsdale, AZ, USA) were mechanically exfoliated (Scotch$^{\textregistered}$ Magic$^{\tiny {\rm TM}}$, 3M Pty.~Ltd., North Ryde, NSW, Australia) and dry-transferred onto the LiNbO\textsubscript{3} chip. As illustrated in Fig.~\ref{fig:schem}d, two electrical contact pads were then patterned to span an individual nanoflake using standard photolithography (Maskless Aligner MLA 150, Heidelberg Instruments Mikrotechnik GmbH, Heidelberg, Germany) by depositing a 10-nm-thick chromium adhesion layer followed by a 100-nm-thick gold layer through e-beam evaporation (PRO Line PVD 75, Kurt J. Lesker Company, Jefferson Hills, PA, USA) \textit{via} a lift-off process. 

\subsubsection{Characterization}

\noindent {\bf{Electrical measurements}}: The electrical measurements were performed on a source measure unit (Keithley 2450 and KickStart Instrument Control Software version 2.7.0, Tektronix Inc., Beaverton, OR, USA) in a 4-wire configuration, connected to the electrical contact pads using metal probes (T20-50 Solid Tungsten, Everbeing, Hsinchu, Taiwan) while the chip was mounted on a probe station (C-2 Mini, Everbeing, Hsinchu, Taiwan). 

\noindent {\bf{Illumination}}: Light emitting diodes (LEDs) with nominal wavelengths of 285, 365, 565, 660, and 880~nm (M285L5, M365L2, M565L3, M660L4, M850L3, Thorlabs Inc., Newton, NJ, USA), operated with an LED driver (DC4104, Thorlabs Inc., Newton, NJ, USA), were employed as the illumination sources in the study. The illumination power was measured with a thermal sensor (S120VC, Thorlabs Inc., Newton, NJ, USA) connected to a power and energy meter console (PM100D, Thorlabs Inc., Newton, NJ, USA). The experimental data were plotted with data analysis software (Origin 2023, OriginLab, Northampton, MA, USA).

\noindent {\bf{Piezoresponse force microscopy (PFM) measurements:}} Exfoliated SnS\textsubscript{2} flakes were transferred onto a SiO\textsubscript{2}/Si substrate (300~nm oxide thickness) coated with 2~nm chromium and 10~nm gold through electron beam evaporation. The electromechanical response of the SnS\textsubscript{2} flake of a specific thickness was then  measured using an atomic force microscope (AFM; MFP-3D, Asylum Research, Oxford Instruments, High Wycombe, UK) in Dual AC Resonance Tracking Piezo Force Microscopy (DART-PFM) contact mode, in which the shift in contact resonance frequency was 276~kHz. Igor Pro 6.32A software (WaveMetrics Inc. Portland, OR, USA) was used for data acquisition and analysis. A 25~nm radius conductive Pt/Ir coated AFM tip (SCM-PIT-V2, Bruker AFM Probes, Camarillo, CA, USA) with a resonance frequency of 75~kHz and spring constant of 3~N/m  was employed for the DART-PFM measurements.

\subsection{Simulation procedures}

A SnS\textsubscript{2} bulk supercell of symmetry ($P\overline{3}m1$) was constructed using crystallography information provided by the Crystallography Open Database (COD).\citep{grazulis2012} Hybrid density functional theory (DFT) calculations were performed using the Gaussian basis set \textit{ab initio} package CRYSTAL14.\citep{dovesi2014crystal14,dovesi2014crystal142} The Screened--Coulomb hybrid functional HSE06,\citep{heyd2003hybrid,krukau2006influence} which uses a Perdew--Burke--Ernzerhof (PBE) functional with 25\% Hartree-Fock exact exchange, was used to calculate the slab energies and electronic density of states. For all atoms, a double zeta valance basis set (DZV), with polarization functions, was used to model the electrons.\citep{laun2018consistent} Band structure and energy calculations were conducted using a $9 \times 9 \times 1$ Monkhorst-Pack k-point mesh. The bulk crystal was geometry optimized (atoms and cell), and then a slab of thickness 20.48~nm (as a good representation of the exfoliated nanoflake) with a (001) surface was cut from the bulk phase and relaxed. Van der Waals dispersion corrections to the energy were performed using the D3 method proposed by Grimme.\citep{grimme2010consistent} The following high symmetry points corresponding to the in-plane paths of a hexagonal Brillouin zone was used for the band structure plots: $\Gamma$(000), M(1/2,0,0), K(1/3,1/3,0), $\Gamma$(000).\citep{aroyo2014brillouin}


\section*{Acknowledgement}
The authors are grateful for access to, and the technical assistance associated with, the use of the equipment and facilities in the RMIT School of Engineering laboratories, the RMIT Micro/Nano Research Facility (MNRF), and the RMIT Microscopy \& Microanalysis Facility (RMMF). ARR, LYY, SW and SB acknowledge funding through various Australian Research Council (ARC) Discovery Projects (ARR, LYY: DP180102110; SW: DP220100020; SB: DP210103428) whereas KC acknowledges support from the ARC Centre of Excellence for Transformative Meta-Optical Systems (CE200100010). PR is grateful for an ARC {\mbox \protect{DECRA}} Fellowship (DE200100279) as well as the RMIT University Vice-Chancellor’s Research Fellowship.


\section*{Author contributions}
HA, LYY and ARR conceived the original research idea, developed the methodology and designed the devices. HA carried out the experimental investigations.
HA, PR, RK, MXL, and SB carried out the materials characterization. SR performed the simulations. HA, ARR, LYY and GRN analysed the results. 
The original draft of the manuscript was written by HA, SR, ARR, and LYY, which was reviewed by PR, SR, SW, SB, KC, and GRN.

\section*{Declaration of interests}
The authors declare no competing interests.

\section*{Data availability}
The data that support the findings of this study are available from the corresponding author upon reasonable request.

\section*{Supporting information}
Supplementary information is available online.

\bibliography{SnS2}

\providecommand{\latin}[1]{#1}
\makeatletter
\providecommand{\doi}
  {\begingroup\let\do\@makeother\dospecials
  \catcode`\{=1 \catcode`\}=2 \doi@aux}
\providecommand{\doi@aux}[1]{\endgroup\texttt{#1}}
\makeatother
\providecommand*\mcitethebibliography{\thebibliography}
\csname @ifundefined\endcsname{endmcitethebibliography}
  {\let\endmcitethebibliography\endthebibliography}{}
\begin{mcitethebibliography}{92}
\providecommand*\natexlab[1]{#1}
\providecommand*\mciteSetBstSublistMode[1]{}
\providecommand*\mciteSetBstMaxWidthForm[2]{}
\providecommand*\mciteBstWouldAddEndPuncttrue
  {\def\EndOfBibitem{\unskip.}}
\providecommand*\mciteBstWouldAddEndPunctfalse
  {\let\EndOfBibitem\relax}
\providecommand*\mciteSetBstMidEndSepPunct[3]{}
\providecommand*\mciteSetBstSublistLabelBeginEnd[3]{}
\providecommand*\EndOfBibitem{}
\mciteSetBstSublistMode{f}
\mciteSetBstMaxWidthForm{subitem}{(\alph{mcitesubitemcount})}
\mciteSetBstSublistLabelBeginEnd
  {\mcitemaxwidthsubitemform\space}
  {\relax}
  {\relax}

\bibitem[Wang \latin{et~al.}(2012)Wang, Kalantar-Zadeh, Kis, Coleman, and
  Strano]{wang2012electronics}
Wang,~Q.~H.; Kalantar-Zadeh,~K.; Kis,~A.; Coleman,~J.~N.; Strano,~M.~S.
  Electronics and optoelectronics of two-dimensional transition metal
  dichalcogenides. \emph{Nat. Nanotechnol.} \textbf{2012}, \emph{7},
  699--712\relax
\mciteBstWouldAddEndPuncttrue
\mciteSetBstMidEndSepPunct{\mcitedefaultmidpunct}
{\mcitedefaultendpunct}{\mcitedefaultseppunct}\relax
\EndOfBibitem
\bibitem[Mak and Shan(2016)Mak, and Shan]{mak2016photonics}
Mak,~K.~F.; Shan,~J. Photonics and optoelectronics of 2{D} semiconductor
  transition metal dichalcogenides. \emph{Nat. Photon.} \textbf{2016},
  \emph{10}, 216--226\relax
\mciteBstWouldAddEndPuncttrue
\mciteSetBstMidEndSepPunct{\mcitedefaultmidpunct}
{\mcitedefaultendpunct}{\mcitedefaultseppunct}\relax
\EndOfBibitem
\bibitem[Zhou \latin{et~al.}(2016)Zhou, Zhang, Gan, Li, Xiong, and
  Zhai]{zhou2016booming}
Zhou,~X.; Zhang,~Q.; Gan,~L.; Li,~H.; Xiong,~J.; Zhai,~T. Booming development
  of group {IV}--{VI} semiconductors: fresh blood of 2{D} family. \emph{Adv.
  Sci.} \textbf{2016}, \emph{3}, 1600177\relax
\mciteBstWouldAddEndPuncttrue
\mciteSetBstMidEndSepPunct{\mcitedefaultmidpunct}
{\mcitedefaultendpunct}{\mcitedefaultseppunct}\relax
\EndOfBibitem
\bibitem[Wang \latin{et~al.}(2019)Wang, Zhong, Zhang, Zheng, Zhang, and
  Zhang]{wang2019broadband}
Wang,~B.; Zhong,~S.~P.; Zhang,~Z.~B.; Zheng,~Z.~Q.; Zhang,~Y.~P.; Zhang,~H.
  Broadband photodetectors based on 2{D} group {IVA} metal chalcogenides
  semiconductors. \emph{Appl. Mater. Today} \textbf{2019}, \emph{15},
  115--138\relax
\mciteBstWouldAddEndPuncttrue
\mciteSetBstMidEndSepPunct{\mcitedefaultmidpunct}
{\mcitedefaultendpunct}{\mcitedefaultseppunct}\relax
\EndOfBibitem
\bibitem[Wang \latin{et~al.}(2019)Wang, Zhang, Gao, Luo, Su, Han, Liu, Li, and
  Zhai]{wang20192d}
Wang,~F.; Zhang,~Y.; Gao,~Y.; Luo,~P.; Su,~J.; Han,~W.; Liu,~K.; Li,~H.;
  Zhai,~T. 2{D} metal chalcogenides for {IR} photodetection. \emph{Small}
  \textbf{2019}, \emph{15}, 1901347\relax
\mciteBstWouldAddEndPuncttrue
\mciteSetBstMidEndSepPunct{\mcitedefaultmidpunct}
{\mcitedefaultendpunct}{\mcitedefaultseppunct}\relax
\EndOfBibitem
\bibitem[Qiu and Huang(2021)Qiu, and Huang]{qiu2021photodetectors}
Qiu,~Q.; Huang,~Z. Photodetectors of 2{D} materials from ultraviolet to
  terahertz waves. \emph{Adv. Mater.} \textbf{2021}, \emph{33}, 2008126\relax
\mciteBstWouldAddEndPuncttrue
\mciteSetBstMidEndSepPunct{\mcitedefaultmidpunct}
{\mcitedefaultendpunct}{\mcitedefaultseppunct}\relax
\EndOfBibitem
\bibitem[Sett \latin{et~al.}(2021)Sett, Parappurath, Gill, Chauhan, and
  Ghosh]{sett2021engineering}
Sett,~S.; Parappurath,~A.; Gill,~N.~K.; Chauhan,~N.; Ghosh,~A. Engineering
  sensitivity and spectral range of photodetection in van der {W}aals materials
  and hybrids. \emph{Nano Express} \textbf{2021}, \emph{3}, 014001\relax
\mciteBstWouldAddEndPuncttrue
\mciteSetBstMidEndSepPunct{\mcitedefaultmidpunct}
{\mcitedefaultendpunct}{\mcitedefaultseppunct}\relax
\EndOfBibitem
\bibitem[Li \latin{et~al.}(2021)Li, Yao, Haidry, Luan, Chen, Zhang, Xu, Deng,
  {Duc Hoa}, Zhou, and Ou]{li2021recent}
Li,~Z.; Yao,~Z.; Haidry,~A.~A.; Luan,~Y.; Chen,~Y.; Zhang,~B.~Y.; Xu,~K.;
  Deng,~R.; {Duc Hoa},~N.; Zhou,~J.; Ou,~J.~Z. Recent advances of atomically
  thin 2{D} heterostructures in sensing applications. \emph{Nano Today}
  \textbf{2021}, \emph{40}, 101287\relax
\mciteBstWouldAddEndPuncttrue
\mciteSetBstMidEndSepPunct{\mcitedefaultmidpunct}
{\mcitedefaultendpunct}{\mcitedefaultseppunct}\relax
\EndOfBibitem
\bibitem[Huang \latin{et~al.}(2014)Huang, Sutter, Sadowski, Cotlet, Monti,
  Racke, Neupane, Wickramaratne, Lake, Parkinson, and Sutter]{huang2014tin}
Huang,~Y.; Sutter,~E.; Sadowski,~J.~T.; Cotlet,~M.; Monti,~O.~L.; Racke,~D.~A.;
  Neupane,~M.~R.; Wickramaratne,~D.; Lake,~R.~K.; Parkinson,~B.~A.; Sutter,~P.
  Tin Disulfide - an emerging layered metal dichalcogenide semiconductor:
  materials properties and device characteristics. \emph{ACS Nano}
  \textbf{2014}, \emph{8}, 10743--10755\relax
\mciteBstWouldAddEndPuncttrue
\mciteSetBstMidEndSepPunct{\mcitedefaultmidpunct}
{\mcitedefaultendpunct}{\mcitedefaultseppunct}\relax
\EndOfBibitem
\bibitem[Su \latin{et~al.}(2015)Su, Hadjiev, Loya, Zhang, Lei, Maharjan, Dong,
  Ajayan, Lou, and Peng]{su2015chemical}
Su,~G.; Hadjiev,~V.~G.; Loya,~P.~E.; Zhang,~J.; Lei,~S.; Maharjan,~S.;
  Dong,~P.; Ajayan,~P.~M.; Lou,~J.; Peng,~H. Chemical vapor deposition of thin
  crystals of layered semiconductor {S}n{S}\textsubscript{2} for fast
  photodetection application. \emph{Nano Lett.} \textbf{2015}, \emph{15},
  506--513\relax
\mciteBstWouldAddEndPuncttrue
\mciteSetBstMidEndSepPunct{\mcitedefaultmidpunct}
{\mcitedefaultendpunct}{\mcitedefaultseppunct}\relax
\EndOfBibitem
\bibitem[Ahn \latin{et~al.}(2015)Ahn, Lee, Heo, Sung, Kim, Hwang, and
  Jo]{ahn2015deterministic}
Ahn,~J.-H.; Lee,~M.-J.; Heo,~H.; Sung,~J.~H.; Kim,~K.; Hwang,~H.; Jo,~M.-H.
  Deterministic two-dimensional polymorphism growth of hexagonal n-type
  {S}n{S}\textsubscript{2} and orthorhombic p-type {S}n{S} crystals. \emph{Nano
  Lett.} \textbf{2015}, \emph{15}, 3703--3708\relax
\mciteBstWouldAddEndPuncttrue
\mciteSetBstMidEndSepPunct{\mcitedefaultmidpunct}
{\mcitedefaultendpunct}{\mcitedefaultseppunct}\relax
\EndOfBibitem
\bibitem[Giri \latin{et~al.}(2019)Giri, Masroor, Yan, Kushnir, Carl, Doiron,
  Zhang, Zhao, McClelland, Tompsett, Wang, Grimm, Titova, and
  Rao]{giri2019balancing}
Giri,~B.; Masroor,~M.; Yan,~T.; Kushnir,~K.; Carl,~A.~D.; Doiron,~C.;
  Zhang,~H.; Zhao,~Y.; McClelland,~A.; Tompsett,~G.~A.; Wang,~D.; Grimm,~R.~L.;
  Titova,~L.~V.; Rao,~P.~M. Balancing light absorption and charge transport in
  vertical {S}n{S}\textsubscript{2} nanoflake photoanodes with stepped layers
  and large intrinsic mobility. \emph{Adv. Energy Mater.} \textbf{2019},
  \emph{9}, 1901236\relax
\mciteBstWouldAddEndPuncttrue
\mciteSetBstMidEndSepPunct{\mcitedefaultmidpunct}
{\mcitedefaultendpunct}{\mcitedefaultseppunct}\relax
\EndOfBibitem
\bibitem[De \latin{et~al.}(2012)De, Manongdo, See, Zhang, Guloy, and
  Peng]{de2012high}
De,~D.; Manongdo,~J.; See,~S.; Zhang,~V.; Guloy,~A.; Peng,~H. High on/off ratio
  field effect transistors based on exfoliated crystalline
  {S}n{S}\textsubscript{2} nano-membranes. \emph{Nanotechnology} \textbf{2012},
  \emph{24}, 025202\relax
\mciteBstWouldAddEndPuncttrue
\mciteSetBstMidEndSepPunct{\mcitedefaultmidpunct}
{\mcitedefaultendpunct}{\mcitedefaultseppunct}\relax
\EndOfBibitem
\bibitem[Xia \latin{et~al.}(2015)Xia, Zhu, Wang, Huang, Huang, and
  Meng]{xia2015large}
Xia,~J.; Zhu,~D.; Wang,~L.; Huang,~B.; Huang,~X.; Meng,~X.-M. Large-scale
  growth of two-dimensional {S}n{S}\textsubscript{2} crystals driven by screw
  dislocations and application to photodetectors. \emph{Adv. Funct. Mater.}
  \textbf{2015}, \emph{25}, 4255--4261\relax
\mciteBstWouldAddEndPuncttrue
\mciteSetBstMidEndSepPunct{\mcitedefaultmidpunct}
{\mcitedefaultendpunct}{\mcitedefaultseppunct}\relax
\EndOfBibitem
\bibitem[Yang \latin{et~al.}(2017)Yang, Zheng, Wang, Xu, Pan, Zou, Zhang, Qi,
  Liu, Feng, Hu, Miao, Sun, Duan, and Pan]{yang2017van}
Yang,~T.; Zheng,~B.; Wang,~Z.; Xu,~T.; Pan,~C.; Zou,~J.; Zhang,~X.; Qi,~Z.;
  Liu,~H.; Feng,~Y.; Hu,~W.; Miao,~F.; Sun,~L.; Duan,~X.; Pan,~A. Van der
  {W}aals epitaxial growth and optoelectronics of large-scale
  {W}{S}e\textsubscript{2}/{S}n{S}\textsubscript{2} vertical bilayer p--n
  junctions. \emph{Nat. Commun.} \textbf{2017}, \emph{8}, 1--9\relax
\mciteBstWouldAddEndPuncttrue
\mciteSetBstMidEndSepPunct{\mcitedefaultmidpunct}
{\mcitedefaultendpunct}{\mcitedefaultseppunct}\relax
\EndOfBibitem
\bibitem[Huang \latin{et~al.}(2015)Huang, Deng, Xu, Wang, Wang, Wang, Wang,
  Zhan, Li, Luo, and He]{huang2015highly}
Huang,~Y.; Deng,~H.-X.; Xu,~K.; Wang,~Z.-X.; Wang,~Q.-S.; Wang,~F.-M.;
  Wang,~F.; Zhan,~X.-Y.; Li,~S.-S.; Luo,~J.-W.; He,~J. Highly sensitive and
  fast phototransistor based on large size {CVD}-grown {S}n{S}\textsubscript{2}
  nanosheets. \emph{Nanoscale} \textbf{2015}, \emph{7}, 14093--14099\relax
\mciteBstWouldAddEndPuncttrue
\mciteSetBstMidEndSepPunct{\mcitedefaultmidpunct}
{\mcitedefaultendpunct}{\mcitedefaultseppunct}\relax
\EndOfBibitem
\bibitem[Tao \latin{et~al.}(2015)Tao, Wu, Wang, and Wang]{tao2015flexible}
Tao,~Y.; Wu,~X.; Wang,~W.; Wang,~J. Flexible photodetector from ultraviolet to
  near infrared based on a {S}n{S}\textsubscript{2} nanosheet microsphere film.
  \emph{J. Mater. Chem. C} \textbf{2015}, \emph{3}, 1347--1353\relax
\mciteBstWouldAddEndPuncttrue
\mciteSetBstMidEndSepPunct{\mcitedefaultmidpunct}
{\mcitedefaultendpunct}{\mcitedefaultseppunct}\relax
\EndOfBibitem
\bibitem[Wu \latin{et~al.}(2016)Wu, Tao, Wu, and Wu]{wu2016ultrathin}
Wu,~J.-J.; Tao,~Y.-R.; Wu,~Y.; Wu,~X.-C. Ultrathin {S}n{S}\textsubscript{2}
  nanosheets of ultrasonic synthesis and their photoresponses from ultraviolet
  to near-infrared. \emph{Sens. Actuators B: Chem.} \textbf{2016}, \emph{231},
  211--217\relax
\mciteBstWouldAddEndPuncttrue
\mciteSetBstMidEndSepPunct{\mcitedefaultmidpunct}
{\mcitedefaultendpunct}{\mcitedefaultseppunct}\relax
\EndOfBibitem
\bibitem[Yang \latin{et~al.}(2016)Yang, Li, Hu, Deng, Dong, Yang, Qiao, Yuan,
  and Song]{yang2016controllable}
Yang,~D.; Li,~B.; Hu,~C.; Deng,~H.; Dong,~D.; Yang,~X.; Qiao,~K.; Yuan,~S.;
  Song,~H. Controllable Growth Orientation of {S}n{S}\textsubscript{2} Flakes
  for Low-Noise, High-Photoswitching Ratio, and Ultrafast Phototransistors.
  \emph{Adv. Opt. Mater.} \textbf{2016}, \emph{4}, 419--426\relax
\mciteBstWouldAddEndPuncttrue
\mciteSetBstMidEndSepPunct{\mcitedefaultmidpunct}
{\mcitedefaultendpunct}{\mcitedefaultseppunct}\relax
\EndOfBibitem
\bibitem[Zhou \latin{et~al.}(2016)Zhou, Zhang, Gan, Li, and
  Zhai]{zhou2016large}
Zhou,~X.; Zhang,~Q.; Gan,~L.; Li,~H.; Zhai,~T. Large-size growth of ultrathin
  {S}n{S}\textsubscript{2} nanosheets and high performance for
  phototransistors. \emph{Adv. Funct. Mater.} \textbf{2016}, \emph{26},
  4405--4413\relax
\mciteBstWouldAddEndPuncttrue
\mciteSetBstMidEndSepPunct{\mcitedefaultmidpunct}
{\mcitedefaultendpunct}{\mcitedefaultseppunct}\relax
\EndOfBibitem
\bibitem[Fan \latin{et~al.}(2016)Fan, Li, Lu, Deng, Wei, and
  Li]{fan2016wavelength}
Fan,~C.; Li,~Y.; Lu,~F.; Deng,~H.-X.; Wei,~Z.; Li,~J. Wavelength dependent
  {UV}-{V}is photodetectors from {S}n{S}\textsubscript{2} flakes. \emph{RSC
  Adv.} \textbf{2016}, \emph{6}, 422--427\relax
\mciteBstWouldAddEndPuncttrue
\mciteSetBstMidEndSepPunct{\mcitedefaultmidpunct}
{\mcitedefaultendpunct}{\mcitedefaultseppunct}\relax
\EndOfBibitem
\bibitem[Gao \latin{et~al.}(2016)Gao, Chen, Zeng, Ge, Yang, Song, and
  Tang]{gao2016broadband}
Gao,~L.; Chen,~C.; Zeng,~K.; Ge,~C.; Yang,~D.; Song,~H.; Tang,~J. Broadband,
  sensitive and spectrally distinctive {S}n{S}\textsubscript{2} nanosheet/PbS
  colloidal quantum dot hybrid photodetector. \emph{Light Sci. Appl.}
  \textbf{2016}, \emph{5}, e16126--e16126\relax
\mciteBstWouldAddEndPuncttrue
\mciteSetBstMidEndSepPunct{\mcitedefaultmidpunct}
{\mcitedefaultendpunct}{\mcitedefaultseppunct}\relax
\EndOfBibitem
\bibitem[Li \latin{et~al.}(2017)Li, Xing, Zhong, Huang, Lei, Zhang, Li, and
  Wei]{li2017two}
Li,~B.; Xing,~T.; Zhong,~M.; Huang,~L.; Lei,~N.; Zhang,~J.; Li,~J.; Wei,~Z. A
  two-dimensional {F}e-doped {S}n{S}\textsubscript{2} magnetic semiconductor.
  \emph{Nat. Commun.} \textbf{2017}, \emph{8}, 1--7\relax
\mciteBstWouldAddEndPuncttrue
\mciteSetBstMidEndSepPunct{\mcitedefaultmidpunct}
{\mcitedefaultendpunct}{\mcitedefaultseppunct}\relax
\EndOfBibitem
\bibitem[Jia \latin{et~al.}(2018)Jia, Tang, Pan, Long, and
  Gu]{jia2018thickness}
Jia,~X.; Tang,~C.; Pan,~R.; Long,~Y.; Gu,~J.,~C.and~Li Thickness-dependently
  enhanced photodetection performance of vertically grown
  {S}n{S}\textsubscript{2} nanoflakes with large size and high production.
  \emph{ACS Appl. Mater. Interfaces} \textbf{2018}, \emph{10},
  18073--18081\relax
\mciteBstWouldAddEndPuncttrue
\mciteSetBstMidEndSepPunct{\mcitedefaultmidpunct}
{\mcitedefaultendpunct}{\mcitedefaultseppunct}\relax
\EndOfBibitem
\bibitem[Liu \latin{et~al.}(2019)Liu, Liu, Chen, Miao, Liu, Li, Tang, Chen,
  Liu, Li, Wei, and Duan]{liu2019tunable}
Liu,~J.; Liu,~X.; Chen,~Z.; Miao,~L.; Liu,~X.; Li,~B.; Tang,~L.; Chen,~K.;
  Liu,~Y.; Li,~J.; Wei,~Z.; Duan,~X. Tunable {S}chottky barrier width and
  enormously enhanced photoresponsivity in {S}b doped {S}n{S}\textsubscript{2}
  monolayer. \emph{Nano Res.} \textbf{2019}, \emph{12}, 463--468\relax
\mciteBstWouldAddEndPuncttrue
\mciteSetBstMidEndSepPunct{\mcitedefaultmidpunct}
{\mcitedefaultendpunct}{\mcitedefaultseppunct}\relax
\EndOfBibitem
\bibitem[Tian \latin{et~al.}(2020)Tian, Meng, Yang, Fan, Yuan, An, Sun, Zhang,
  Wang, Zheng, Wei, and Li]{tian2020visible}
Tian,~H.; Meng,~X.; Yang,~J.; Fan,~C.; Yuan,~S.; An,~X.; Sun,~C.; Zhang,~Y.;
  Wang,~M.; Zheng,~H.; Wei,~Z.; Li,~E. Visible phototransistors based on
  vertical nanolayered heterostructures of {S}n{S}/{S}n{S}\textsubscript{2}
  p--n and {S}n{S}e\textsubscript{2}/{S}n{S}\textsubscript{2} n--n nanoflakes.
  \emph{ACS Appl. Nano Mater.} \textbf{2020}, \emph{3}, 6847--6854\relax
\mciteBstWouldAddEndPuncttrue
\mciteSetBstMidEndSepPunct{\mcitedefaultmidpunct}
{\mcitedefaultendpunct}{\mcitedefaultseppunct}\relax
\EndOfBibitem
\bibitem[Yu \latin{et~al.}(2020)Yu, Suleiman, Zheng, Zhou, and
  Zhai]{yu2020giant}
Yu,~J.; Suleiman,~A.~A.; Zheng,~Z.; Zhou,~X.; Zhai,~T. Giant-enhanced
  {S}n{S}\textsubscript{2} photodetectors with broadband response through
  oxygen plasma treatment. \emph{Adv. Funct. Mater.} \textbf{2020}, \emph{30},
  2001650\relax
\mciteBstWouldAddEndPuncttrue
\mciteSetBstMidEndSepPunct{\mcitedefaultmidpunct}
{\mcitedefaultendpunct}{\mcitedefaultseppunct}\relax
\EndOfBibitem
\bibitem[Lei \latin{et~al.}(2020)Lei, Luo, Yang, Cai, Qi, Gu, and
  Zheng]{lei2020thermal}
Lei,~Y.; Luo,~J.; Yang,~X.; Cai,~T.; Qi,~R.; Gu,~L.; Zheng,~Z. Thermal
  evaporation of large-area {S}n{S}\textsubscript{2} thin films with a
  {UV}-to-{NIR} photoelectric response for flexible photodetector applications.
  \emph{ACS Appl. Mater. Interfaces} \textbf{2020}, \emph{12},
  24940--24950\relax
\mciteBstWouldAddEndPuncttrue
\mciteSetBstMidEndSepPunct{\mcitedefaultmidpunct}
{\mcitedefaultendpunct}{\mcitedefaultseppunct}\relax
\EndOfBibitem
\bibitem[Fan \latin{et~al.}(2021)Fan, Liu, Yuan, Meng, An, Jing, Sun, Zhang,
  Zhang, Wang, Zheng, and Li]{fan2021enhanced}
Fan,~C.; Liu,~Z.; Yuan,~S.; Meng,~X.; An,~X.; Jing,~Y.; Sun,~C.; Zhang,~Y.;
  Zhang,~Z.; Wang,~M.; Zheng,~H.; Li,~E. Enhanced photodetection performance of
  photodetectors based on indium-doped tin disulfide few layers. \emph{ACS
  Appl. Mater. Interfaces} \textbf{2021}, \emph{13}, 35889--35896\relax
\mciteBstWouldAddEndPuncttrue
\mciteSetBstMidEndSepPunct{\mcitedefaultmidpunct}
{\mcitedefaultendpunct}{\mcitedefaultseppunct}\relax
\EndOfBibitem
\bibitem[Fu \latin{et~al.}(2021)Fu, Mo, Ostrikov, Gu, Nan, and
  Xiao]{fu2021controllable}
Fu,~Q.; Mo,~H.; Ostrikov,~K.~K.; Gu,~X.; Nan,~H.; Xiao,~S. Controllable
  synthesis of {S}n{S}\textsubscript{2} flakes and
  {M}o{S}\textsubscript{2}/{S}n{S}\textsubscript{2} heterostructures by
  confined-space chemical vapor deposition. \emph{CrystEngComm} \textbf{2021},
  \emph{23}, 2563--2571\relax
\mciteBstWouldAddEndPuncttrue
\mciteSetBstMidEndSepPunct{\mcitedefaultmidpunct}
{\mcitedefaultendpunct}{\mcitedefaultseppunct}\relax
\EndOfBibitem
\bibitem[Shooshtari \latin{et~al.}(2021)Shooshtari, Esfandiar, Orooji,
  Samadpour, and Rahighi]{shooshtari2021ultrafast}
Shooshtari,~L.; Esfandiar,~A.; Orooji,~Y.; Samadpour,~M.; Rahighi,~R. Ultrafast
  and stable planar photodetector based on {S}n{S}\textsubscript{2}
  nanosheets/perovskite structure. \emph{Sci. Rep.} \textbf{2021}, \emph{11},
  1--15\relax
\mciteBstWouldAddEndPuncttrue
\mciteSetBstMidEndSepPunct{\mcitedefaultmidpunct}
{\mcitedefaultendpunct}{\mcitedefaultseppunct}\relax
\EndOfBibitem
\bibitem[Luo \latin{et~al.}(2022)Luo, Song, Lu, Hu, Lv, Li, Li, Deng, Yan,
  Jiang, and Xia]{luo2022phase}
Luo,~J.; Song,~X.; Lu,~Y.; Hu,~Y.; Lv,~X.; Li,~L.; Li,~X.; Deng,~J.; Yan,~Y.;
  Jiang,~Y.; Xia,~C. Phase-controlled synthesis of {S}n{S}\textsubscript{2} and
  {S}n{S} flakes and photodetection properties. \emph{J. Phys.: Condens.
  Matter} \textbf{2022}, \emph{34}, 285701\relax
\mciteBstWouldAddEndPuncttrue
\mciteSetBstMidEndSepPunct{\mcitedefaultmidpunct}
{\mcitedefaultendpunct}{\mcitedefaultseppunct}\relax
\EndOfBibitem
\bibitem[Ying \latin{et~al.}(2019)Ying, Li, Wu, Yao, Xi, Su, Jin, Xu, He, and
  Zhang]{ying2019high}
Ying,~H.; Li,~X.; Wu,~Y.; Yao,~Y.; Xi,~J.; Su,~W.; Jin,~C.; Xu,~M.; He,~Z.;
  Zhang,~Q. High-performance ultra-violet phototransistors based on {CVT}-grown
  high quality {S}n{S}\textsubscript{2} flakes. \emph{Nanoscale Adv.}
  \textbf{2019}, \emph{1}, 3973--3979\relax
\mciteBstWouldAddEndPuncttrue
\mciteSetBstMidEndSepPunct{\mcitedefaultmidpunct}
{\mcitedefaultendpunct}{\mcitedefaultseppunct}\relax
\EndOfBibitem
\bibitem[Khimani \latin{et~al.}(2019)Khimani, Chaki, Deshpande, Chauhan, and
  Tailor]{khimani2019alloy}
Khimani,~A.~J.; Chaki,~S.~H.; Deshpande,~M.~P.; Chauhan,~S.~M.; Tailor,~J.~P.
  Alloy engineering to promote photodetection in
  {I}n\textsubscript{x}{S}n\textsubscript{1-x}{S}\textsubscript{2} and
  {S}b\textsubscript{x}{S}n\textsubscript{1-x}{S}\textsubscript{2} ternary
  alloys. \emph{Mater. Lett.} \textbf{2019}, \emph{236}, 187--189\relax
\mciteBstWouldAddEndPuncttrue
\mciteSetBstMidEndSepPunct{\mcitedefaultmidpunct}
{\mcitedefaultendpunct}{\mcitedefaultseppunct}\relax
\EndOfBibitem
\bibitem[Shu \latin{et~al.}(2020)Shu, Peng, Huang, Xu, Suleiman, Zhang, Bai,
  Zhou, and Zhai]{shu2020growth}
Shu,~Z.; Peng,~Q.; Huang,~P.; Xu,~Z.; Suleiman,~A.~A.; Zhang,~X.; Bai,~X.;
  Zhou,~X.; Zhai,~T. Growth of ultrathin ternary teallite
  ({P}b{S}n{S}\textsubscript{2}) flakes for highly anisotropic optoelectronics.
  \emph{Matter} \textbf{2020}, \emph{2}, 977--987\relax
\mciteBstWouldAddEndPuncttrue
\mciteSetBstMidEndSepPunct{\mcitedefaultmidpunct}
{\mcitedefaultendpunct}{\mcitedefaultseppunct}\relax
\EndOfBibitem
\bibitem[Yuan \latin{et~al.}(2019)Yuan, Fan, Tian, Zhang, Zhang, Zhong, Liu,
  Wang, and Li]{yuan2019enhanced}
Yuan,~S.; Fan,~C.; Tian,~H.; Zhang,~Y.; Zhang,~Z.; Zhong,~M.; Liu,~H.;
  Wang,~M.; Li,~E. Enhanced photoresponse of indium-doped tin disulfide
  nanosheets. \emph{ACS Appl. Mater. Interfaces} \textbf{2019}, \emph{12},
  2607--2614\relax
\mciteBstWouldAddEndPuncttrue
\mciteSetBstMidEndSepPunct{\mcitedefaultmidpunct}
{\mcitedefaultendpunct}{\mcitedefaultseppunct}\relax
\EndOfBibitem
\bibitem[Delsing \latin{et~al.}(2019)Delsing, Cleland, Schuetz, Knörzer,
  Giedke, Cirac, Srinivasan, Wu, Balram, Bäuerle, Meunier, Ford, Santos,
  Cerda-Méndez, Wang, Krenner, Nysten, Weiß, Nash, Thevenard, Gourdon,
  Rovillain, Marangolo, Duquesne, Fischerauer, Ruile, Reiner, Paschke,
  Denysenko, Volkmer, Wixforth, Bruus, Wiklund, Reboud, Cooper, Fu, Brugger,
  Rehfeldt, and Westerhausen]{delsing2019roadmap}
Delsing,~P. \latin{et~al.}  The 2019 surface acoustic waves roadmap. \emph{J.
  Phys. D: Appl. Phys.} \textbf{2019}, \emph{52}, 353001\relax
\mciteBstWouldAddEndPuncttrue
\mciteSetBstMidEndSepPunct{\mcitedefaultmidpunct}
{\mcitedefaultendpunct}{\mcitedefaultseppunct}\relax
\EndOfBibitem
\bibitem[Zhang \latin{et~al.}(2020)Zhang, Bachman, Ozcelik, and
  Huang]{zhang2020acoustic}
Zhang,~P.; Bachman,~H.; Ozcelik,~A.; Huang,~T.~J. Acoustic microfluidics.
  \emph{Annu. Rev. Anal. Chem.} \textbf{2020}, \emph{13}, 17--43\relax
\mciteBstWouldAddEndPuncttrue
\mciteSetBstMidEndSepPunct{\mcitedefaultmidpunct}
{\mcitedefaultendpunct}{\mcitedefaultseppunct}\relax
\EndOfBibitem
\bibitem[Rezk \latin{et~al.}(2021)Rezk, Ahmed, Ramesan, and Yeo]{rezk2021high}
Rezk,~A.~R.; Ahmed,~H.; Ramesan,~S.; Yeo,~L.~Y. High frequency sonoprocessing:
  a new field of cavitation-free acoustic materials synthesis, processing, and
  manipulation. \emph{Adv. Sci.} \textbf{2021}, \emph{8}, 2001983\relax
\mciteBstWouldAddEndPuncttrue
\mciteSetBstMidEndSepPunct{\mcitedefaultmidpunct}
{\mcitedefaultendpunct}{\mcitedefaultseppunct}\relax
\EndOfBibitem
\bibitem[Marqus \latin{et~al.}(2018)Marqus, Ahmed, Ahmed, Xu, Rezk, and
  Yeo]{marqus2018increasing}
Marqus,~S.; Ahmed,~H.; Ahmed,~M.; Xu,~C.; Rezk,~A.~R.; Yeo,~L.~Y. Increasing
  exfoliation yield in the synthesis of {M}o{S}\textsubscript{2} quantum dots
  for optoelectronic and other applications through a continuous multicycle
  acoustomicrofluidic approach. \emph{ACS Appl. Nano Mater.} \textbf{2018},
  \emph{1}, 2503--2508\relax
\mciteBstWouldAddEndPuncttrue
\mciteSetBstMidEndSepPunct{\mcitedefaultmidpunct}
{\mcitedefaultendpunct}{\mcitedefaultseppunct}\relax
\EndOfBibitem
\bibitem[Ahmed \latin{et~al.}(2018)Ahmed, Rezk, Carey, Wang, Mohiuddin, Berean,
  Russo, Kalantar-Zadeh, and Yeo]{ahmed2018ultrafast}
Ahmed,~H.; Rezk,~A.~R.; Carey,~B.~J.; Wang,~Y.; Mohiuddin,~M.; Berean,~K.~J.;
  Russo,~S.~P.; Kalantar-Zadeh,~K.; Yeo,~L.~Y. Ultrafast acoustofluidic
  exfoliation of stratified crystals. \emph{Adv. Mater.} \textbf{2018},
  \emph{30}, 1704756\relax
\mciteBstWouldAddEndPuncttrue
\mciteSetBstMidEndSepPunct{\mcitedefaultmidpunct}
{\mcitedefaultendpunct}{\mcitedefaultseppunct}\relax
\EndOfBibitem
\bibitem[Ahmed \latin{et~al.}(2019)Ahmed, Ahmed, Rezk, and Yeo]{ahmed2019rapid}
Ahmed,~M.; Ahmed,~H.; Rezk,~A.~R.; Yeo,~L.~Y. Rapid dry exfoliation method for
  tuneable production of molybdenum disulphide quantum dots and large
  micron-dimension sheets. \emph{Nanoscale} \textbf{2019}, \emph{11},
  11626--11633\relax
\mciteBstWouldAddEndPuncttrue
\mciteSetBstMidEndSepPunct{\mcitedefaultmidpunct}
{\mcitedefaultendpunct}{\mcitedefaultseppunct}\relax
\EndOfBibitem
\bibitem[Alijani \latin{et~al.}(2021)Alijani, Rezk, Khosravi~Farsani, Ahmed,
  Halim, Reineck, Murdoch, El-Ghazaly, Rosen, and
  Yeo]{alijani2021acoustomicrofluidic}
Alijani,~H.; Rezk,~A.~R.; Khosravi~Farsani,~M.~M.; Ahmed,~H.; Halim,~J.;
  Reineck,~P.; Murdoch,~B.~J.; El-Ghazaly,~A.; Rosen,~J.; Yeo,~L.~Y.
  Acoustomicrofluidic Synthesis of Pristine Ultrathin
  {T}i\textsubscript{3}{C}\textsubscript{2}{T}\textsubscript{z} MXene
  Nanosheets and Quantum Dots. \emph{ACS Nano} \textbf{2021}, \emph{15},
  12099--12108\relax
\mciteBstWouldAddEndPuncttrue
\mciteSetBstMidEndSepPunct{\mcitedefaultmidpunct}
{\mcitedefaultendpunct}{\mcitedefaultseppunct}\relax
\EndOfBibitem
\bibitem[El-Ghazaly \latin{et~al.}(2021)El-Ghazaly, Ahmed, Rezk, Halim,
  Persson, Yeo, and Rosen]{ghazaly2021ultrafast}
El-Ghazaly,~A.; Ahmed,~H.; Rezk,~A.~R.; Halim,~J.; Persson,~P. O.~{\AA}.;
  Yeo,~L.~Y.; Rosen,~J. Ultrafast, one-step, salt-solution-based acoustic
  synthesis of {T}i\textsubscript{3}{C}\textsubscript{2} {M}Xene. \emph{ACS
  Nano} \textbf{2021}, \emph{15}, 4287--4293\relax
\mciteBstWouldAddEndPuncttrue
\mciteSetBstMidEndSepPunct{\mcitedefaultmidpunct}
{\mcitedefaultendpunct}{\mcitedefaultseppunct}\relax
\EndOfBibitem
\bibitem[Ahmed \latin{et~al.}(2023)Ahmed, Alijani, El-Ghazaly, Halim,
  J.~Murdoch, Massahud, Rezk, Rosen, and Yeo]{ahmed2022recovery}
Ahmed,~H.; Alijani,~H.; El-Ghazaly,~A.; Halim,~J.; J.~Murdoch,~B.;
  Massahud,~E.; Rezk,~A.~R.; Rosen,~J.; Yeo,~L.~Y. Recovery of Oxidized
  Two-Dimensional Titanium Carbide
  {T}i\textsubscript{3}{C}\textsubscript{2}{T}\textsubscript{z} {M}Xene Films
  Through High Frequency Nanoscale Electromechanical Vibration. \emph{Nat.
  Commun.} \textbf{2023}, \emph{accepted for publication}\relax
\mciteBstWouldAddEndPuncttrue
\mciteSetBstMidEndSepPunct{\mcitedefaultmidpunct}
{\mcitedefaultendpunct}{\mcitedefaultseppunct}\relax
\EndOfBibitem
\bibitem[Rotter \latin{et~al.}(1998)Rotter, Wixforth, Ruile, Bernklau, and
  Riechert]{rotter1998giant}
Rotter,~M.; Wixforth,~A.; Ruile,~W.; Bernklau,~D.; Riechert,~H. Giant
  acoustoelectric effect in {G}a{A}s/{L}i{N}b{O}\textsubscript{3} hybrids.
  \emph{Appl. Phys. Lett.} \textbf{1998}, \emph{73}, 2128--2130\relax
\mciteBstWouldAddEndPuncttrue
\mciteSetBstMidEndSepPunct{\mcitedefaultmidpunct}
{\mcitedefaultendpunct}{\mcitedefaultseppunct}\relax
\EndOfBibitem
\bibitem[Kinzel \latin{et~al.}(2011)Kinzel, Rudolph, Bichler, Abstreiter,
  Finley, Koblm{\"u}ller, Wixforth, and Krenner]{kinzel2011directional}
Kinzel,~J.~B.; Rudolph,~D.; Bichler,~M.; Abstreiter,~G.; Finley,~J.~J.;
  Koblm{\"u}ller,~G.; Wixforth,~A.; Krenner,~H.~J. Directional and dynamic
  modulation of the optical emission of an individual {G}a{A}s nanowire using
  surface acoustic waves. \emph{Nano Lett.} \textbf{2011}, \emph{11},
  1512--1517\relax
\mciteBstWouldAddEndPuncttrue
\mciteSetBstMidEndSepPunct{\mcitedefaultmidpunct}
{\mcitedefaultendpunct}{\mcitedefaultseppunct}\relax
\EndOfBibitem
\bibitem[Zheng \latin{et~al.}(2016)Zheng, Zhang, Feng, Yu, Zhang, Sun, Liu,
  Duan, Pang, and Zhang]{zheng2016acoustic}
Zheng,~S.; Zhang,~H.; Feng,~Z.; Yu,~Y.; Zhang,~R.; Sun,~C.; Liu,~J.; Duan,~X.;
  Pang,~W.; Zhang,~D. Acoustic charge transport induced by the surface acoustic
  wave in chemical doped graphene. \emph{Appl. Phys. Lett.} \textbf{2016},
  \emph{109}, 183110\relax
\mciteBstWouldAddEndPuncttrue
\mciteSetBstMidEndSepPunct{\mcitedefaultmidpunct}
{\mcitedefaultendpunct}{\mcitedefaultseppunct}\relax
\EndOfBibitem
\bibitem[Rezk \latin{et~al.}(2015)Rezk, Walia, Ramanathan, Nili, Ou, Bansal,
  Friend, Bhaskaran, Yeo, and Sriram]{rezk2015acoustic}
Rezk,~A.~R.; Walia,~S.; Ramanathan,~R.; Nili,~H.; Ou,~J.~Z.; Bansal,~V.;
  Friend,~J.~R.; Bhaskaran,~M.; Yeo,~L.~Y.; Sriram,~S. Acoustic--Excitonic
  Coupling for Dynamic Photoluminescence Manipulation of Quasi-2{D}
  {M}o{S}\textsubscript{2} Nanoflakes. \emph{Adv. Opt. Mater.} \textbf{2015},
  \emph{3}, 888--894\relax
\mciteBstWouldAddEndPuncttrue
\mciteSetBstMidEndSepPunct{\mcitedefaultmidpunct}
{\mcitedefaultendpunct}{\mcitedefaultseppunct}\relax
\EndOfBibitem
\bibitem[Rezk \latin{et~al.}(2016)Rezk, Carey, Chrimes, Lau, Gibson, Zheng,
  Fuhrer, Yeo, and Kalantar-Zadeh]{rezk2016acoustically}
Rezk,~A.~R.; Carey,~B.; Chrimes,~A.~F.; Lau,~D. W.~M.; Gibson,~B.~C.;
  Zheng,~C.; Fuhrer,~M.~S.; Yeo,~L.~Y.; Kalantar-Zadeh,~K. Acoustically-driven
  trion and exciton modulation in piezoelectric two-dimensional
  {M}o{S}\textsubscript{2}. \emph{Nano Lett.} \textbf{2016}, \emph{16},
  849--855\relax
\mciteBstWouldAddEndPuncttrue
\mciteSetBstMidEndSepPunct{\mcitedefaultmidpunct}
{\mcitedefaultendpunct}{\mcitedefaultseppunct}\relax
\EndOfBibitem
\bibitem[Peng \latin{et~al.}(2022)Peng, Ripin, Ye, Zhu, Wu, Lee, Li, Taniguchi,
  Watanabe, Cao, Xu, and Li]{peng2022long}
Peng,~R.; Ripin,~A.; Ye,~Y.; Zhu,~J.; Wu,~C.; Lee,~S.; Li,~H.; Taniguchi,~T.;
  Watanabe,~K.; Cao,~T.; Xu,~X.; Li,~M. Long-range transport of 2{D} excitons
  with acoustic waves. \emph{Nat. Commun.} \textbf{2022}, \emph{13}, 1--7\relax
\mciteBstWouldAddEndPuncttrue
\mciteSetBstMidEndSepPunct{\mcitedefaultmidpunct}
{\mcitedefaultendpunct}{\mcitedefaultseppunct}\relax
\EndOfBibitem
\bibitem[Rotter \latin{et~al.}(1999)Rotter, Wixforth, Govorov, Ruile, Bernklau,
  and Riechert]{rotter1999nonlinear}
Rotter,~M.; Wixforth,~A.; Govorov,~A.~O.; Ruile,~W.; Bernklau,~D.; Riechert,~H.
  Nonlinear acoustoelectric interactions in
  {G}a{A}s/{L}i{N}b{O}\textsubscript{3} structures. \emph{Appl. Phys. Lett.}
  \textbf{1999}, \emph{75}, 965--967\relax
\mciteBstWouldAddEndPuncttrue
\mciteSetBstMidEndSepPunct{\mcitedefaultmidpunct}
{\mcitedefaultendpunct}{\mcitedefaultseppunct}\relax
\EndOfBibitem
\bibitem[Govorov \latin{et~al.}(2000)Govorov, Kalameitsev, Rotter, Wixforth,
  Kotthaus, Hoffmann, and Botkin]{govorov2000nonlinear}
Govorov,~A.~O.; Kalameitsev,~A.~V.; Rotter,~M.; Wixforth,~A.; Kotthaus,~J.~P.;
  Hoffmann,~K.-H.; Botkin,~N. Nonlinear acoustoelectric transport in a
  two-dimensional electron system. \emph{Phys. Rev. B} \textbf{2000},
  \emph{62}, 2659\relax
\mciteBstWouldAddEndPuncttrue
\mciteSetBstMidEndSepPunct{\mcitedefaultmidpunct}
{\mcitedefaultendpunct}{\mcitedefaultseppunct}\relax
\EndOfBibitem
\bibitem[Esslinger \latin{et~al.}(1992)Esslinger, Wixforth, Winkler, Kotthaus,
  Nickel, Schlapp, and L{\"o}sch]{esslinger1992acoustoelectric}
Esslinger,~A.; Wixforth,~A.; Winkler,~R.~W.; Kotthaus,~J.~P.; Nickel,~H.;
  Schlapp,~W.; L{\"o}sch,~R. Acoustoelectric study of localized states in the
  quantized hall effect. \emph{Solid State Commun.} \textbf{1992}, \emph{84},
  939--942\relax
\mciteBstWouldAddEndPuncttrue
\mciteSetBstMidEndSepPunct{\mcitedefaultmidpunct}
{\mcitedefaultendpunct}{\mcitedefaultseppunct}\relax
\EndOfBibitem
\bibitem[Rocke \latin{et~al.}(1994)Rocke, Manus, Wixforth, Sundaram, English,
  and Gossard]{rocke1994voltage}
Rocke,~C.; Manus,~S.; Wixforth,~A.; Sundaram,~M.; English,~J.~H.;
  Gossard,~A.~C. Voltage tunable acoustoelectric interaction in
  {G}a{A}s/{A}l{G}a{A}s heterojunctions. \emph{Appl. Phys. Lett.}
  \textbf{1994}, \emph{65}, 2422--2424\relax
\mciteBstWouldAddEndPuncttrue
\mciteSetBstMidEndSepPunct{\mcitedefaultmidpunct}
{\mcitedefaultendpunct}{\mcitedefaultseppunct}\relax
\EndOfBibitem
\bibitem[Ebbecke \latin{et~al.}(2004)Ebbecke, Strobl, and
  Wixforth]{ebbecke2004acoustoelectric}
Ebbecke,~J.; Strobl,~C.~J.; Wixforth,~A. Acoustoelectric current transport
  through single-walled carbon nanotubes. \emph{Phys. Rev. B} \textbf{2004},
  \emph{70}, 233401\relax
\mciteBstWouldAddEndPuncttrue
\mciteSetBstMidEndSepPunct{\mcitedefaultmidpunct}
{\mcitedefaultendpunct}{\mcitedefaultseppunct}\relax
\EndOfBibitem
\bibitem[Bandhu \latin{et~al.}(2013)Bandhu, Lawton, and
  Nash]{bandhu2013macroscopic}
Bandhu,~L.; Lawton,~L.~M.; Nash,~G.~R. Macroscopic acoustoelectric charge
  transport in graphene. \emph{Appl. Phys. Lett.} \textbf{2013}, \emph{103},
  133101\relax
\mciteBstWouldAddEndPuncttrue
\mciteSetBstMidEndSepPunct{\mcitedefaultmidpunct}
{\mcitedefaultendpunct}{\mcitedefaultseppunct}\relax
\EndOfBibitem
\bibitem[Lane \latin{et~al.}(2018)Lane, Zhang, Khasawneh, Zhou, Henriksen, and
  Pollanen]{lane2018flip}
Lane,~J.~R.; Zhang,~L.; Khasawneh,~M.~A.; Zhou,~B.~N.; Henriksen,~E.~A.;
  Pollanen,~J. Flip-chip gate-tunable acoustoelectric effect in graphene.
  \emph{J. Appl. Phys.} \textbf{2018}, \emph{124}, 194302\relax
\mciteBstWouldAddEndPuncttrue
\mciteSetBstMidEndSepPunct{\mcitedefaultmidpunct}
{\mcitedefaultendpunct}{\mcitedefaultseppunct}\relax
\EndOfBibitem
\bibitem[Zhao \latin{et~al.}(2022)Zhao, Sharma, Liang, Glasenapp, Mourokh,
  Kovalev, Huber, Prada, Tiemann, and Blick]{zhao2022acoustically}
Zhao,~P.; Sharma,~C.~H.; Liang,~R.; Glasenapp,~C.; Mourokh,~L.; Kovalev,~V.~M.;
  Huber,~P.; Prada,~M.; Tiemann,~L.; Blick,~R.~H. Acoustically Induced Giant
  Synthetic Hall Voltages in Graphene. \emph{Phys. Rev. Lett.} \textbf{2022},
  \emph{128}, 256601\relax
\mciteBstWouldAddEndPuncttrue
\mciteSetBstMidEndSepPunct{\mcitedefaultmidpunct}
{\mcitedefaultendpunct}{\mcitedefaultseppunct}\relax
\EndOfBibitem
\bibitem[Zheng \latin{et~al.}(2018)Zheng, Wu, and Zhang]{zheng2018anomalous}
Zheng,~S.; Wu,~E.; Zhang,~H. Anomalous acoustoelectric currents in few-layer
  black phosphorus nanocrystals. \emph{IEEE Trans. Nanotechnol.} \textbf{2018},
  \emph{17}, 590--595\relax
\mciteBstWouldAddEndPuncttrue
\mciteSetBstMidEndSepPunct{\mcitedefaultmidpunct}
{\mcitedefaultendpunct}{\mcitedefaultseppunct}\relax
\EndOfBibitem
\bibitem[Preciado \latin{et~al.}(2015)Preciado, Sch{\"{u}}lein, Nguyen,
  Barroso, Isarraraz, Von~Son, Lu, Michailow, M{\"o}ller, Klee, Mann, Wixforth,
  Bartels, and Krenner]{preciado2015scalable}
Preciado,~E.; Sch{\"{u}}lein,~F. J.~R.; Nguyen,~A.~E.; Barroso,~D.;
  Isarraraz,~M.; Von~Son,~G.; Lu,~I.-H.; Michailow,~W.; M{\"o}ller,~B.;
  Klee,~V.; Mann,~J.; Wixforth,~A.; Bartels,~L.; Krenner,~H.~J. Scalable
  fabrication of a hybrid field-effect and acousto-electric device by direct
  growth of monolayer {M}o{S}\textsubscript{2}/{L}i{N}b{O}\textsubscript{3}.
  \emph{Nat. Commun.} \textbf{2015}, \emph{6}, 1--8\relax
\mciteBstWouldAddEndPuncttrue
\mciteSetBstMidEndSepPunct{\mcitedefaultmidpunct}
{\mcitedefaultendpunct}{\mcitedefaultseppunct}\relax
\EndOfBibitem
\bibitem[Poole \latin{et~al.}(2015)Poole, Bandhu, and
  Nash]{poole2015acoustoelectric}
Poole,~T.; Bandhu,~L.; Nash,~G.~R. Acoustoelectric photoresponse in graphene.
  \emph{Appl. Phys. Lett.} \textbf{2015}, \emph{106}, 133107\relax
\mciteBstWouldAddEndPuncttrue
\mciteSetBstMidEndSepPunct{\mcitedefaultmidpunct}
{\mcitedefaultendpunct}{\mcitedefaultseppunct}\relax
\EndOfBibitem
\bibitem[Poole and Nash(2018)Poole, and Nash]{poole2018acoustoelectric}
Poole,~T.; Nash,~G.~R. Acoustoelectric photoresponse of graphene nanoribbons.
  \emph{J. Phys. D: Appl. Phys.} \textbf{2018}, \emph{51}, 154001\relax
\mciteBstWouldAddEndPuncttrue
\mciteSetBstMidEndSepPunct{\mcitedefaultmidpunct}
{\mcitedefaultendpunct}{\mcitedefaultseppunct}\relax
\EndOfBibitem
\bibitem[Song \latin{et~al.}(2013)Song, Li, Gao, Xu, Ueno, Tang, Cheng, and
  Tsukagoshi]{song2013high}
Song,~H.; Li,~S.; Gao,~L.; Xu,~Y.; Ueno,~K.; Tang,~J.; Cheng,~Y.;
  Tsukagoshi,~K. High-performance top-gated monolayer {S}n{S}\textsubscript{2}
  field-effect transistors and their integrated logic circuits.
  \emph{Nanoscale} \textbf{2013}, \emph{5}, 9666--9670\relax
\mciteBstWouldAddEndPuncttrue
\mciteSetBstMidEndSepPunct{\mcitedefaultmidpunct}
{\mcitedefaultendpunct}{\mcitedefaultseppunct}\relax
\EndOfBibitem
\bibitem[Weinreich(1956)]{weinreich1956acoustodynamic}
Weinreich,~G. Acoustodynamic effects in semiconductors. \emph{Phys. Rev.}
  \textbf{1956}, \emph{104}, 321--324\relax
\mciteBstWouldAddEndPuncttrue
\mciteSetBstMidEndSepPunct{\mcitedefaultmidpunct}
{\mcitedefaultendpunct}{\mcitedefaultseppunct}\relax
\EndOfBibitem
\bibitem[Rotter \latin{et~al.}(1999)Rotter, Kalameitsev, Govorov, Ruile, and
  Wixforth]{rotter1999charge}
Rotter,~M.; Kalameitsev,~A.~V.; Govorov,~A.~O.; Ruile,~W.; Wixforth,~A. Charge
  conveyance and nonlinear acoustoelectric phenomena for intense surface
  acoustic waves on a semiconductor quantum well. \emph{Phys. Rev. Lett.}
  \textbf{1999}, \emph{82}, 2171--2174\relax
\mciteBstWouldAddEndPuncttrue
\mciteSetBstMidEndSepPunct{\mcitedefaultmidpunct}
{\mcitedefaultendpunct}{\mcitedefaultseppunct}\relax
\EndOfBibitem
\bibitem[Zhan \latin{et~al.}(2020)Zhan, Zheng, Xiao, and Zhao]{zhan2020phonon}
Zhan,~S.; Zheng,~L.; Xiao,~Y.; Zhao,~L.-D. Phonon and carrier transport
  properties in low-cost and environmentally friendly {S}n{S}\textsubscript{2}:
  a promising thermoelectric material. \emph{Chem. Mater.} \textbf{2020},
  \emph{32}, 10348--10356\relax
\mciteBstWouldAddEndPuncttrue
\mciteSetBstMidEndSepPunct{\mcitedefaultmidpunct}
{\mcitedefaultendpunct}{\mcitedefaultseppunct}\relax
\EndOfBibitem
\bibitem[Zheng \latin{et~al.}(2018)Zheng, Wu, Feng, Zhang, Xie, Yu, Zhang, Li,
  Liu, Pang, Zhang, and Zhang]{zheng2018acoustically}
Zheng,~S.; Wu,~E.; Feng,~Z.; Zhang,~R.; Xie,~Y.; Yu,~Y.; Zhang,~R.; Li,~Q.;
  Liu,~J.; Pang,~W.; Zhang,~H.; Zhang,~D. Acoustically enhanced photodetection
  by a black phosphorus--{M}o{S}\textsubscript{2} van der {W}aals
  heterojunction p--n diode. \emph{Nanoscale} \textbf{2018}, \emph{10},
  10148--10153\relax
\mciteBstWouldAddEndPuncttrue
\mciteSetBstMidEndSepPunct{\mcitedefaultmidpunct}
{\mcitedefaultendpunct}{\mcitedefaultseppunct}\relax
\EndOfBibitem
\bibitem[V{\"o}lk \latin{et~al.}(2010)V{\"o}lk, Sch{\"u}lein, Knall, Reuter,
  Wieck, Truong, Kim, Petroff, Wixforth, and Krenner]{volk2010enhanced}
V{\"o}lk,~S.; Sch{\"u}lein,~F. J.~R.; Knall,~F.; Reuter,~D.; Wieck,~A.~D.;
  Truong,~T.~A.; Kim,~H.; Petroff,~P.~M.; Wixforth,~A.; Krenner,~H.~J. Enhanced
  sequential carrier capture into individual quantum dots and quantum posts
  controlled by surface acoustic waves. \emph{Nano Lett.} \textbf{2010},
  \emph{10}, 3399--3407\relax
\mciteBstWouldAddEndPuncttrue
\mciteSetBstMidEndSepPunct{\mcitedefaultmidpunct}
{\mcitedefaultendpunct}{\mcitedefaultseppunct}\relax
\EndOfBibitem
\bibitem[Hern{\'a}ndez-M{\'\i}nguez
  \latin{et~al.}(2012)Hern{\'a}ndez-M{\'\i}nguez, M\"{o}ller, Breuer,
  Pf\"{u}ller, Somaschini, Lazic, Brandt, Garc{\'\i}a-Crist{\'o}bal,
  de~Lima~Jr, Cantarero, Geelhaar, Riechert, and
  Santos]{hernandez2012acoustically}
Hern{\'a}ndez-M{\'\i}nguez,~A.; M\"{o}ller,~M.; Breuer,~S.; Pf\"{u}ller,~C.;
  Somaschini,~C.; Lazic,~S.; Brandt,~O.; Garc{\'\i}a-Crist{\'o}bal,~A.;
  de~Lima~Jr,~M.~M.; Cantarero,~A.; Geelhaar,~L.; Riechert,~H.; Santos,~P.~V.
  Acoustically driven photon antibunching in nanowires. \emph{Nano Lett.}
  \textbf{2012}, \emph{12}, 252--258\relax
\mciteBstWouldAddEndPuncttrue
\mciteSetBstMidEndSepPunct{\mcitedefaultmidpunct}
{\mcitedefaultendpunct}{\mcitedefaultseppunct}\relax
\EndOfBibitem
\bibitem[Wei{\ss} \latin{et~al.}(2016)Wei{\ss}, Kapfinger, Reichert, Finley,
  Wixforth, and Krenner]{weiss2016surface}
Wei{\ss},~M.; Kapfinger,~S.; Reichert,~T.; Finley,~J.~J.;
  Wixforth,~M.,~A.and~Kaniber; Krenner,~H.~J. Surface acoustic wave regulated
  single photon emission from a coupled quantum dot--nanocavity system.
  \emph{Appl. Phys. Lett.} \textbf{2016}, \emph{109}, 033105\relax
\mciteBstWouldAddEndPuncttrue
\mciteSetBstMidEndSepPunct{\mcitedefaultmidpunct}
{\mcitedefaultendpunct}{\mcitedefaultseppunct}\relax
\EndOfBibitem
\bibitem[Rocke \latin{et~al.}(1997)Rocke, Zimmermann, Wixforth, Kotthaus,
  B{\"o}hm, and Weimann]{rocke1997acoustically}
Rocke,~C.; Zimmermann,~S.; Wixforth,~A.; Kotthaus,~J.~P.; B{\"o}hm,~G.;
  Weimann,~G. Acoustically driven storage of light in a quantum well.
  \emph{Phys. Rev. Lett.} \textbf{1997}, \emph{78}, 4099\relax
\mciteBstWouldAddEndPuncttrue
\mciteSetBstMidEndSepPunct{\mcitedefaultmidpunct}
{\mcitedefaultendpunct}{\mcitedefaultseppunct}\relax
\EndOfBibitem
\bibitem[Parmenter(1953)]{parmenter1953acousto}
Parmenter,~R.~H. The acousto-electric effect. \emph{Phys. Rev.} \textbf{1953},
  \emph{89}, 990--998\relax
\mciteBstWouldAddEndPuncttrue
\mciteSetBstMidEndSepPunct{\mcitedefaultmidpunct}
{\mcitedefaultendpunct}{\mcitedefaultseppunct}\relax
\EndOfBibitem
\bibitem[Dovesi \latin{et~al.}(2014)Dovesi, Orlando, Erba, Zicovich-Wilson,
  Civalleri, Casassa, Maschio, Ferrabone, De~La~Pierre, D'Arco, Noël, Causà,
  Rérat, and Kirtman]{dovesi2014crystal14}
Dovesi,~R.; Orlando,~R.; Erba,~A.; Zicovich-Wilson,~C.~M.; Civalleri,~B.;
  Casassa,~S.; Maschio,~L.; Ferrabone,~M.; De~La~Pierre,~M.; D'Arco,~P.;
  Noël,~Y.; Causà,~M.; Rérat,~M.; Kirtman,~B. CRYSTAL14: A program for the
  ab initio investigation of crystalline solids. 2014\relax
\mciteBstWouldAddEndPuncttrue
\mciteSetBstMidEndSepPunct{\mcitedefaultmidpunct}
{\mcitedefaultendpunct}{\mcitedefaultseppunct}\relax
\EndOfBibitem
\bibitem[Dovesi \latin{et~al.}(2014)Dovesi, Saunders, Roetti, Orlando,
  Zicovich-Wilson, Pascale, Civalleri, Doll, Harrison, Bush, D’Arco, Llunel,
  Caus'a, and No\"{e}l]{dovesi2014crystal142}
Dovesi,~R.; Saunders,~V.~R.; Roetti,~C.; Orlando,~R.; Zicovich-Wilson,~C.~M.;
  Pascale,~F.; Civalleri,~B.; Doll,~K.; Harrison,~N.~M.; Bush,~I.~J.;
  D’Arco,~P.; Llunel,~M.,~l; Caus'a,~M.; No\"{e}l,~Y. CRYSTAL14 User's
  Manual. \url{http://www.crystal.unito.it}, 2014\relax
\mciteBstWouldAddEndPuncttrue
\mciteSetBstMidEndSepPunct{\mcitedefaultmidpunct}
{\mcitedefaultendpunct}{\mcitedefaultseppunct}\relax
\EndOfBibitem
\bibitem[Gu \latin{et~al.}(2007)Gu, Srot, Sigle, Koch, van Aken, Scholz, Thapa,
  Kirchner, Jetter, and R{\"u}hle]{gu2007band}
Gu,~L.; Srot,~V.; Sigle,~W.; Koch,~C.; van Aken,~P.; Scholz,~F.; Thapa,~S.~B.;
  Kirchner,~C.; Jetter,~M.; R{\"u}hle,~M. Band-gap measurements of direct and
  indirect semiconductors using monochromated electrons. \emph{Phys. Rev. B}
  \textbf{2007}, \emph{75}, 195214\relax
\mciteBstWouldAddEndPuncttrue
\mciteSetBstMidEndSepPunct{\mcitedefaultmidpunct}
{\mcitedefaultendpunct}{\mcitedefaultseppunct}\relax
\EndOfBibitem
\bibitem[Bunge \latin{et~al.}(2016)Bunge, Beckers, and Gries]{bunge2016polymer}
Bunge,~C.-A.; Beckers,~M.; Gries,~T. \emph{Polymer optical fibres: fibre types,
  materials, fabrication, characterisation and applications}; Woodhead
  Publishing, 2016\relax
\mciteBstWouldAddEndPuncttrue
\mciteSetBstMidEndSepPunct{\mcitedefaultmidpunct}
{\mcitedefaultendpunct}{\mcitedefaultseppunct}\relax
\EndOfBibitem
\bibitem[Cheng \latin{et~al.}(2010)Cheng, Ko, Chen, Peng, Luo, Liu, and
  Tseng]{cheng2010competitiveness}
Cheng,~T.-H.; Ko,~C.-Y.; Chen,~C.-Y.; Peng,~K.-L.; Luo,~G.-L.; Liu,~C.~W.;
  Tseng,~H.-H. Competitiveness between direct and indirect radiative
  transitions of Ge. \emph{Appl. Phys. Lett.} \textbf{2010}, \emph{96},
  091105\relax
\mciteBstWouldAddEndPuncttrue
\mciteSetBstMidEndSepPunct{\mcitedefaultmidpunct}
{\mcitedefaultendpunct}{\mcitedefaultseppunct}\relax
\EndOfBibitem
\bibitem[Chaves \latin{et~al.}(2020)Chaves, Azadani, Alsalman, Da~Costa,
  Frisenda, Chaves, Song, Kim, He, Zhou, Castellanos-Gomez, Peeters, Liu,
  Hinkle, Oh, Ye, Koester, Lee, Avouris, Wang, and Low]{chaves2020bandgap}
Chaves,~A. \latin{et~al.}  Bandgap engineering of two-dimensional semiconductor
  materials. \emph{npj 2D Mater. Appl.} \textbf{2020}, \emph{4}, 1--21\relax
\mciteBstWouldAddEndPuncttrue
\mciteSetBstMidEndSepPunct{\mcitedefaultmidpunct}
{\mcitedefaultendpunct}{\mcitedefaultseppunct}\relax
\EndOfBibitem
\bibitem[Huang \latin{et~al.}(2021)Huang, Lee, Nakayama, Schrock, Cao, Cho,
  Bazan, and Nguyen]{huang2021understanding}
Huang,~J.; Lee,~J.; Nakayama,~H.; Schrock,~M.; Cao,~D.~X.; Cho,~K.;
  Bazan,~G.~C.; Nguyen,~T.-Q. Understanding and countering
  illumination-sensitive dark current: toward organic photodetectors with
  reliable high detectivity. \emph{ACS Nano} \textbf{2021}, \emph{15},
  1753--1763\relax
\mciteBstWouldAddEndPuncttrue
\mciteSetBstMidEndSepPunct{\mcitedefaultmidpunct}
{\mcitedefaultendpunct}{\mcitedefaultseppunct}\relax
\EndOfBibitem
\bibitem[Satterthwaite \latin{et~al.}(2018)Satterthwaite, Yalamarthy,
  Scandrette, Newaz, and Senesky]{satterthwaite2018high}
Satterthwaite,~P.~F.; Yalamarthy,~A.~S.; Scandrette,~N.~A.; Newaz,~A. K.~M.;
  Senesky,~D.~G. High responsivity, low dark current ultraviolet photodetectors
  based on two-dimensional electron gas interdigitated transducers. \emph{ACS
  Photonics} \textbf{2018}, \emph{5}, 4277--4282\relax
\mciteBstWouldAddEndPuncttrue
\mciteSetBstMidEndSepPunct{\mcitedefaultmidpunct}
{\mcitedefaultendpunct}{\mcitedefaultseppunct}\relax
\EndOfBibitem
\bibitem[Gonzalez and Oleynik(2016)Gonzalez, and Oleynik]{gonzalez2016layer}
Gonzalez,~J.~M.; Oleynik,~I.~I. Layer-dependent properties of
  SnS\textsubscript{2} and SnSe\textsubscript{2} two-dimensional materials.
  \emph{Phys. Rev. B} \textbf{2016}, \emph{94}, 125443\relax
\mciteBstWouldAddEndPuncttrue
\mciteSetBstMidEndSepPunct{\mcitedefaultmidpunct}
{\mcitedefaultendpunct}{\mcitedefaultseppunct}\relax
\EndOfBibitem
\bibitem[Seo \latin{et~al.}(2017)Seo, Shin, Ham, Lee, Lee, Choi, and
  Jeon]{seo2017thickness}
Seo,~W.; Shin,~S.; Ham,~G.; Lee,~J.; Lee,~S.; Choi,~H.; Jeon,~H.
  Thickness-dependent structure and properties of SnS\textsubscript{2} thin
  films prepared by atomic layer deposition. \emph{Jpn. J. Appl. Phys.}
  \textbf{2017}, \emph{56}, 031201\relax
\mciteBstWouldAddEndPuncttrue
\mciteSetBstMidEndSepPunct{\mcitedefaultmidpunct}
{\mcitedefaultendpunct}{\mcitedefaultseppunct}\relax
\EndOfBibitem
\bibitem[Wang \latin{et~al.}(2020)Wang, Vu, Lu, Xu, Liu, Ou, and
  Li]{wang2020piezoelectric}
Wang,~Y.; Vu,~L.-M.; Lu,~T.; Xu,~C.; Liu,~Y.; Ou,~J.~Z.; Li,~Y. Piezoelectric
  responses of mechanically exfoliated two-dimensional {S}n{S}\textsubscript{2}
  nanosheets. \emph{ACS Appl. Mater. Interfaces} \textbf{2020}, \emph{12},
  51662--51668\relax
\mciteBstWouldAddEndPuncttrue
\mciteSetBstMidEndSepPunct{\mcitedefaultmidpunct}
{\mcitedefaultendpunct}{\mcitedefaultseppunct}\relax
\EndOfBibitem
\bibitem[Liang \latin{et~al.}(2019)Liang, Wang, Zhang, Wei, Lim, Zhu, Hu, Wei,
  Lee, Sow, Zhang, and Wee]{liang2019high}
Liang,~Q.; Wang,~Q.; Zhang,~Q.; Wei,~J.; Lim,~S.~X.; Zhu,~R.; Hu,~J.; Wei,~W.;
  Lee,~C.; Sow,~C.; Zhang,~W.; Wee,~A. T.~S. High-performance, room
  temperature, ultra-broadband photodetectors based on air-stable \ch{PdSe2}.
  \emph{Adv. Mater.} \textbf{2019}, \emph{31}, 1807609\relax
\mciteBstWouldAddEndPuncttrue
\mciteSetBstMidEndSepPunct{\mcitedefaultmidpunct}
{\mcitedefaultendpunct}{\mcitedefaultseppunct}\relax
\EndOfBibitem
\bibitem[Gra{\v z}ulis \latin{et~al.}(2012)Gra{\v z}ulis, Da{\v s}kevi{\v c},
  Merkys, Chateigner, Lutterotti, Quirós, Serebryanaya, Moeck, Downs, and
  Le~Bail]{grazulis2012}
Gra{\v z}ulis,~S.; Da{\v s}kevi{\v c},~A.; Merkys,~A.; Chateigner,~D.;
  Lutterotti,~L.; Quirós,~M.; Serebryanaya,~N.~R.; Moeck,~P.; Downs,~R.~T.;
  Le~Bail,~A. {C}rystallography {O}pen {D}atabase ({COD}): an open-access
  collection of crystal structures and platform for world-wide collaboration.
  \emph{Nucleic Acids Res.} \textbf{2012}, \emph{40}, D420--D427\relax
\mciteBstWouldAddEndPuncttrue
\mciteSetBstMidEndSepPunct{\mcitedefaultmidpunct}
{\mcitedefaultendpunct}{\mcitedefaultseppunct}\relax
\EndOfBibitem
\bibitem[Heyd \latin{et~al.}(2003)Heyd, Scuseria, and
  Ernzerhof]{heyd2003hybrid}
Heyd,~J.; Scuseria,~G.~E.; Ernzerhof,~M. Hybrid functionals based on a screened
  Coulomb potential. \emph{J. Chem. Phys.} \textbf{2003}, \emph{118},
  8207--8215\relax
\mciteBstWouldAddEndPuncttrue
\mciteSetBstMidEndSepPunct{\mcitedefaultmidpunct}
{\mcitedefaultendpunct}{\mcitedefaultseppunct}\relax
\EndOfBibitem
\bibitem[Krukau \latin{et~al.}(2006)Krukau, Vydrov, Izmaylov, and
  Scuseria]{krukau2006influence}
Krukau,~A.~V.; Vydrov,~O.~A.; Izmaylov,~A.~F.; Scuseria,~G.~E. Influence of the
  exchange screening parameter on the performance of screened hybrid
  functionals. \emph{J. Chem. Phys.} \textbf{2006}, \emph{125}, 224106\relax
\mciteBstWouldAddEndPuncttrue
\mciteSetBstMidEndSepPunct{\mcitedefaultmidpunct}
{\mcitedefaultendpunct}{\mcitedefaultseppunct}\relax
\EndOfBibitem
\bibitem[Laun \latin{et~al.}(2018)Laun, Vilela~Oliveira, and
  Bredow]{laun2018consistent}
Laun,~J.; Vilela~Oliveira,~D.; Bredow,~T. Consistent gaussian basis sets of
  double-and triple-zeta valence with polarization quality of the fifth period
  for solid-state calculations. \emph{J. Comput. Chem.} \textbf{2018},
  \emph{39}, 1285--1290\relax
\mciteBstWouldAddEndPuncttrue
\mciteSetBstMidEndSepPunct{\mcitedefaultmidpunct}
{\mcitedefaultendpunct}{\mcitedefaultseppunct}\relax
\EndOfBibitem
\bibitem[Grimme \latin{et~al.}(2010)Grimme, Antony, Ehrlich, and
  Krieg]{grimme2010consistent}
Grimme,~S.; Antony,~J.; Ehrlich,~S.; Krieg,~H. A consistent and accurate ab
  initio parametrization of density functional dispersion correction
  ({DFT}-{D}) for the 94 elements H-Pu. \emph{J. Chem. Phys.} \textbf{2010},
  \emph{132}, 154104\relax
\mciteBstWouldAddEndPuncttrue
\mciteSetBstMidEndSepPunct{\mcitedefaultmidpunct}
{\mcitedefaultendpunct}{\mcitedefaultseppunct}\relax
\EndOfBibitem
\bibitem[Aroyo \latin{et~al.}(2014)Aroyo, Orobengoa, de~la Flor, Tasci,
  Perez-Mato, and Wondratschek]{aroyo2014brillouin}
Aroyo,~M.~I.; Orobengoa,~D.; de~la Flor,~G.; Tasci,~E.~S.; Perez-Mato,~J.~M.;
  Wondratschek,~H. Brillouin-zone database on the Bilbao Crystallographic
  Server. \emph{Acta Crystallogr., Sect. A: Found. Crystallogr.} \textbf{2014},
  \emph{70}, 126--137\relax
\mciteBstWouldAddEndPuncttrue
\mciteSetBstMidEndSepPunct{\mcitedefaultmidpunct}
{\mcitedefaultendpunct}{\mcitedefaultseppunct}\relax
\EndOfBibitem
\end{mcitethebibliography}


\providecommand{\latin}[1]{#1}
\makeatletter
\providecommand{\doi}
  {\begingroup\let\do\@makeother\dospecials
  \catcode`\{=1 \catcode`\}=2 \doi@aux}
\providecommand{\doi@aux}[1]{\endgroup\texttt{#1}}
\makeatother
\providecommand*\mcitethebibliography{\thebibliography}
\csname @ifundefined\endcsname{endmcitethebibliography}
  {\let\endmcitethebibliography\endthebibliography}{}
\begin{mcitethebibliography}{44}
\providecommand*\natexlab[1]{#1}
\providecommand*\mciteSetBstSublistMode[1]{}
\providecommand*\mciteSetBstMaxWidthForm[2]{}
\providecommand*\mciteBstWouldAddEndPuncttrue
  {\def\EndOfBibitem{\unskip.}}
\providecommand*\mciteBstWouldAddEndPunctfalse
  {\let\EndOfBibitem\relax}
\providecommand*\mciteSetBstMidEndSepPunct[3]{}
\providecommand*\mciteSetBstSublistLabelBeginEnd[3]{}
\providecommand*\EndOfBibitem{}
\mciteSetBstSublistMode{f}
\mciteSetBstMaxWidthForm{subitem}{(\alph{mcitesubitemcount})}
\mciteSetBstSublistLabelBeginEnd
  {\mcitemaxwidthsubitemform\space}
  {\relax}
  {\relax}

\bibitem[Weinreich(1956)]{weinreich1956acoustodynamic}
Weinreich,~G. Acoustodynamic effects in semiconductors. \emph{Phys. Rev.}
  \textbf{1956}, \emph{104}, 321--324\relax
\mciteBstWouldAddEndPuncttrue
\mciteSetBstMidEndSepPunct{\mcitedefaultmidpunct}
{\mcitedefaultendpunct}{\mcitedefaultseppunct}\relax
\EndOfBibitem
\bibitem[Rotter \latin{et~al.}(1998)Rotter, Wixforth, Ruile, Bernklau, and
  Riechert]{rotter1998giant}
Rotter,~M.; Wixforth,~A.; Ruile,~W.; Bernklau,~D.; Riechert,~H. Giant
  acoustoelectric effect in {G}a{A}s/{L}i{N}b{O}\textsubscript{3} hybrids.
  \emph{Appl. Phys. Lett.} \textbf{1998}, \emph{73}, 2128--2130\relax
\mciteBstWouldAddEndPuncttrue
\mciteSetBstMidEndSepPunct{\mcitedefaultmidpunct}
{\mcitedefaultendpunct}{\mcitedefaultseppunct}\relax
\EndOfBibitem
\bibitem[Rotter \latin{et~al.}(1999)Rotter, Kalameitsev, Govorov, Ruile, and
  Wixforth]{rotter1999charge}
Rotter,~M.; Kalameitsev,~A.~V.; Govorov,~A.~O.; Ruile,~W.; Wixforth,~A. Charge
  conveyance and nonlinear acoustoelectric phenomena for intense surface
  acoustic waves on a semiconductor quantum well. \emph{Phys. Rev. Lett.}
  \textbf{1999}, \emph{82}, 2171--2174\relax
\mciteBstWouldAddEndPuncttrue
\mciteSetBstMidEndSepPunct{\mcitedefaultmidpunct}
{\mcitedefaultendpunct}{\mcitedefaultseppunct}\relax
\EndOfBibitem
\bibitem[Mitta \latin{et~al.}(2020)Mitta, Choi, Nipane, Ali, Kim, Teherani,
  Hone, and Yoo]{mitta2020electrical}
Mitta,~S.~B.; Choi,~M.~S.; Nipane,~A.; Ali,~F.; Kim,~C.; Teherani,~J.~T.;
  Hone,~J.; Yoo,~W.~J. Electrical characterization of 2{D} materials-based
  field-effect transistors. \emph{2D Mater.} \textbf{2020}, \emph{8},
  012002\relax
\mciteBstWouldAddEndPuncttrue
\mciteSetBstMidEndSepPunct{\mcitedefaultmidpunct}
{\mcitedefaultendpunct}{\mcitedefaultseppunct}\relax
\EndOfBibitem
\bibitem[Zhang \latin{et~al.}(2018)Zhang, Balaji, Mehta, Heyns, Caymax, Radu,
  Vandervorst, and Delabie]{zhang2018formation}
Zhang,~H.; Balaji,~Y.; Mehta,~A.~N.; Heyns,~M.; Caymax,~M.; Radu,~I.;
  Vandervorst,~W.; Delabie,~A. Formation mechanism of 2{D}
  {S}n{S}\textsubscript{2} and {S}n{S} by chemical vapor deposition using
  {S}n{C}l\textsubscript{4} and {H}\textsubscript{2}{S}. \emph{J. Mater. Chem.
  C} \textbf{2018}, \emph{6}, 6172--6178\relax
\mciteBstWouldAddEndPuncttrue
\mciteSetBstMidEndSepPunct{\mcitedefaultmidpunct}
{\mcitedefaultendpunct}{\mcitedefaultseppunct}\relax
\EndOfBibitem
\bibitem[Tian \latin{et~al.}(2020)Tian, Meng, Yang, Fan, Yuan, An, Sun, Zhang,
  Wang, Zheng, Wei, and Li]{tian2020visible}
Tian,~H.; Meng,~X.; Yang,~J.; Fan,~C.; Yuan,~S.; An,~X.; Sun,~C.; Zhang,~Y.;
  Wang,~M.; Zheng,~H.; Wei,~Z.; Li,~E. Visible phototransistors based on
  vertical nanolayered heterostructures of {S}n{S}/{S}n{S}\textsubscript{2}
  p--n and {S}n{S}e\textsubscript{2}/{S}n{S}\textsubscript{2} n--n nanoflakes.
  \emph{ACS Appl. Nano Mater.} \textbf{2020}, \emph{3}, 6847--6854\relax
\mciteBstWouldAddEndPuncttrue
\mciteSetBstMidEndSepPunct{\mcitedefaultmidpunct}
{\mcitedefaultendpunct}{\mcitedefaultseppunct}\relax
\EndOfBibitem
\bibitem[Long \latin{et~al.}(2019)Long, Wang, Fang, and Hu]{long2019progress}
Long,~M.; Wang,~P.; Fang,~H.; Hu,~W. Progress, challenges, and opportunities
  for 2{D} material based photodetectors. \emph{Adv. Funct. Mater.}
  \textbf{2019}, \emph{29}, 1803807\relax
\mciteBstWouldAddEndPuncttrue
\mciteSetBstMidEndSepPunct{\mcitedefaultmidpunct}
{\mcitedefaultendpunct}{\mcitedefaultseppunct}\relax
\EndOfBibitem
\bibitem[Sett \latin{et~al.}(2021)Sett, Parappurath, Gill, Chauhan, and
  Ghosh]{sett2021engineering}
Sett,~S.; Parappurath,~A.; Gill,~N.~K.; Chauhan,~N.; Ghosh,~A. Engineering
  sensitivity and spectral range of photodetection in van der {W}aals materials
  and hybrids. \emph{Nano Express} \textbf{2021}, \emph{3}, 014001\relax
\mciteBstWouldAddEndPuncttrue
\mciteSetBstMidEndSepPunct{\mcitedefaultmidpunct}
{\mcitedefaultendpunct}{\mcitedefaultseppunct}\relax
\EndOfBibitem
\bibitem[Krishnamurthi \latin{et~al.}(2020)Krishnamurthi, Khan, Ahmed,
  Zavabeti, Tawfik, Jain, Spencer, Balendhran, Crozier, Li, Fu, Mohiuddin, Low,
  Shabbir, Boes, Mitchell, McConville, Li, Kalantar-Zadeh, Mahmood, and
  Walia]{krishnamurthi2020liquid}
Krishnamurthi,~V. \latin{et~al.}  Liquid-Metal Synthesized Ultrathin {SnS}
  Layers for High-Performance Broadband Photodetectors. \emph{Adv. Mater.}
  \textbf{2020}, \emph{32}, 2004247\relax
\mciteBstWouldAddEndPuncttrue
\mciteSetBstMidEndSepPunct{\mcitedefaultmidpunct}
{\mcitedefaultendpunct}{\mcitedefaultseppunct}\relax
\EndOfBibitem
\bibitem[Huang \latin{et~al.}(2015)Huang, Deng, Xu, Wang, Wang, Wang, Wang,
  Zhan, Li, Luo, and He]{huang2015highly}
Huang,~Y.; Deng,~H.-X.; Xu,~K.; Wang,~Z.-X.; Wang,~Q.-S.; Wang,~F.-M.;
  Wang,~F.; Zhan,~X.-Y.; Li,~S.-S.; Luo,~J.-W.; He,~J. Highly sensitive and
  fast phototransistor based on large size {CVD}-grown {S}n{S}\textsubscript{2}
  nanosheets. \emph{Nanoscale} \textbf{2015}, \emph{7}, 14093--14099\relax
\mciteBstWouldAddEndPuncttrue
\mciteSetBstMidEndSepPunct{\mcitedefaultmidpunct}
{\mcitedefaultendpunct}{\mcitedefaultseppunct}\relax
\EndOfBibitem
\bibitem[Su \latin{et~al.}(2015)Su, Hadjiev, Loya, Zhang, Lei, Maharjan, Dong,
  Ajayan, Lou, and Peng]{su2015chemical}
Su,~G.; Hadjiev,~V.~G.; Loya,~P.~E.; Zhang,~J.; Lei,~S.; Maharjan,~S.;
  Dong,~P.; Ajayan,~P.~M.; Lou,~J.; Peng,~H. Chemical vapor deposition of thin
  crystals of layered semiconductor {S}n{S}\textsubscript{2} for fast
  photodetection application. \emph{Nano Lett.} \textbf{2015}, \emph{15},
  506--513\relax
\mciteBstWouldAddEndPuncttrue
\mciteSetBstMidEndSepPunct{\mcitedefaultmidpunct}
{\mcitedefaultendpunct}{\mcitedefaultseppunct}\relax
\EndOfBibitem
\bibitem[Tao \latin{et~al.}(2015)Tao, Wu, Wang, and Wang]{tao2015flexible}
Tao,~Y.; Wu,~X.; Wang,~W.; Wang,~J. Flexible photodetector from ultraviolet to
  near infrared based on a {S}n{S}\textsubscript{2} nanosheet microsphere film.
  \emph{J. Mater. Chem. C} \textbf{2015}, \emph{3}, 1347--1353\relax
\mciteBstWouldAddEndPuncttrue
\mciteSetBstMidEndSepPunct{\mcitedefaultmidpunct}
{\mcitedefaultendpunct}{\mcitedefaultseppunct}\relax
\EndOfBibitem
\bibitem[Xia \latin{et~al.}(2015)Xia, Zhu, Wang, Huang, Huang, and
  Meng]{xia2015large}
Xia,~J.; Zhu,~D.; Wang,~L.; Huang,~B.; Huang,~X.; Meng,~X.-M. Large-scale
  growth of two-dimensional {S}n{S}\textsubscript{2} crystals driven by screw
  dislocations and application to photodetectors. \emph{Adv. Funct. Mater.}
  \textbf{2015}, \emph{25}, 4255--4261\relax
\mciteBstWouldAddEndPuncttrue
\mciteSetBstMidEndSepPunct{\mcitedefaultmidpunct}
{\mcitedefaultendpunct}{\mcitedefaultseppunct}\relax
\EndOfBibitem
\bibitem[Fan \latin{et~al.}(2016)Fan, Li, Lu, Deng, Wei, and
  Li]{fan2016wavelength}
Fan,~C.; Li,~Y.; Lu,~F.; Deng,~H.-X.; Wei,~Z.; Li,~J. Wavelength dependent
  {UV}-{V}is photodetectors from {S}n{S}\textsubscript{2} flakes. \emph{RSC
  Adv.} \textbf{2016}, \emph{6}, 422--427\relax
\mciteBstWouldAddEndPuncttrue
\mciteSetBstMidEndSepPunct{\mcitedefaultmidpunct}
{\mcitedefaultendpunct}{\mcitedefaultseppunct}\relax
\EndOfBibitem
\bibitem[Yang \latin{et~al.}(2016)Yang, Li, Hu, Deng, Dong, Yang, Qiao, Yuan,
  and Song]{yang2016controllable}
Yang,~D.; Li,~B.; Hu,~C.; Deng,~H.; Dong,~D.; Yang,~X.; Qiao,~K.; Yuan,~S.;
  Song,~H. Controllable Growth Orientation of {S}n{S}\textsubscript{2} Flakes
  for Low-Noise, High-Photoswitching Ratio, and Ultrafast Phototransistors.
  \emph{Adv. Opt. Mater.} \textbf{2016}, \emph{4}, 419--426\relax
\mciteBstWouldAddEndPuncttrue
\mciteSetBstMidEndSepPunct{\mcitedefaultmidpunct}
{\mcitedefaultendpunct}{\mcitedefaultseppunct}\relax
\EndOfBibitem
\bibitem[Zhou \latin{et~al.}(2016)Zhou, Zhang, Gan, Li, and
  Zhai]{zhou2016large}
Zhou,~X.; Zhang,~Q.; Gan,~L.; Li,~H.; Zhai,~T. Large-size growth of ultrathin
  {S}n{S}\textsubscript{2} nanosheets and high performance for
  phototransistors. \emph{Adv. Funct. Mater.} \textbf{2016}, \emph{26},
  4405--4413\relax
\mciteBstWouldAddEndPuncttrue
\mciteSetBstMidEndSepPunct{\mcitedefaultmidpunct}
{\mcitedefaultendpunct}{\mcitedefaultseppunct}\relax
\EndOfBibitem
\bibitem[Wu \latin{et~al.}(2016)Wu, Tao, Wu, and Wu]{wu2016ultrathin}
Wu,~J.-J.; Tao,~Y.-R.; Wu,~Y.; Wu,~X.-C. Ultrathin {S}n{S}\textsubscript{2}
  nanosheets of ultrasonic synthesis and their photoresponses from ultraviolet
  to near-infrared. \emph{Sens. Actuators B: Chem.} \textbf{2016}, \emph{231},
  211--217\relax
\mciteBstWouldAddEndPuncttrue
\mciteSetBstMidEndSepPunct{\mcitedefaultmidpunct}
{\mcitedefaultendpunct}{\mcitedefaultseppunct}\relax
\EndOfBibitem
\bibitem[Gao \latin{et~al.}(2016)Gao, Chen, Zeng, Ge, Yang, Song, and
  Tang]{gao2016broadband}
Gao,~L.; Chen,~C.; Zeng,~K.; Ge,~C.; Yang,~D.; Song,~H.; Tang,~J. Broadband,
  sensitive and spectrally distinctive {S}n{S}\textsubscript{2} nanosheet/PbS
  colloidal quantum dot hybrid photodetector. \emph{Light Sci. Appl.}
  \textbf{2016}, \emph{5}, e16126--e16126\relax
\mciteBstWouldAddEndPuncttrue
\mciteSetBstMidEndSepPunct{\mcitedefaultmidpunct}
{\mcitedefaultendpunct}{\mcitedefaultseppunct}\relax
\EndOfBibitem
\bibitem[Li \latin{et~al.}(2017)Li, Xing, Zhong, Huang, Lei, Zhang, Li, and
  Wei]{li2017two}
Li,~B.; Xing,~T.; Zhong,~M.; Huang,~L.; Lei,~N.; Zhang,~J.; Li,~J.; Wei,~Z. A
  two-dimensional {F}e-doped {S}n{S}\textsubscript{2} magnetic semiconductor.
  \emph{Nat. Commun.} \textbf{2017}, \emph{8}, 1--7\relax
\mciteBstWouldAddEndPuncttrue
\mciteSetBstMidEndSepPunct{\mcitedefaultmidpunct}
{\mcitedefaultendpunct}{\mcitedefaultseppunct}\relax
\EndOfBibitem
\bibitem[Jia \latin{et~al.}(2018)Jia, Tang, Pan, Long, and
  Gu]{jia2018thickness}
Jia,~X.; Tang,~C.; Pan,~R.; Long,~Y.; Gu,~J.,~C.and~Li Thickness-dependently
  enhanced photodetection performance of vertically grown
  {S}n{S}\textsubscript{2} nanoflakes with large size and high production.
  \emph{ACS Appl. Mater. Interfaces} \textbf{2018}, \emph{10},
  18073--18081\relax
\mciteBstWouldAddEndPuncttrue
\mciteSetBstMidEndSepPunct{\mcitedefaultmidpunct}
{\mcitedefaultendpunct}{\mcitedefaultseppunct}\relax
\EndOfBibitem
\bibitem[Liu \latin{et~al.}(2019)Liu, Liu, Chen, Miao, Liu, Li, Tang, Chen,
  Liu, Li, Wei, and Duan]{liu2019tunable}
Liu,~J.; Liu,~X.; Chen,~Z.; Miao,~L.; Liu,~X.; Li,~B.; Tang,~L.; Chen,~K.;
  Liu,~Y.; Li,~J.; Wei,~Z.; Duan,~X. Tunable {S}chottky barrier width and
  enormously enhanced photoresponsivity in {S}b doped {S}n{S}\textsubscript{2}
  monolayer. \emph{Nano Res.} \textbf{2019}, \emph{12}, 463--468\relax
\mciteBstWouldAddEndPuncttrue
\mciteSetBstMidEndSepPunct{\mcitedefaultmidpunct}
{\mcitedefaultendpunct}{\mcitedefaultseppunct}\relax
\EndOfBibitem
\bibitem[Yu \latin{et~al.}(2020)Yu, Suleiman, Zheng, Zhou, and
  Zhai]{yu2020giant}
Yu,~J.; Suleiman,~A.~A.; Zheng,~Z.; Zhou,~X.; Zhai,~T. Giant-enhanced
  {S}n{S}\textsubscript{2} photodetectors with broadband response through
  oxygen plasma treatment. \emph{Adv. Funct. Mater.} \textbf{2020}, \emph{30},
  2001650\relax
\mciteBstWouldAddEndPuncttrue
\mciteSetBstMidEndSepPunct{\mcitedefaultmidpunct}
{\mcitedefaultendpunct}{\mcitedefaultseppunct}\relax
\EndOfBibitem
\bibitem[Lei \latin{et~al.}(2020)Lei, Luo, Yang, Cai, Qi, Gu, and
  Zheng]{lei2020thermal}
Lei,~Y.; Luo,~J.; Yang,~X.; Cai,~T.; Qi,~R.; Gu,~L.; Zheng,~Z. Thermal
  evaporation of large-area {S}n{S}\textsubscript{2} thin films with a
  {UV}-to-{NIR} photoelectric response for flexible photodetector applications.
  \emph{ACS Appl. Mater. Interfaces} \textbf{2020}, \emph{12},
  24940--24950\relax
\mciteBstWouldAddEndPuncttrue
\mciteSetBstMidEndSepPunct{\mcitedefaultmidpunct}
{\mcitedefaultendpunct}{\mcitedefaultseppunct}\relax
\EndOfBibitem
\bibitem[Fan \latin{et~al.}(2021)Fan, Liu, Yuan, Meng, An, Jing, Sun, Zhang,
  Zhang, Wang, Zheng, and Li]{fan2021enhanced}
Fan,~C.; Liu,~Z.; Yuan,~S.; Meng,~X.; An,~X.; Jing,~Y.; Sun,~C.; Zhang,~Y.;
  Zhang,~Z.; Wang,~M.; Zheng,~H.; Li,~E. Enhanced photodetection performance of
  photodetectors based on indium-doped tin disulfide few layers. \emph{ACS
  Appl. Mater. Interfaces} \textbf{2021}, \emph{13}, 35889--35896\relax
\mciteBstWouldAddEndPuncttrue
\mciteSetBstMidEndSepPunct{\mcitedefaultmidpunct}
{\mcitedefaultendpunct}{\mcitedefaultseppunct}\relax
\EndOfBibitem
\bibitem[Fu \latin{et~al.}(2021)Fu, Mo, Ostrikov, Gu, Nan, and
  Xiao]{fu2021controllable}
Fu,~Q.; Mo,~H.; Ostrikov,~K.~K.; Gu,~X.; Nan,~H.; Xiao,~S. Controllable
  synthesis of {S}n{S}\textsubscript{2} flakes and
  {M}o{S}\textsubscript{2}/{S}n{S}\textsubscript{2} heterostructures by
  confined-space chemical vapor deposition. \emph{CrystEngComm} \textbf{2021},
  \emph{23}, 2563--2571\relax
\mciteBstWouldAddEndPuncttrue
\mciteSetBstMidEndSepPunct{\mcitedefaultmidpunct}
{\mcitedefaultendpunct}{\mcitedefaultseppunct}\relax
\EndOfBibitem
\bibitem[Shooshtari \latin{et~al.}(2021)Shooshtari, Esfandiar, Orooji,
  Samadpour, and Rahighi]{shooshtari2021ultrafast}
Shooshtari,~L.; Esfandiar,~A.; Orooji,~Y.; Samadpour,~M.; Rahighi,~R. Ultrafast
  and stable planar photodetector based on {S}n{S}\textsubscript{2}
  nanosheets/perovskite structure. \emph{Sci. Rep.} \textbf{2021}, \emph{11},
  1--15\relax
\mciteBstWouldAddEndPuncttrue
\mciteSetBstMidEndSepPunct{\mcitedefaultmidpunct}
{\mcitedefaultendpunct}{\mcitedefaultseppunct}\relax
\EndOfBibitem
\bibitem[Luo \latin{et~al.}(2022)Luo, Song, Lu, Hu, Lv, Li, Li, Deng, Yan,
  Jiang, and Xia]{luo2022phase}
Luo,~J.; Song,~X.; Lu,~Y.; Hu,~Y.; Lv,~X.; Li,~L.; Li,~X.; Deng,~J.; Yan,~Y.;
  Jiang,~Y.; Xia,~C. Phase-controlled synthesis of {S}n{S}\textsubscript{2} and
  {S}n{S} flakes and photodetection properties. \emph{J. Phys.: Condens.
  Matter} \textbf{2022}, \emph{34}, 285701\relax
\mciteBstWouldAddEndPuncttrue
\mciteSetBstMidEndSepPunct{\mcitedefaultmidpunct}
{\mcitedefaultendpunct}{\mcitedefaultseppunct}\relax
\EndOfBibitem
\bibitem[Wang \latin{et~al.}(2020)Wang, Wang, Wang, Ye, He, Wu, Peng, Wu, Chen,
  Zhong, Xie, Cui, Shen, Zhang, Gu, Luo, Wang, Chen, Zhou, Pan, Zhou, Zhang,
  and Hu]{wang2020noble}
Wang,~Z. \latin{et~al.}  A noble metal dichalcogenide for high-performance
  field-effect transistors and broadband photodetectors. \emph{Adv. Func.
  Mater.} \textbf{2020}, \emph{30}, 1907945\relax
\mciteBstWouldAddEndPuncttrue
\mciteSetBstMidEndSepPunct{\mcitedefaultmidpunct}
{\mcitedefaultendpunct}{\mcitedefaultseppunct}\relax
\EndOfBibitem
\bibitem[Wang \latin{et~al.}(2020)Wang, Wu, Wang, Luo, Zhong, Ge, Zhang, Peng,
  Ye, Li, Ge, Ye, He, Chen, Xu, Yu, Wang, Hu, Zhou, Shan, Long, Wang, Zhou, and
  Hu]{wang2020air}
Wang,~Y. \latin{et~al.}  Air-Stable Low-Symmetry Narrow-Bandgap {2D} Sulfide
  Niobium for Polarization Photodetection. \emph{Adv. Mater.} \textbf{2020},
  \emph{32}, 2005037\relax
\mciteBstWouldAddEndPuncttrue
\mciteSetBstMidEndSepPunct{\mcitedefaultmidpunct}
{\mcitedefaultendpunct}{\mcitedefaultseppunct}\relax
\EndOfBibitem
\bibitem[Wang \latin{et~al.}(2020)Wang, Yu, Tong, Sun, Zhang, Xu, Sun, and
  Tsang]{wang2020high}
Wang,~Y.; Yu,~Z.; Tong,~Y.; Sun,~B.; Zhang,~Z.; Xu,~J.-B.; Sun,~X.;
  Tsang,~H.~K. High-speed infrared two-dimensional platinum diselenide
  photodetectors. \emph{App. Phys. Lett.} \textbf{2020}, \emph{116}\relax
\mciteBstWouldAddEndPuncttrue
\mciteSetBstMidEndSepPunct{\mcitedefaultmidpunct}
{\mcitedefaultendpunct}{\mcitedefaultseppunct}\relax
\EndOfBibitem
\bibitem[Huang \latin{et~al.}(2016)Huang, Wang, Hu, Liao, Wang, Wang, Gong,
  Chen, Wu, Luo, Shen, Lin, Sun, Meng, Chen, and Chu]{huang2016highly}
Huang,~H. \latin{et~al.}  Highly sensitive visible to infrared \ch{MoTe2}
  photodetectors enhanced by the photogating effect. \emph{Nanotechnology}
  \textbf{2016}, \emph{27}, 445201\relax
\mciteBstWouldAddEndPuncttrue
\mciteSetBstMidEndSepPunct{\mcitedefaultmidpunct}
{\mcitedefaultendpunct}{\mcitedefaultseppunct}\relax
\EndOfBibitem
\bibitem[Buscema \latin{et~al.}(2014)Buscema, Groenendijk, Blanter, Steele, Van
  Der~Zant, and Castellanos-Gomez]{buscema2014fast}
Buscema,~M.; Groenendijk,~D.~J.; Blanter,~S.~I.; Steele,~G.~A.; Van
  Der~Zant,~H.~S.; Castellanos-Gomez,~A. Fast and broadband photoresponse of
  few-layer black phosphorus field-effect transistors. \emph{Nano Lett.}
  \textbf{2014}, \emph{14}, 3347--3352\relax
\mciteBstWouldAddEndPuncttrue
\mciteSetBstMidEndSepPunct{\mcitedefaultmidpunct}
{\mcitedefaultendpunct}{\mcitedefaultseppunct}\relax
\EndOfBibitem
\bibitem[Wang \latin{et~al.}(2015)Wang, Klots, Prasai, Yang, Bolotin, and
  Valentine]{wang2015hot}
Wang,~W.; Klots,~A.; Prasai,~D.; Yang,~Y.; Bolotin,~K.~I.; Valentine,~J. Hot
  electron-based near-infrared photodetection using bilayer \ch{MoS2}.
  \emph{Nano Lett.} \textbf{2015}, \emph{15}, 7440--7444\relax
\mciteBstWouldAddEndPuncttrue
\mciteSetBstMidEndSepPunct{\mcitedefaultmidpunct}
{\mcitedefaultendpunct}{\mcitedefaultseppunct}\relax
\EndOfBibitem
\bibitem[Liang \latin{et~al.}(2019)Liang, Wang, Zhang, Wei, Lim, Zhu, Hu, Wei,
  Lee, Sow, Zhang, and Wee]{liang2019high}
Liang,~Q.; Wang,~Q.; Zhang,~Q.; Wei,~J.; Lim,~S.~X.; Zhu,~R.; Hu,~J.; Wei,~W.;
  Lee,~C.; Sow,~C.; Zhang,~W.; Wee,~A. T.~S. High-performance, room
  temperature, ultra-broadband photodetectors based on air-stable \ch{PdSe2}.
  \emph{Adv. Mater.} \textbf{2019}, \emph{31}, 1807609\relax
\mciteBstWouldAddEndPuncttrue
\mciteSetBstMidEndSepPunct{\mcitedefaultmidpunct}
{\mcitedefaultendpunct}{\mcitedefaultseppunct}\relax
\EndOfBibitem
\bibitem[Tan \latin{et~al.}(2020)Tan, Amani, Zhao, Hettick, Song, Lien, Li,
  Yeh, Shrestha, Crozier, Scott, and Javey]{tan2020evaporated}
Tan,~C.; Amani,~M.; Zhao,~C.; Hettick,~M.; Song,~X.; Lien,~D.-H.; Li,~H.;
  Yeh,~M.; Shrestha,~V.~R.; Crozier,~K.~B.; Scott,~M.~C.; Javey,~A. Evaporated
  Se\textsubscript{x}Te\textsubscript{1-x} thin films with tunable bandgaps for
  short-wave infrared photodetectors. \emph{Adv. Mater.} \textbf{2020},
  \emph{32}, 2001329\relax
\mciteBstWouldAddEndPuncttrue
\mciteSetBstMidEndSepPunct{\mcitedefaultmidpunct}
{\mcitedefaultendpunct}{\mcitedefaultseppunct}\relax
\EndOfBibitem
\bibitem[Mueller \latin{et~al.}(2010)Mueller, Xia, and
  Avouris]{mueller2010graphene}
Mueller,~T.; Xia,~F.; Avouris,~P. Graphene photodetectors for high-speed
  optical communications. \emph{Nat. Photon.} \textbf{2010}, \emph{4},
  297--301\relax
\mciteBstWouldAddEndPuncttrue
\mciteSetBstMidEndSepPunct{\mcitedefaultmidpunct}
{\mcitedefaultendpunct}{\mcitedefaultseppunct}\relax
\EndOfBibitem
\bibitem[Guo \latin{et~al.}(2018)Guo, Liu, Ma, Zhu, Lee, Lu, Zhao, Xu, Lee, Wu,
  Kovnir, Huang, and Duan]{guo2018few}
Guo,~J.; Liu,~Y.; Ma,~Y.; Zhu,~E.; Lee,~S.; Lu,~Z.; Zhao,~Z.; Xu,~C.;
  Lee,~S.-J.; Wu,~H.; Kovnir,~K.; Huang,~Y.; Duan,~X. Few-layer {GeAs}
  field-effect transistors and infrared photodetectors. \emph{Adv. Mater.}
  \textbf{2018}, \emph{30}, 1705934\relax
\mciteBstWouldAddEndPuncttrue
\mciteSetBstMidEndSepPunct{\mcitedefaultmidpunct}
{\mcitedefaultendpunct}{\mcitedefaultseppunct}\relax
\EndOfBibitem
\bibitem[Jiang \latin{et~al.}(2017)Jiang, Zang, Sun, Zheng, Liu, Gong, Fang,
  Cheng, and He]{jiang2017broadband}
Jiang,~T.; Zang,~Y.; Sun,~H.; Zheng,~X.; Liu,~Y.; Gong,~Y.; Fang,~L.;
  Cheng,~X.; He,~K. Broadband High-Responsivity Photodetectors Based on
  Large-Scale Topological Crystalline Insulator {SnTe} Ultrathin Film Grown by
  Molecular Beam Epitaxy. \emph{Adv. Opt. Mater.} \textbf{2017}, \emph{5},
  1600727\relax
\mciteBstWouldAddEndPuncttrue
\mciteSetBstMidEndSepPunct{\mcitedefaultmidpunct}
{\mcitedefaultendpunct}{\mcitedefaultseppunct}\relax
\EndOfBibitem
\bibitem[Pezeshki \latin{et~al.}(2016)Pezeshki, Shokouh, Nazari, Oh, and
  Im]{pezeshki2016electric}
Pezeshki,~A.; Shokouh,~S. H.~H.; Nazari,~T.; Oh,~K.; Im,~S. Electric and
  photovoltaic behavior of a few-layer $\alpha$ch{MoTe2}/\ch{MoS2}
  dichalcogenide heterojunction. \emph{Adv. Mater.} \textbf{2016}, \emph{28},
  3216--3222\relax
\mciteBstWouldAddEndPuncttrue
\mciteSetBstMidEndSepPunct{\mcitedefaultmidpunct}
{\mcitedefaultendpunct}{\mcitedefaultseppunct}\relax
\EndOfBibitem
\bibitem[Zeng \latin{et~al.}(2018)Zeng, Lin, Li, Zhang, Zhang, Xie, Mak, Chai,
  Lau, Luo, and Tsang]{zeng2018fast}
Zeng,~L.-H.; Lin,~S.-H.; Li,~Z.-J.; Zhang,~Z.-X.; Zhang,~T.-F.; Xie,~C.;
  Mak,~C.-H.; Chai,~Y.; Lau,~S.~P.; Luo,~L.-B.; Tsang,~Y.~H. Fast, self-driven,
  air-Stable, and broadband photodetector based on vertically aligned
  \ch{PtSe2}/GaAs heterojunction. \emph{Adv. Func. Mater.} \textbf{2018},
  \emph{28}, 1705970\relax
\mciteBstWouldAddEndPuncttrue
\mciteSetBstMidEndSepPunct{\mcitedefaultmidpunct}
{\mcitedefaultendpunct}{\mcitedefaultseppunct}\relax
\EndOfBibitem
\bibitem[An \latin{et~al.}(2013)An, Liu, Jung, and Kar]{an2013tunable}
An,~X.; Liu,~F.; Jung,~Y.~J.; Kar,~S. Tunable graphene--silicon heterojunctions
  for ultrasensitive photodetection. \emph{Nano Lett.} \textbf{2013},
  \emph{13}, 909--916\relax
\mciteBstWouldAddEndPuncttrue
\mciteSetBstMidEndSepPunct{\mcitedefaultmidpunct}
{\mcitedefaultendpunct}{\mcitedefaultseppunct}\relax
\EndOfBibitem
\bibitem[Long \latin{et~al.}(2016)Long, Liu, Wang, Gao, Xia, Luo, Wang, Zeng,
  Fu, Xu, Zhou, Lv, Yao, Lu, Chen, Ni, You, Zhang, Qin, Shi, Hu, Xing, and
  Miao]{long2016broadband}
Long,~M. \latin{et~al.}  Broadband photovoltaic detectors based on an
  atomically thin heterostructure. \emph{Nano Lett.} \textbf{2016}, \emph{16},
  2254--2259\relax
\mciteBstWouldAddEndPuncttrue
\mciteSetBstMidEndSepPunct{\mcitedefaultmidpunct}
{\mcitedefaultendpunct}{\mcitedefaultseppunct}\relax
\EndOfBibitem
\bibitem[John \latin{et~al.}(2020)John, Dhyani, Maity, Mukherjee, Ray, Kumar,
  and Das]{john2020broadband}
John,~J.~W.; Dhyani,~V.; Maity,~S.; Mukherjee,~S.; Ray,~S.~K.; Kumar,~V.;
  Das,~S. Broadband infrared photodetector based on nanostructured
  \ch{MoSe2}--{Si} heterojunction extended up to 2.5 $\mu$m spectral range.
  \emph{Nanotechnology} \textbf{2020}, \emph{31}, 455208\relax
\mciteBstWouldAddEndPuncttrue
\mciteSetBstMidEndSepPunct{\mcitedefaultmidpunct}
{\mcitedefaultendpunct}{\mcitedefaultseppunct}\relax
\EndOfBibitem
\end{mcitethebibliography}

\end{document}


\clearpage

\subsection*{High-Resolution X-Ray Photoelectron Spectroscopy (XPS) of Bulk SnS\textsubscript{2}}
The pristine SnS\textsubscript{2} bulk crystal prior to its exfoliation was examined through x-ray photoelectron spectroscopy (XPS; AXIS Supra, Kratos Analytical Ltd., Manchester, UK). Peak fitting was carried out using CasaXPS software (Version 2.3.16 RP 1.6, Casa Software Ltd., Teignmouth, UK).

\begin{figure*}[!ht]
\centering
  \includegraphics[width=0.7\textwidth]{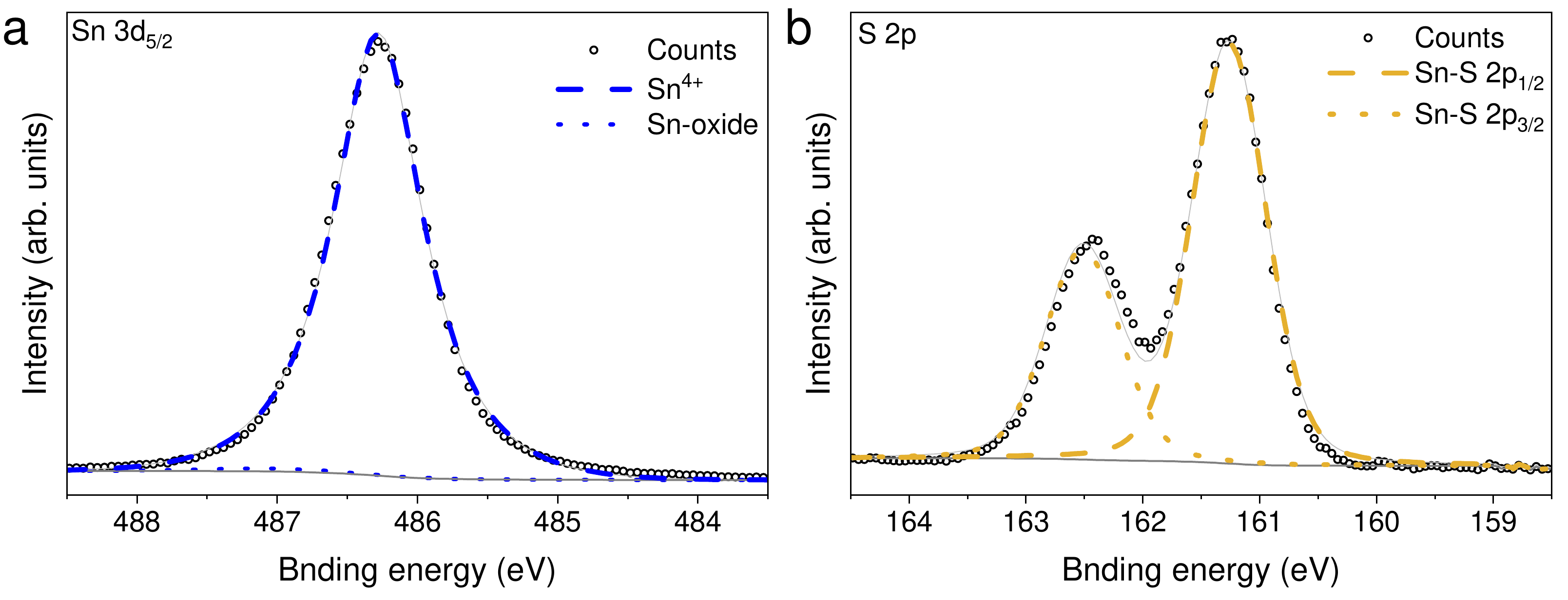}
  \caption{High-resolution XPS spectra of SnS\textsubscript{2} in (a) the Sn~3d\textsubscript{5/2}, and, (b) the S~2p regions.}
  \label{fig:sxps}
\end{figure*}

\clearpage
\subsection*{Atomic Force Microscopy (AFM) Characterization of the SnS\textsubscript{2} Nanoflake}
The thickness of the tape exfoliated SnS\textsubscript{2} nanoflake was measured with an atomic force microscope (AFM; Dimension Icon, Bruker Corp., Billerica, MA, USA) operated using the supplied software. The AFM topography image was also analyzed using the supplied software NanoScope Analysis 2.0 (Bruker Corp., Billerica, MA, USA). 

\begin{figure*}[!ht]
\centering
  \includegraphics[width=0.7\textwidth]{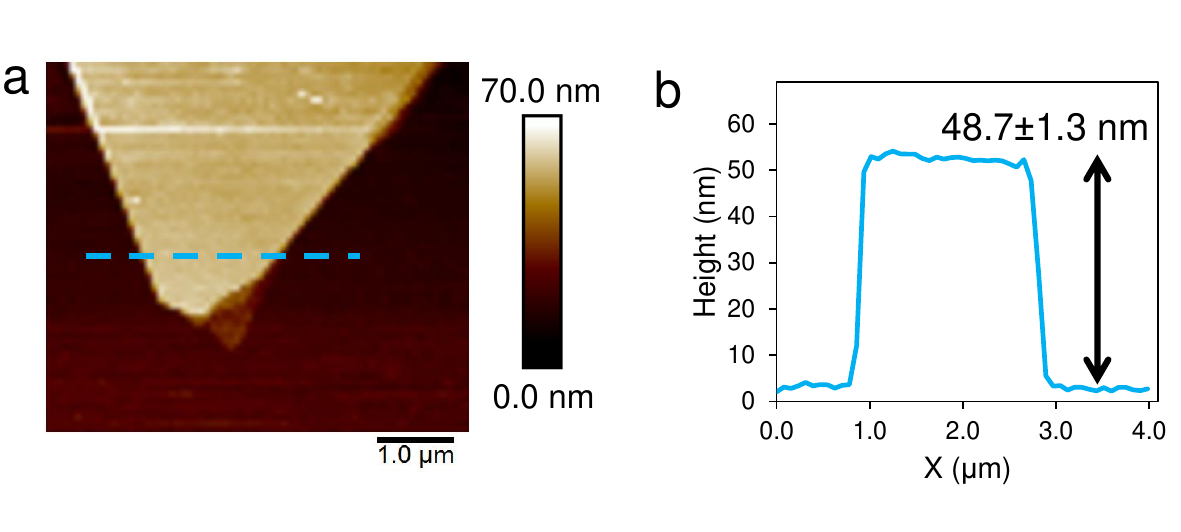}
  \caption{(a) AFM image, and, (b) height profile corresponding to the dashed blue line along the SnS\textsubscript{2} nanoflake.}
  \label{fig:safm}
\end{figure*}

\clearpage
\subsection*{Acoustoelectric (AE) Current Calculations}
The AE current $j$ can be predicted by the well-known Weinreich relation:\cite{weinreich1956acoustodynamic,rotter1998giant,rotter1999charge}
\begin{equation}
\label{eq:ae}
j =  \frac{K_{\rm eff}^2 \pi \mu P_{\rm e} f}{\nu_{\rm SAW}^2}  \left[  \frac{\sigma_{\rm 2D}/\sigma_{\rm M}}{1+(\sigma_{\rm 2D}/\sigma_{\rm M})^2} \right],
\end{equation}
where $K_{\rm eff}^2$ is the piezoelectric coupling coefficient (5.6\% for LiNbO\textsubscript{3}), $\nu_{\rm SAW}$ the SAW velocity when the surface is shorted (approximately 3990~m/s in 128$^{\circ}$ Y-cut X-propagating LiNbO\textsubscript{3}), $\mu$ the electron mobility in the material, which we specify through the field-effect mobility $\mu_{\rm FE}= 0.7$ cm$^2$/Vs (see detailed calculations below), $P_{\rm e} = P/Q_{\rm SAW}$ the input power of the SAW, wherein $Q_{\rm SAW} \approx 300$ is the SAW device quality factor and $P$ the total input power, and, $f$ the SAW central frequency. For 128$^{\circ}$ Y-cut X-propagating LiNbO\textsubscript{3}, $\sigma_{M}$ is approximately given by $\sigma_{M} = \nu_{0}\varepsilon_{0}\left[\left(\varepsilon_{xx}\varepsilon_{zz}\right)^{1/2}+1\right] \approx 1.25 \times 10^{-6}$~S, where $\varepsilon_{0} = 8.85 \times 10^{-12}$~F/m is the permittivity of free space, and, $\varepsilon_{xx} \approx 85$ and $\varepsilon_{zz} \approx 29.5$ are the dielectric constants of LiNbO\textsubscript{3} at constant stress. $\sigma_{\rm 2D}$ is the electrical conductivity of SnS\textsubscript{2}, which from the slope of the $I$--$V$ curve in Fig.~2a, had a calculated value of 10.8~nS. 

\begin{figure*}[!ht]
\centering
  \includegraphics[width=0.5\textwidth]{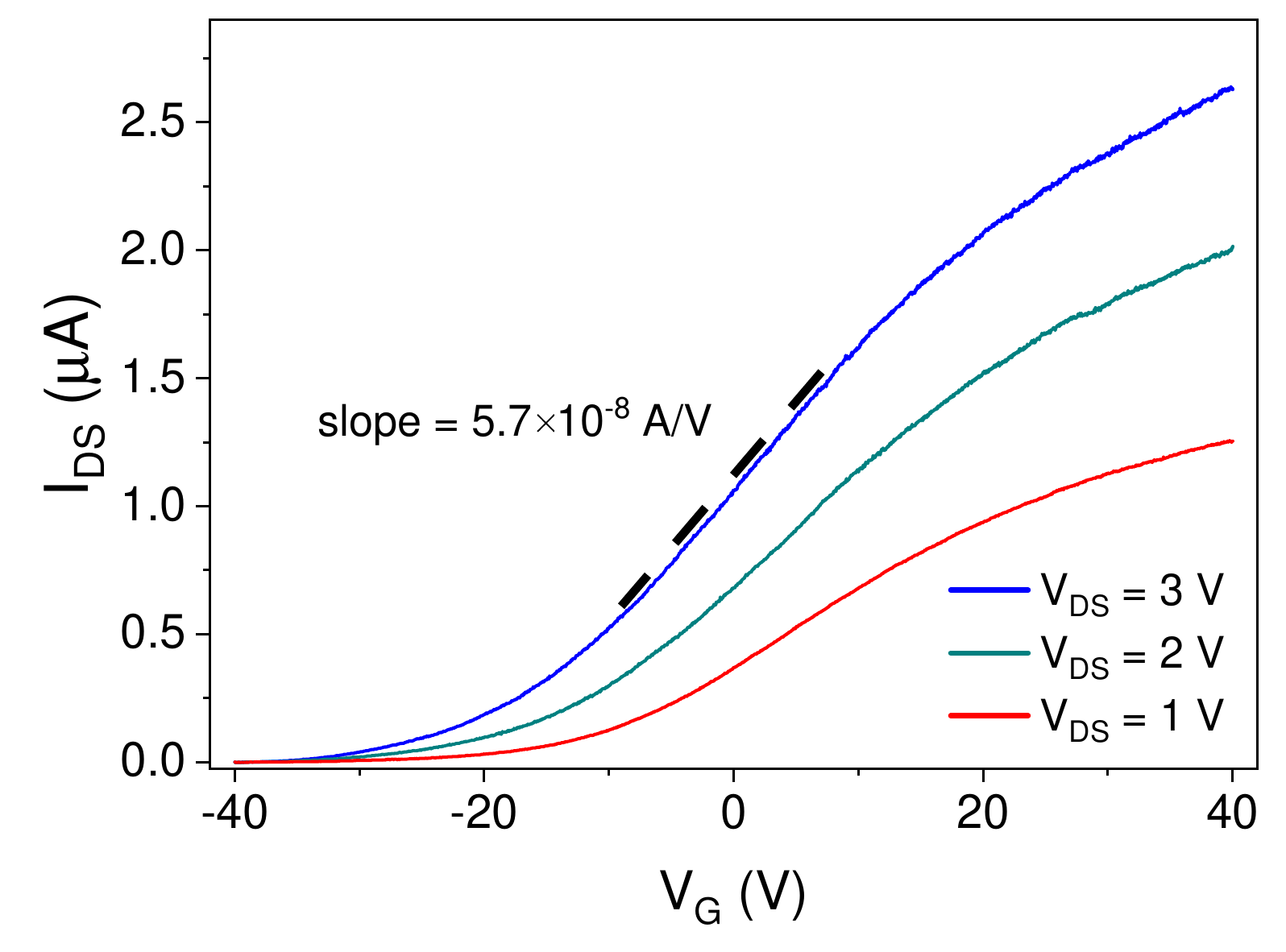}
  \caption{Transfer curves for the SnS\textsubscript{2} nanoflake for $V_{\rm DS}$~= 1, 2, and 3~V, from which the largest slope was calculated to be $5.7 \times 10^{-8}$~A/V. }
  \label{fig:smu}
\end{figure*}

\noindent \textbf{Mobility calculations:} To calculate the electron mobility in SnS\textsubscript{2} in Eq.~(\ref{eq:ae}), SnS\textsubscript{2} was tape exfoliated in the same way as the rest of the experiments and transferred onto a SiO\textsubscript{2}/Si substrate (with an oxide thickness of 300~nm). The back gate of Si was accessed by scratching off the SiO\textsubscript{2} layer. Similar electrical contact pads were then deposited on the nanoflake using the same method as previously described in the Methods section in the main manuscript. To calculate the electron mobility in SnS\textsubscript{2}, the transfer characteristics of the nanoflake were plotted, as shown in Fig.~\ref{fig:smu}, from which the field-effect mobility can be calculated:\cite{mitta2020electrical}
\begin{equation}
\mu_{\rm FE} = \frac{\Delta I_{\rm DS}}{\Delta V_{\rm G}} \times \frac{L}{WCV_{\rm DS}},
\end{equation}
where $\Delta I_{\rm DS} / \Delta V_{\rm G}$ is the largest slope along the transfer curve, $L$ and $W$ the length and width of the channel, respectively, and $C$ the capacitance per unit area, which can be determined from $\varepsilon_0 \varepsilon_r / d$, in which $\varepsilon_r=3.9$ is the relative permittivity of SiO\textsubscript{2} and $d$ the thickness of the gate oxide (300~nm). Given $\Delta I_{\rm DS} / \Delta V_{\rm G}$ values of $5.7 \times 10^{-8}$~A/V,  $L = 4.3\ \upmu$m, $W = 10.8\ \upmu$m, $C = 1.15 \times 10^{-4}$ F, and $V_{\rm DS} = 3$ V, we obtain $\mu_{\rm FE} = 0.7$ cm$^2$/Vs, which is comparable to mobility values in SnS\textsubscript{2} reported in the literature.\cite{zhang2018formation,tian2020visible} 
 
\clearpage

\subsection*{Electrical response of LiNbO\textsubscript{3}}

\begin{figure*}[!ht]
\centering
  \includegraphics[width=\textwidth]{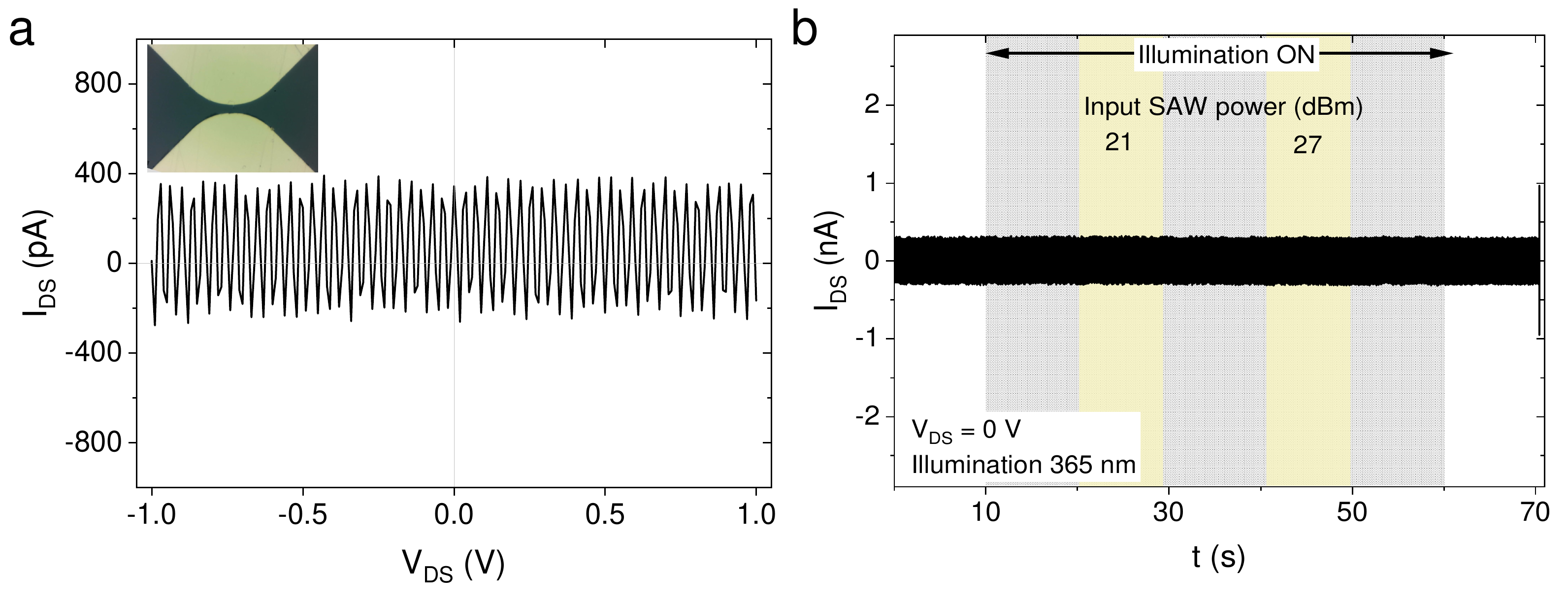}
  \caption{(a) $I$--$V$ characteristics of LiNbO\textsubscript{3}, and its (b) transient response to 365~nm illumination and SAW excitation (21 and 27~dBm input power). The inset in (a) shows an optical micrograph of bare LiNbO\textsubscript{3} patterned with source and drain contact pads.}
  \label{fig:empty}
\end{figure*}

To eliminate the possibility of LiNbO\textsubscript{3} contributing to the electrical signal, a similar experiment was performed on LiNbO\textsubscript{3} in the absence of the  SnS\textsubscript{2}  flake. The fabricated device, which is shown in the inset of Fig.~\ref{fig:empty}a, wherein source and drain contact pads are fabricated on bare LiNbO\textsubscript{3} substrate, shows no electrical conductivity according to its current--voltage ($I$--$V$) plot in Fig.~\ref{fig:empty}a. To show the lack of responsivity of LiNbO\textsubscript{3} to illumination or SAW excitation, current--time ($I$--$t$) measurements were also performed. From the results in Fig.~\ref{fig:empty}b, we are able to conclude that exposing the LiNbO\textsubscript{3} substrate to 365~nm illumination does not produce any photocurrent. Similarly, exposure of the substrate to the SAW irradiation at input powers of 21 and 27~dBm did not produce any acoustophotoelectric response either. As such, it is reasonable to assert that the LiNbO\textsubscript{3} does not contribute to the electrical signal in our experiments.

\clearpage
\subsection*{Acoustophotoelectric Current Measurements in Other 2D Materials}
MoS\textsubscript{2} (SPI Supplies, Structure Probe, Inc., West Chester, PA, USA) and SnS (2D Semiconductors, Scottsdale, AZ, USA) were exfoliated, transferred onto LiNbO\textsubscript{3} substrates, and electrically characterized using the same methods as those employed for SnS\textsubscript{2}. 

\begin{figure*}[!ht]
\centering
  \includegraphics[width=\textwidth]{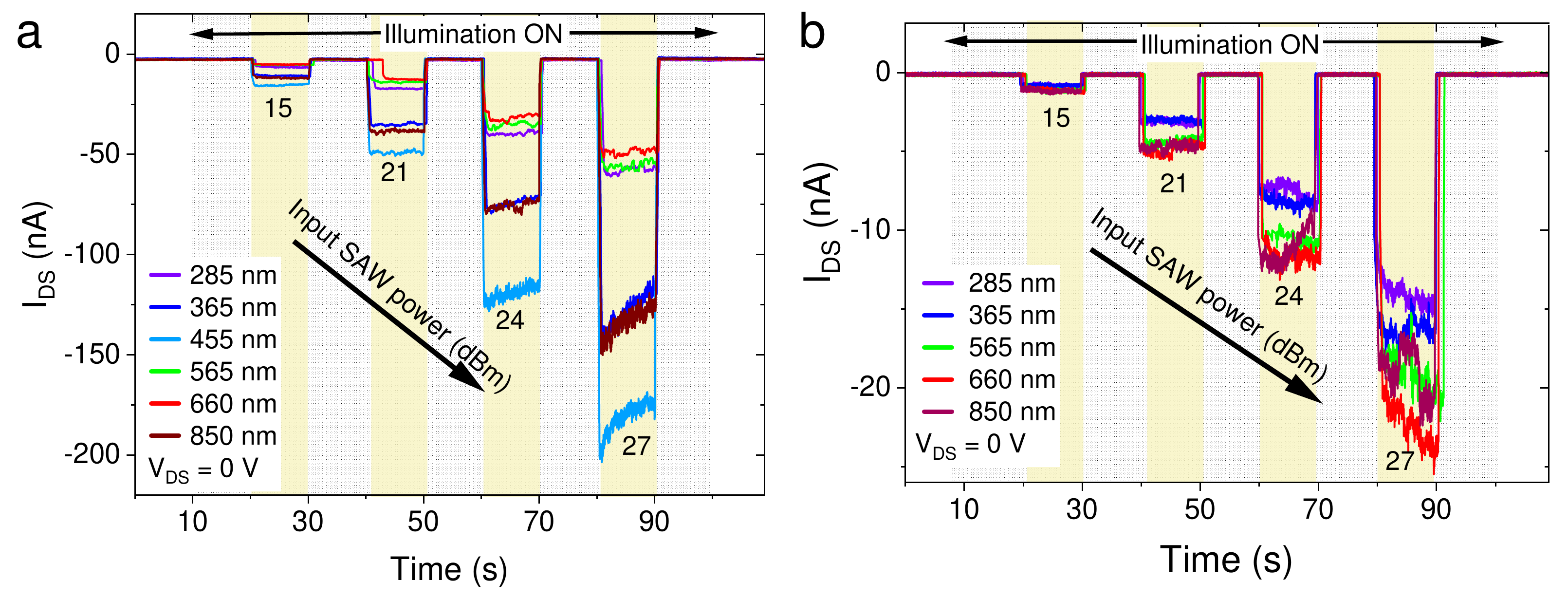}
  \caption{Acoustophotoelectric current in (a) MoS\textsubscript{2}, and, (b) SnS, with $V_{\rm DS} = 0$ V under different illumination wavelengths with 3~mW/cm$^{2}$ intensity.}
  \label{fig:mos2sns}
\end{figure*}

\clearpage
\subsection*{Photoluminesence (PL) Measurements of the SnS\textsubscript{2} Nanoflake}

Following exfoliation of bulk SnS\textsubscript{2} in the same way as was carried out for the rest of the experiments, and its subsequent transfer onto a Si substrate (prime grade, P-doped, 1--100~$\Omega\cdot$cm resistivity, UniversityWafer Inc., South Boston, MA, USA), the nanoflake was then excited with a 400~nm laser operating at 80~MHz, 30~nm bandwidth  and 660~pW illumination power. The photoluminescence (PL) signal was filtered with a 500~nm optical long-pass filter (Edmund Optics, Barrington, NJ, USA) and detected using a spectrometer (Spectra Pro, Princeton Instruments, Trenton, NJ, USA) fitted with a CCD camera (Pixis:100, Princeton Instruments, Trenton, NJ, USA) or an avalanche photodiode (SPCM-AQRH-14, Excelitas Technologies Corp., Waltham, MA, USA). 

\begin{figure*}[!ht]
\centering
  \includegraphics[width=0.7\textwidth]{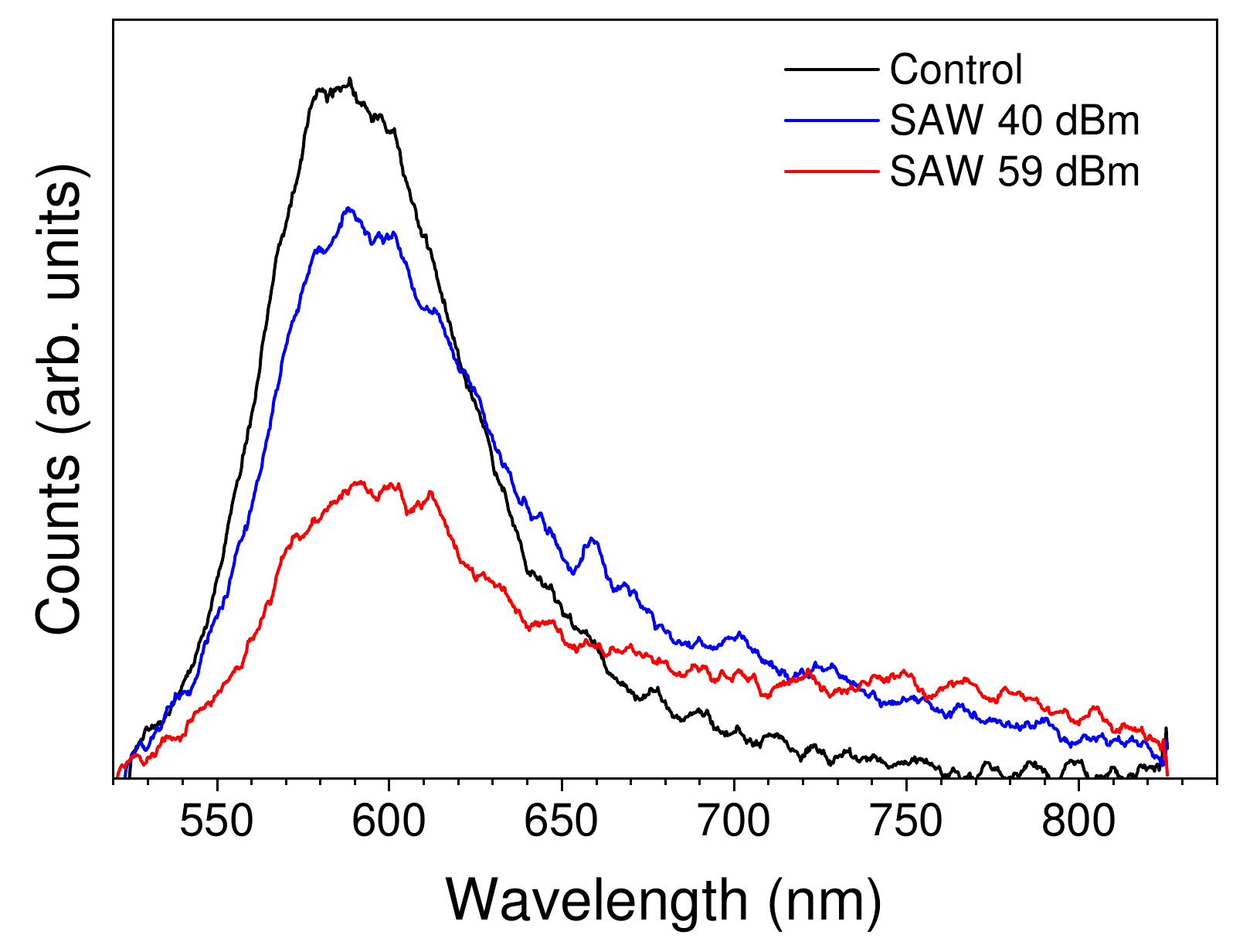}
  \caption{Quenching and broadening of the PL spectra of the SnS\textsubscript{2} nanoflake under SAW excitation at two different input powers.}
  \label{fig:spl}
\end{figure*}

\clearpage
\subsection*{Photoconductivity of SnS\textsubscript{2}}
To contrast the acoustophotoelectric effect with the typical photoconductive effect, the response of SnS\textsubscript{2} under 1~V of external DC bias to different illumination wavelengths was examined. 

\begin{figure*}[!ht]
\centering
  \includegraphics[width=0.5\textwidth]{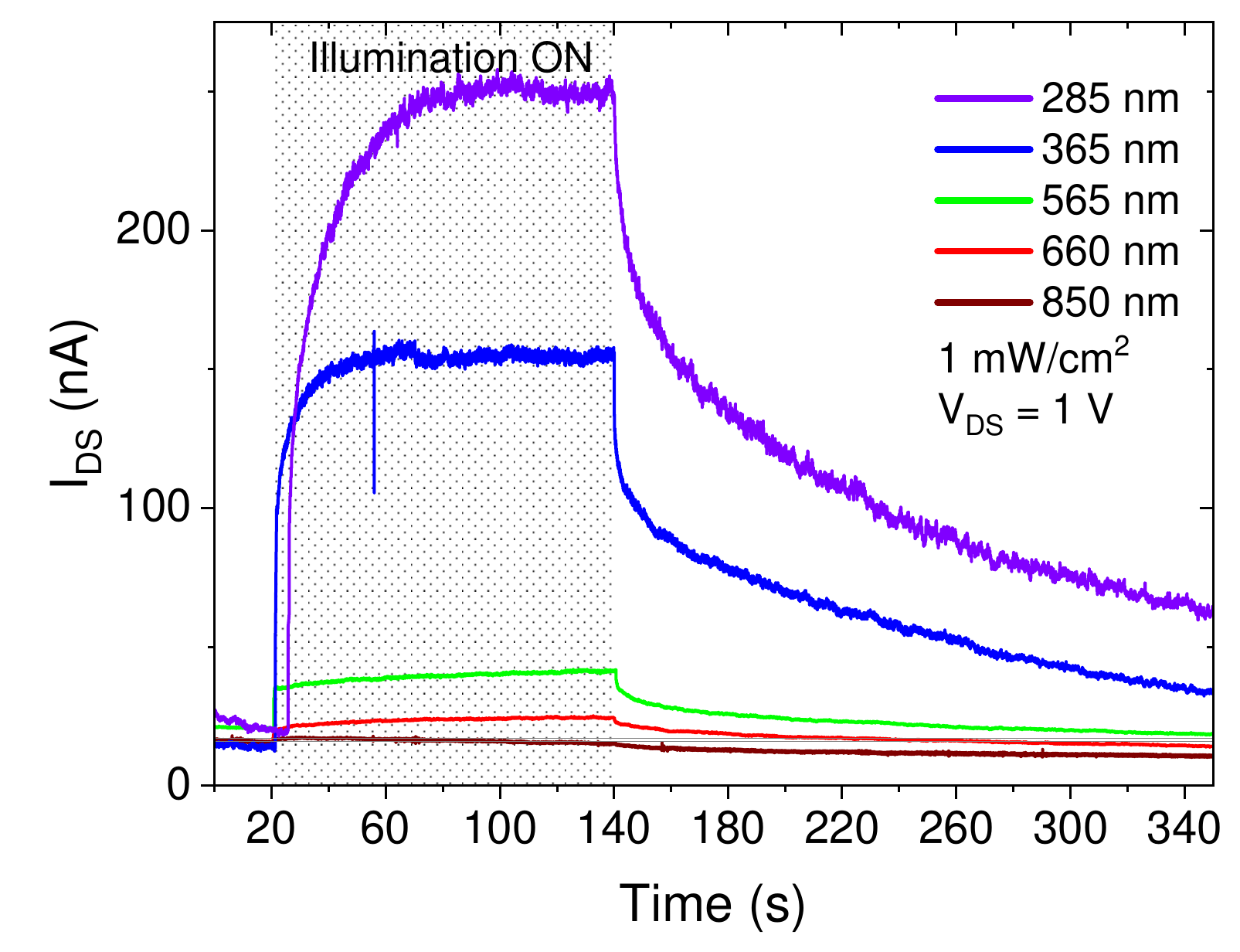}
  \caption{Transient current ($I_{\rm DS}$--$t$) response of the SnS\textsubscript{2} photodetector with a 1~V applied DC bias under different illumination wavelengths with 1~mW/cm$^{2}$ intensity. The shaded region denotes the time window when the nanoflake was
  illuminated.}
  \label{fig:sdc1}
\end{figure*}

\clearpage
\subsection*{Optical Absorbance of SnS\textsubscript{2}}

The optical absorbance of the SnS\textsubscript{2} nanoflake on the  128$^{\circ}$ Y-cut X-propagating LiNbO\textsubscript{3} substrate was measured with a 
microspectrometer (CRAIC Technologies Inc., San Dimas, CA, USA).

\begin{figure*}[!ht]
\centering
  \includegraphics[width=0.75\textwidth]{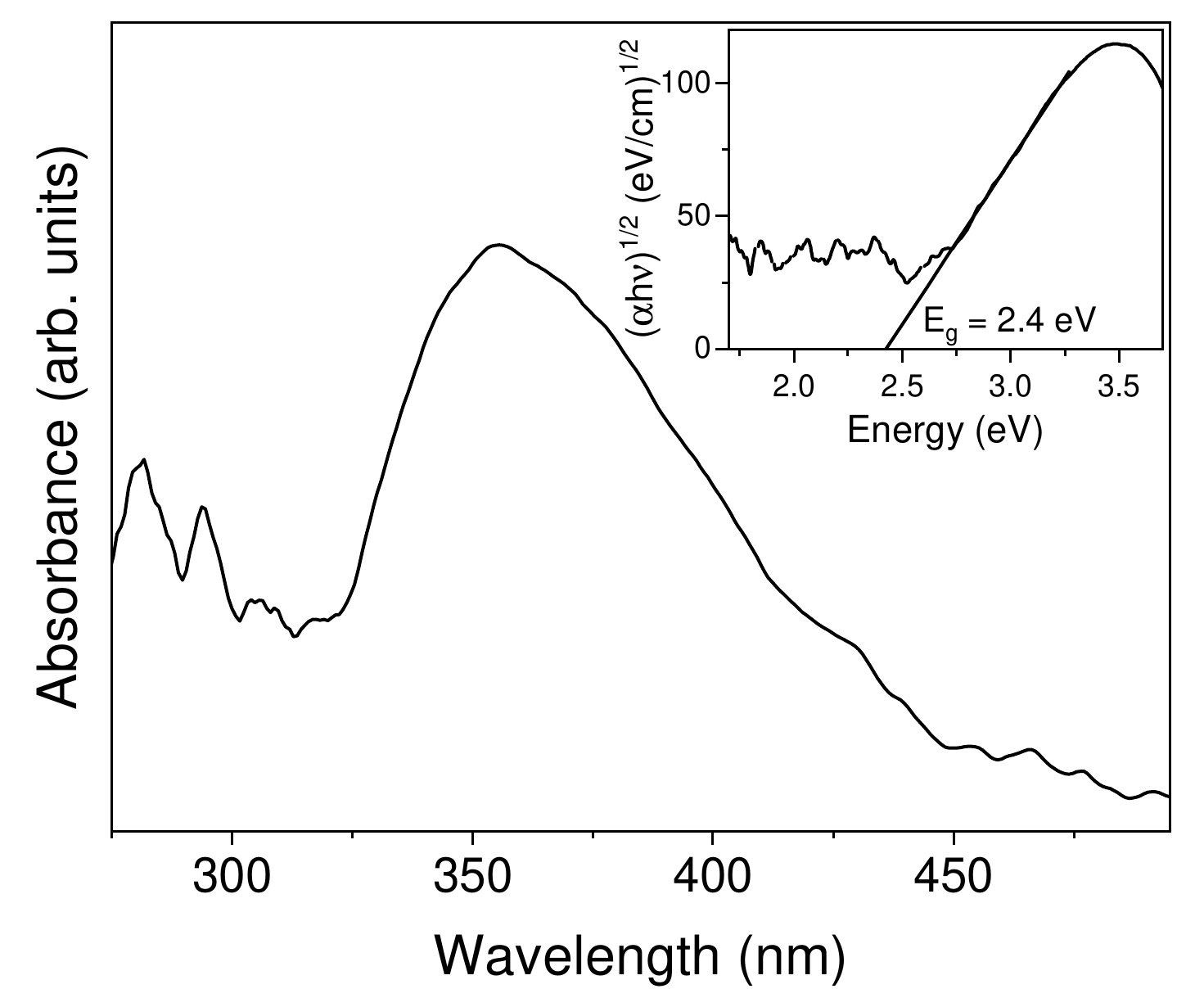}
  \caption{Optical absorbance of SnS\textsubscript{2} in the absence of the SAW. The inset shows the corresponding Tauc plot and calculated optical bandgap; $\alpha$, $h$ and $\nu$ are the absorption coefficient, Plank's constant, and light frequency, respectively.}
  \label{fig:abs}
\end{figure*}

\clearpage
 \subsection*{Photodetector Characterization}
 \label{sec:photo}

\noindent \textbf{External quantum efficiency}: The external quantum efficiency EQE, which describes the photoelectric conversion capability of optoelectronic devices and is defined as the number of electron--hole pairs produced by the device per unit time of photon illumination, can be calculated from 
\begin{equation}
{\rm EQE} = \frac{hc}{e\lambda} R_{\lambda},
\end{equation}
where $c$ is the speed of light, $e$ the electronic charge and $\lambda$ the wavelength of the incident light. The parameter EQE$^*$~=~EQE$_{\lambda + {\rm SAW}}$/EQE$_{\lambda}$ is then used to quantify the enhancement in EQE under the SAW excitation.

\sloppy

\begin{figure*}[!h]
\centering
  \includegraphics[width=\textwidth]{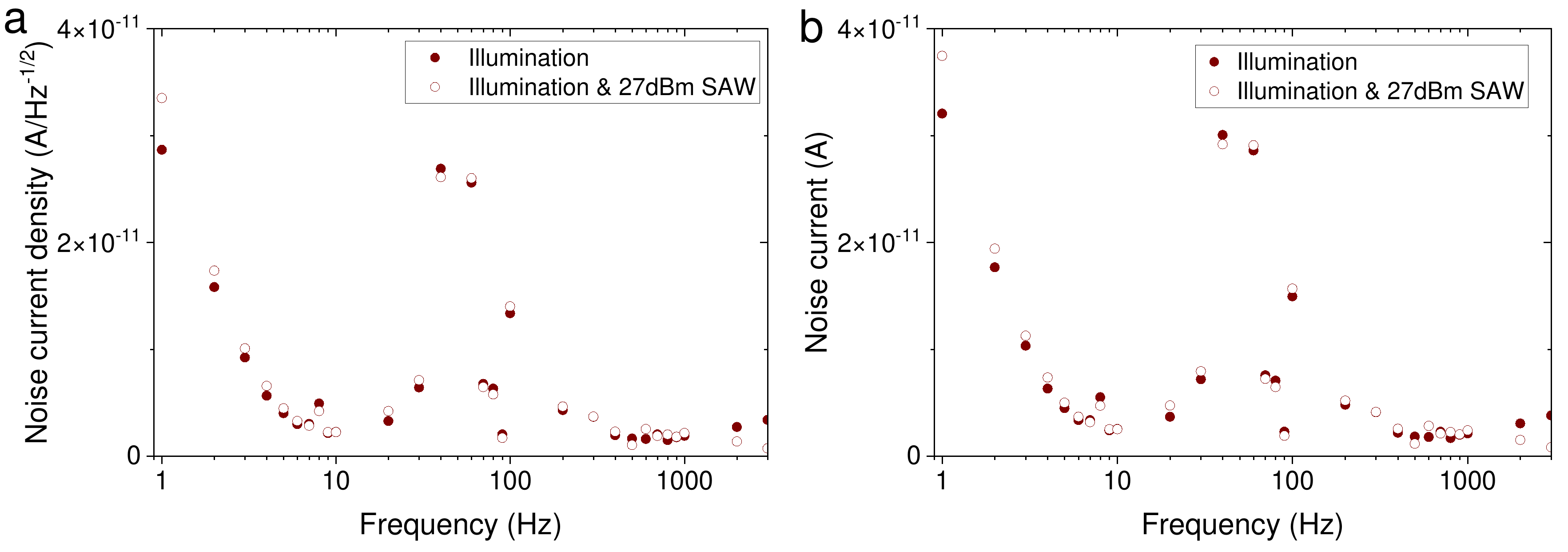}
  \caption{(a) Noise current density, and, (b) noise current as a function of frequency of the SnS\textsubscript{2} nanoflake under illumination, both in the absence and in the presence of the SAW excitation at an input power of 27~dBm, which shows no significant noise generation by the SAWs.}
  \label{fig:ncd}
\end{figure*}

\noindent \textbf{Noise characteristics}: To calculate the noise levels in the dark current produced by the SnS\textsubscript{2} photodetector, a low noise transimpedance amplifier (SR570, Stanford Research Systems, Sunnyvale, CA, USA) was used together with a lock-in amplifier (SR860, Stanford Research Systems, Sunnyvale, CA, USA) against an internal sinusoidal reference; a low-pass filter (10 kHz) was employed to eliminate high frequency ambient noise. Sixteen measurements of the signal amplitude were acquired at each frequency and averaged. Current noise values were obtained by multiplying the signal amplitude by the transimpedance amplifier sensitivity (1~$\upmu$A/V). The same method was applied to measure the noise current corresponding to the SAW excitation, both under dark conditions and under white source illumination. From the measurements, the noise spectral density for different experimental conditions were obtained and plotted in Fig.~\ref{fig:ncd}. In addition, according to the noise characteristic measurements shown in Figure~\ref{fig:ncd}, the noise current spectral density at 1~Hz bandwidth for the illumination only case, and for the case of illumination in the presence of 27~dBm SAW, are $2.87 \times 10^{-11}$ and $3.35 \times 10^{-11}$~A/Hz$^{1/2}$, respectively. The corresponding noise current values are $3.21\times10^{-11}$ and $3.75\times10^{-11}$~A, respectively. The high noise current density and noise current values at 50~Hz are associated with the frequency of the mains electricity supply. 

\fussy

\noindent\textbf{Noise equivalent power}: The noise equivalent power NEP (W$\cdot$Hz$^{-1/2}$), defined as the minimum light signal power that can be distinguished from the total noise by a photodetector with signal-to-noise ratio of unity at 1~Hz bandwidth,\cite{long2019progress} can be calculated from
\begin{equation}
{\rm NEP} = \frac{i_{\rm N}}{R_{\lambda}},
\end{equation}
where $i_{\rm N}$ is the noise current spectra at 1~Hz bandwidth, which is dependent on the photodetection mechanism and the associated uncertainties.\cite{sett2021engineering}

\noindent \textbf{Specific detectivity}: The specific detectivity $D^*$  (cm$\cdot$Hz$^{1/2}$/W or Jones) is a useful parameter for comparing the detection performance of photodetectors with different materials and geometries, and can be defined by\cite{long2019progress} 
\begin{equation}
D^* = \frac{ \sqrt{ A B} } {\rm NEP},
\end{equation}
where $A$ is the active area of the photodetector and $B$ its bandwidth. The higher the value of $D^*$, the stronger the detection capability. To quantify the enhancement of $D^*$ by the SAW excitation, we define $D^{**} = D^*_{\lambda + {\rm SAW}}/D^*_{\lambda}$. 
It should be noted that it is common in the literature to consider shot noise as the main noise, such that the equation above can be simplified to $D^* = \sqrt{A/(2eI_{\rm Dark})}R$, which results in an overestimation of $D^*$.\cite{krishnamurthi2020liquid}
\noindent \textbf{Response time}: The rise $\tau_{r}$ and fall $\tau_{f}$ times of the photodetector are defined as the time for the photocurrent to increase from 10 to 90\% of its peak value, and vice versa.

\clearpage

\begin{figure*}[!ht]
\centering
  \includegraphics[width=\textwidth]{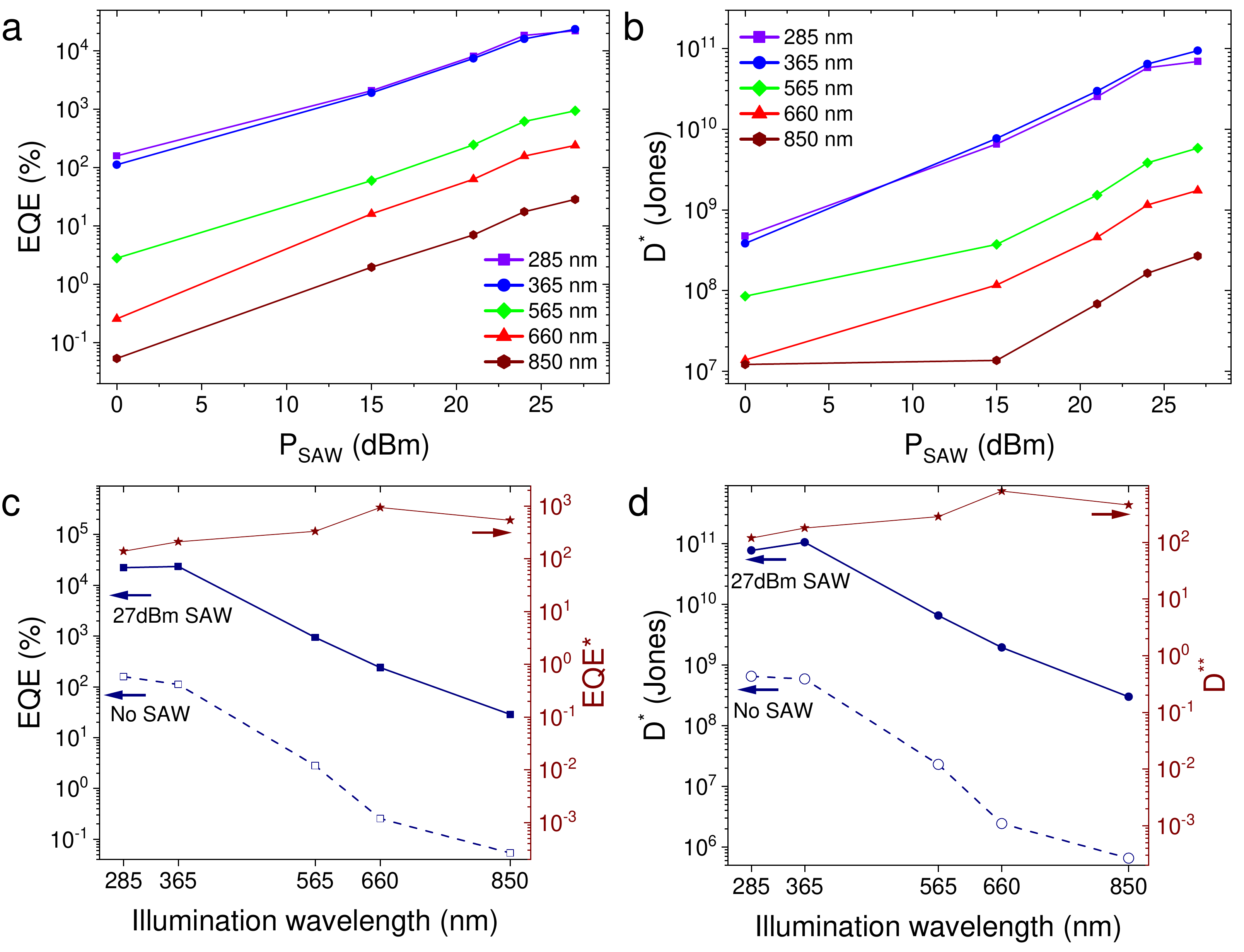}
  \caption{ (a) External quantum efficiency EQE, and, (b) specific detectivity $D^*$ of the SnS\textsubscript{2} photodetector as a function of the input SAW power for different illumination wavelengths with 1~mW/cm$^{2}$ intensity. (c) EQE and EQE$^*$, and, (d) $D^*$ and $D^{**}$, as a function of the illumination wavelength, both in the absence and in the presence of the SAW excitation.}
  \label{fig:spd}
\end{figure*}

\clearpage
\begin{landscape}

\subsection*{Comparison of Photodetector Performance}

{\scriptsize\tabcolsep=0.1pt  
\begin{longtable}{ lcccccccccc }
\caption{Comparison of the figures of merit of the SAW SnS\textsubscript{2} photodetector in this work with SnS\textsubscript{2} photodetectors that have been reported in the literature ($\lambda_{\rm inc}$: incident light wavelength; $P_{\lambda}$: incident light intensity; $V_{\rm DS}$: source-to-drain voltage bias; $I_{\rm D}$: Dark current;  $R$: Responsivity; EQE: external quantum efficiency; NEP: Noise equivalent power, $D^*$: specific detectivity; $\tau_{r}$: rise time; $\tau_{f}$: fall time; ND: no data). Values highlighted in green are the strengths of this work against values reported for other photodetectors under similar operating conditions, highlighted in red.}
\label{tab:lit} \\\toprule
\hline Material & Synthesis method & $\lambda_{\rm inc}$ & $P_{\lambda}$ &$V_{\rm DS}$ & $I_{\rm D}$ & $R$   & EQE  & NEP & $D^*$ & $\tau_{r} / \tau_{f}$ \\ 
 & &  (nm) & (mW/cm$^{2}$) & (V) & & (A/W)& (\%) & (W/Hz$^{-1/2}$) &  (Jones)&  (ms) \\ 
\hline

 \multirow{5}{*}{SnS\textsubscript{2} nanoflake, 27~dBm SAW}    &  & 285 &  & \cellcolor[HTML]{34FF34} &  \cellcolor[HTML]{34FF34}48~pA & \cellcolor[HTML]{34FF34} 50.6	& $2.6 \times 10^{4}$ & 	\cellcolor[HTML]{34FF34} $7.4 \times 10^{-13}$ & 	$6.9 \times 10^{10}$& 86/87  \\
 &  &  365 &  & \cellcolor[HTML]{34FF34} &  \cellcolor[HTML]{34FF34}60~pA & \cellcolor[HTML]{34FF34}68.9& $3.0 \times 10^{4}$ & \cellcolor[HTML]{34FF34} $5.4 \times 10^{-13}$&	$9.4 \times 10^{10}$  & 80/73    \\
 & &  \cellcolor[HTML]{34FF34} 565 &  & \cellcolor[HTML]{34FF34}&   \cellcolor[HTML]{34FF34}2~pA & \cellcolor[HTML]{34FF34} 33.3&	$1.3 \times 10^{3}$ & \cellcolor[HTML]{34FF34} $8.8 \times 10^{-12}$& $5.8 \times 10^{9}$  &  6/7 \\
  & &  \cellcolor[HTML]{34FF34} 660 &  & \cellcolor[HTML]{34FF34}&   \cellcolor[HTML]{34FF34}1~pA& \cellcolor[HTML]{34FF34}20.5 &	$3.5 \times 10^{2}$ & \cellcolor[HTML]{34FF34} $2.9 \times 10^{-11}$&	 \cellcolor[HTML]{34FF34}$1.7 \times 10^{9}$   & \cellcolor[HTML]{34FF34}19/4  \\  
  &  \multirow{-5}{*}{Tape exfoliation} & \cellcolor[HTML]{34FF34} 850 &  \multirow{-5}{*}{1.0} & \cellcolor[HTML]{34FF34} \multirow{-5}{*}{0} &  \cellcolor[HTML]{34FF34}0.1~pA& \cellcolor[HTML]{34FF34}13.6  &	42 & \cellcolor[HTML]{34FF34} $1.9 \times 10^{-10}$&	 \cellcolor[HTML]{34FF34} $2.7 \times 10^{8}$ & \cellcolor[HTML]{34FF34}14/7  \\   \hline
  
SnS\textsubscript{2} nanosheets ($V_{\rm G} = 50$ V) \cite{huang2015highly} & Chemical vapour deposition & 473 & 2.1 &\cellcolor[HTML]{FE0000} 3 &\cellcolor[HTML]{FE0000} $\approx$2~nA & 100 & $3.3 \times 10^{4}$  & ND & ND & 22/11  \\ \hline  
  
SnS\textsubscript{2}  thin crystal\cite{su2015chemical}   & Chemical vapour deposition  & 457 & 41000 &  \cellcolor[HTML]{FE0000} & 5.8~pA & \cellcolor[HTML]{FE0000}0.0052 & 2.4 &ND  & $2 \times 10^{9}$& 0.012/0.017\\  
$V_{\rm G} = 10$ V  &  & 457 & 41000 &\cellcolor[HTML]{FE0000} \multirow{-2}{*}{2} & ND & \cellcolor[HTML]{FE0000}0.008 & ND  & ND & ND & ND   \\ \hline
      
SnS\textsubscript{2} microsphere film\cite{tao2015flexible}  & Solvothermal &  365 & ND & \cellcolor[HTML]{FE0000} &ND & \cellcolor[HTML]{FE0000}$6.41 \times 10^{-7}$ & ND  &ND & ND & ND  \\ 
    &     & 405 & ND & \cellcolor[HTML]{FE0000} &ND & \cellcolor[HTML]{FE0000} $9.36 \times 10^{-9}$ & ND & ND  &ND  &ND  \\ 
    &     & 532 & ND & \cellcolor[HTML]{FE0000}&ND  & \cellcolor[HTML]{FE0000}$3.83 \times 10^{-8}$& ND  &ND & ND &ND   \\ 
    &     & 650 & ND & \cellcolor[HTML]{FE0000} & 32~pA & \cellcolor[HTML]{FE0000}$1.05 \times 10^{-6}$ & ND  &ND &ND  & \cellcolor[HTML]{FE0000}20000/30000  \\ 
   &      & 780 & ND & \cellcolor[HTML]{FE0000}&ND  & \cellcolor[HTML]{FE0000} $5.5\times 10^{-9}$& ND  &ND & ND  & ND  \\              
    &     & 830 & ND & \cellcolor[HTML]{FE0000} \multirow{-6}{*}{5}& 10~pA & \cellcolor[HTML]{FE0000}$ 5.77 \times 10^{-9}$ & ND  & ND& ND & \cellcolor[HTML]{FE0000} 26500/26700  \\        \hline             
        
2D SnS\textsubscript{2}  crystals\cite{xia2015large} & Atmospheric pressure chemical vapour deposition & 450 & 100 & \cellcolor[HTML]{FE0000} 10& \cellcolor[HTML]{FE0000} 13~nA & \cellcolor[HTML]{FE0000}  0.31 & ND  & ND& ND & 42/42 \\  \hline
    
    2D SnS\textsubscript{2}\cite{fan2016wavelength}  &  Chemical vapour deposition & 255 & 0.68 & \cellcolor[HTML]{FE0000} &  ND & \cellcolor[HTML]{FE0000} 10.88 & $5.3 \times 10^{3}$ &  ND & ND& \cellcolor[HTML]{FE0000} 2010/8400   \\
  & & 405 & 310 &\cellcolor[HTML]{FE0000} &  ND & \cellcolor[HTML]{FE0000} 1.568 &  $4.8 \times 10^{2}$ & ND & ND& 42/44  \\
  & &533 & 55.4 &\cellcolor[HTML]{FE0000} \multirow{-3}{*}{1} &ND & \cellcolor[HTML]{FE0000}0.305  & 71  & ND & ND &42/40  \\  \hline
  
SnS\textsubscript{2}  flake \cite{yang2016controllable} & Chemical vapour deposition  & 400 &2.5 & \cellcolor[HTML]{FE0000}1&  \cellcolor[HTML]{FE0000} 80~pA& \cellcolor[HTML]{FE0000} 1.19 & ND &ND  & $2.35 \times 10^{11}$& $<$1/$<$1 \\  \hline    
   
Ultrathin SnS\textsubscript{2} nanosheet\cite{zhou2016large}   & Chemical vapour deposition & 350 & 6.93 & \cellcolor[HTML]{FE0000} 1 &  \cellcolor[HTML]{FE0000} 15~nA&260  & $9.3 \times 10^{4}$ & ND &  $\times$ $10^{10}$& 20/16 \\      
 \hline

SnS\textsubscript{2} nanosheets\cite{wu2016ultrathin} & Ultrasonic chemical method & 254 & 0.006 & \cellcolor[HTML]{FE0000}& ND& \cellcolor[HTML]{FE0000} 1.95 & \cellcolor[HTML]{FE0000}$9.55 \times 10^{2}$ &ND &\cellcolor[HTML]{FE0000} $2.02 \times 10^{10}$&ND \\ 
& & 365 & 0.004 &\cellcolor[HTML]{FE0000} &ND & \cellcolor[HTML]{FE0000} 0.005 & \cellcolor[HTML]{FE0000}1.75 & ND&\cellcolor[HTML]{FE0000}$5.33 \times 10^{7}$ & ND\\
& & 405 & 162 &\cellcolor[HTML]{FE0000} & 0.05~pA & \cellcolor[HTML]{FE0000} 0.0007 & \cellcolor[HTML]{FE0000}0.22 & ND& \cellcolor[HTML]{FE0000}$7.39 \times 10^{6}$&\cellcolor[HTML]{FE0000}400/200 \\
& & 532 & 19.4 &\cellcolor[HTML]{FE0000} & 0.1~pA&\cellcolor[HTML]{FE0000}  0.024 & \cellcolor[HTML]{FE0000}5.52 & ND&\cellcolor[HTML]{FE0000}$2.48 \times 10^{8}$ &\cellcolor[HTML]{FE0000}360 \\
& & 650 & 20.4 &\cellcolor[HTML]{FE0000} & ND& \cellcolor[HTML]{FE0000} 0.011 & \cellcolor[HTML]{FE0000}2.21 & ND&\cellcolor[HTML]{FE0000}$1.17 \times 10^{8}$ &ND \\
& & 780 & 48.3 &\cellcolor[HTML]{FE0000} &ND & \cellcolor[HTML]{FE0000} 0.001 & \cellcolor[HTML]{FE0000}0.2 & ND& \cellcolor[HTML]{FE0000}$1.35 \times 10^{7}$ & ND\\
& & 850 & 12.2 &\cellcolor[HTML]{FE0000} & ND&\cellcolor[HTML]{FE0000}  0.0009 & \cellcolor[HTML]{FE0000}0.15 &ND & \cellcolor[HTML]{FE0000}$9.93 \times 10^{6}$ & ND\\
& & 980 & 48.9 & \cellcolor[HTML]{FE0000} \multirow{-8}{*}{5} & ND & \cellcolor[HTML]{FE0000}  0.0002 &\cellcolor[HTML]{FE0000} 0.0184 & ND& \cellcolor[HTML]{FE0000}$3.5 \times 10^{6}$ &ND \\ \hline

SnS\textsubscript{2} nanosheet/PbS colloidal quantum dot\cite{gao2016broadband} & Chemical vapour transport& 365 &$\approx 10^{-5}$ & \cellcolor[HTML]{FE0000} & \cellcolor[HTML]{FE0000}  $\approx$0.1$\mu$A& >$10^{5}$ &  ND&$7.23\times 10^{-15}$  & $2.4\times 10^{11}$& ND/20 \\
$V_{\rm G} = 30$ V & & 970 &$\approx 10^{-5}$ & \cellcolor[HTML]{FE0000}  \multirow{-2}{*}{1} & \cellcolor[HTML]{FE0000}  $\approx$0.1$\mu$A & >$10^{5}$& ND  &  $7.89\times 10^{-16}$& $2.2\times 10^{12}$& \cellcolor[HTML]{FE0000} ND/290 \\
\hline

SnS\textsubscript{2} monolayer \cite{li2017two}  & Direct vapour transport & 638 & ND & \cellcolor[HTML]{FE0000} & \cellcolor[HTML]{FE0000} 5~nA  & \cellcolor[HTML]{FE0000}0.088 & ND & ND &  ND & 6/6  \\ 
Fe-doped SnS\textsubscript{2} (Fe\textsubscript{0.021}Sn\textsubscript{0.979}S\textsubscript{2})   &   & 638  & ND & \cellcolor[HTML]{FE0000} \multirow{-2}{*}{3} &\cellcolor[HTML]{FE0000} 10~nA  & \cellcolor[HTML]{FE0000}0.206 & ND & ND & ND & 9/9  \\ \hline

SnS\textsubscript{2} nanoflakes\cite{jia2018thickness}  & Chemical vapour deposition  &  405 & 0.5 & \cellcolor[HTML]{FE0000}&  \cellcolor[HTML]{FE0000} 1~nA& 354.4 & $1.1 \times 10^{5}$  & ND&$ 2 \times 10^{10}$ & 0.4/0.56  \\        
SnS\textsubscript{2} nanoflkes coated with HfO\textsubscript{2} &   & 405 & 0.1 & \cellcolor[HTML]{FE0000}\multirow{-2}{*}{1} &\cellcolor[HTML]{FE0000} 17~nA & 1922 & $5.9 \times 10^{5}$ & ND &ND & 5 \\    \hline
      
SnS\textsubscript{2}\cite{liu2019tunable}    & Direct vapour transport & 520& 110 & \cellcolor[HTML]{FE0000}& ND &\cellcolor[HTML]{FE0000}0.034 & 8.1  & ND&ND & 8   \\ 
1.21\%-Sb-SnS\textsubscript{2}  & &  520 &110 & \cellcolor[HTML]{FE0000}\multirow{-2}{*}{1} & ND& 14.634 & $3.5 \times 10^{3}$ &ND &ND & 24  \\ \hline

SnS\textsubscript{2} nanoflakes \cite{tian2020visible}  &  Chemical vapour transport & 405 & 0.64 & \cellcolor[HTML]{FE0000}& \cellcolor[HTML]{FE0000} 1~nA& 12.34 & ND  & ND & $5.09 \times 10^{9}$ & 22.5/24.1 \\  & & 532& 0.39 & \cellcolor[HTML]{FE0000}\multirow{-2}{*}{5} & \cellcolor[HTML]{FE0000} 0.1~nA& 17.26 & $5.29 \times 10^{3}$ & ND &  $7.12 \times 10^{9}$ & 3.3/4.2 \\  \hline 

Single crystal SnS\textsubscript{2} flake\cite{yu2020giant}  & Tape exfoliation   & & ND & \cellcolor[HTML]{FE0000}&\cellcolor[HTML]{FE0000} 250~nA & 385 & $1.3 \times 10^{5}$ &   ND & $4.5 \times 10^{9}$& \cellcolor[HTML]{FE0000}12000/17000  \\ 
O\textsubscript{2} plasma treated SnS\textsubscript{2} flake &   & \multirow{-2}{*}{350}& ND &\cellcolor[HTML]{FE0000}\multirow{-2}{*}{2} & \cellcolor[HTML]{FE0000} 500~nA & 860 & $3.1 \times 10^{5}$ & ND & $1.1 \times 10^{10}$ &\cellcolor[HTML]{FE0000} 700/600  \\ \hline
    
 SnS\textsubscript{2} thin film\cite{lei2020thermal} & Thermal  & 405 & 270 & \cellcolor[HTML]{FE0000}& \cellcolor[HTML]{FE0000} 70~$\mu$A & \cellcolor[HTML]{FE0000}  0.0445 &ND & ND &ND & \cellcolor[HTML]{FE0000}  2100/2700  \\ 
 (Thermally-annealed)&  evaporation& 532 & 120 &\cellcolor[HTML]{FE0000}&  ND & \cellcolor[HTML]{FE0000}0.015 & ND  & ND& ND & ND  \\
  & & 650 & 180 & \cellcolor[HTML]{FE0000}& ND& \cellcolor[HTML]{FE0000}0.006 & ND  & ND& ND & ND  \\ 
 &  & 780 & 150 & \cellcolor[HTML]{FE0000}& ND& \cellcolor[HTML]{FE0000}  0.007 & ND  & ND& ND & ND \\ 
 &  & 980 & 180 &\cellcolor[HTML]{FE0000}  \multirow{-5}{*}{4.5} &  \cellcolor[HTML]{FE0000} 69~$\mu$A & \cellcolor[HTML]{FE0000}  0.003 & ND  & ND& ND & ND \\ \hline
    
SnS\textsubscript{2} \cite{fan2021enhanced}  & Chemical vapour transport &405 &1.54 &\cellcolor[HTML]{FE0000}&\cellcolor[HTML]{FE0000} 4~nA &1.64  & $5.03 \times 10^{2}$  &ND  &$1.95 \times 10^{11}$  & 0.5/0.5   \\  
 &  &532 &10.3 &\cellcolor[HTML]{FE0000} &\cellcolor[HTML]{FE0000} 4.5~nA &\cellcolor[HTML]{FE0000}0.65  &$1.52 \times 10^{2}$   & ND& $7.74 \times 10^{10}$  & 1/2   \\  
1.9\% In-doped SnS\textsubscript{2}   & &405 &1.54 &\cellcolor[HTML]{FE0000} & \cellcolor[HTML]{FE0000} 120~nA &542.8  & $1.67 \times 10^{5}$  &  ND& $1.3 \times 10^{13}$  & \cellcolor[HTML]{FE0000} 280/160   \\
  & &532 &10.3 &\cellcolor[HTML]{FE0000}  \multirow{-4}{*}{5} &\cellcolor[HTML]{FE0000} 70~nA &43.28  & $1.01 \times 10^{4}$ &   ND& $4.11 \times 10^{12}$  & \cellcolor[HTML]{FE0000} 150/310   \\  \hline 

SnS\textsubscript{2} flakes\cite{fu2021controllable} & Chemical vapour deposition & 520 & ND &  \cellcolor[HTML]{FE0000} 1 &  \cellcolor[HTML]{FE0000} $\approx $4~nA& \cellcolor[HTML]{FE0000} 0.016 & ND & ND & ND & 20/22\\ \hline

SnS\textsubscript{2} nanosheets\cite{shooshtari2021ultrafast} & Chemical vapour deposition &445 & 1.05 & \cellcolor[HTML]{FE0000}  & ND &  \cellcolor[HTML]{FE0000}0.14 & \cellcolor[HTML]{FE0000}41 & ND & $1.33 \times 10^{11}$ & 0.34/0.47 \\ 
SnS\textsubscript{2} nanosheets/perovskite &  &445 & 1.05 &   \cellcolor[HTML]{FE0000} \multirow{-2}{*}{-3} & ND & \cellcolor[HTML]{FE0000}1.84 & \cellcolor[HTML]{FE0000}513 & ND & $1.69 \times 10^{11}$ & 0.0207/0.0314  \\  \hline

SnS\textsubscript{2} flakes \cite{luo2022phase}& Chemical vapour deposition & 405 & 0.29& \cellcolor[HTML]{FE0000} 3 & \cellcolor[HTML]{FE0000} 61.5~nA & 1723.7 & $5.3 \times 10^{5}$ & ND & $6.24 \times 10^{12}$ & 13/51 \\ \hline

\bottomrule
\end{longtable}
}

\end{landscape}

\begin{table}[]
\caption{Comparison of photodetector performance in terms of the responsivity $R$ in the NIR range between the SAW \ch{SnS2} photodetector in this work with other reported 2D photodetectors, including both monolithic and heterojunction materials.}
\label{tab:s2}
\begin{tabular}{lcc}
\\\toprule
\hline 
Material & Illumination wavelength (nm) & $R$ (A/W)\\ \hline
\bf Present work (27~dBm SAW) & 850    &  13.6 \\ 
\ch{PtS2} \cite{wang2020noble} & 830& 0.3\\ 
\ch{NbS3} \cite{wang2020air} & 830& 0.025\\
\ch{PtSe2} \cite{wang2020high} & 850& 0.001\\
\ch{MoTe2} \cite{huang2016highly} &1060 & 0.024 \\
BP \cite{buscema2014fast} & 940& 0.0048 \\
\ch{MoS2}  \cite{wang2015hot}& 1070 & 5.2\\  
\ch{PdSe2} \cite{liang2019high}& 1064 & 708\\
Se\textsubscript{0.32}Te\textsubscript{0.68} \cite{tan2020evaporated} &1200 & 1.8\\
Graphene \cite{mueller2010graphene} & 1550 & 0.0061 \\
GaAs \cite{guo2018few} & 1600& 6 \\
SnTe \cite{jiang2017broadband} &2003 & 3.75 \\
\hline
\ch{MoTe2}/\ch{MoS2} \cite{pezeshki2016electric} & 800 & 0.322\\
\ch{PtSe2}/GaAs  \cite{zeng2018fast} & 808 &0.262 \\
Graphene/Si \cite{an2013tunable} & 850-900 & 0.435\\
\ch{MoS2}/Graphene/\ch{WSe2} \cite{long2016broadband}  & 940 & 0.306\\
\ch{MoSe2}/Si \cite{john2020broadband} & 1100 & 0.522 \\

\hline
\bottomrule
\end{tabular}
\end{table}

\bibliography{SnS2}